\documentclass[11pt, a4paper]{amsart}
\usepackage{amssymb,amsmath,amsfonts,latexsym,amsthm,mathtools}
\usepackage{tikz}
\usepackage{enumitem}
\usepackage{booktabs}
\usepackage[margin=1.02in]{geometry}
\usepackage[colorlinks=true,linkcolor=blue,citecolor=blue]{hyperref}
\newtheorem{Theorem}{Theorem}[section]
\newtheorem{Lemma}[Theorem]{Lemma}
\newtheorem{Proposition}[Theorem]{Proposition}
\newtheorem{Corollary}[Theorem]{Corollary}
\newtheorem{Conjecture}[Theorem]{Conjecture}
\newtheorem{theoremletter}{Theorem}

\theoremstyle{definition}
\newtheorem{Definition}[Theorem]{Definition}
\newtheorem{Example}[Theorem]{Example}
\newtheorem{Remark}[Theorem]{Remark}
\newtheorem{Question}[Theorem]{Question}
\numberwithin{equation}{section}

\newcommand{\Mbar}[1]{\overline{M}_{0,#1}}
\def\Mo{\mathcal{M}}
\def\Po{{\mathfrak P}}
\def\pl{{\mathfrak p}_\ell}
\def\m{{\mathfrak m}}
\def\hm{\hat{\mathfrak m}}
\def\R{{\mathbf R}}
\def\Z{{\mathbf Z}}
\def\C{{\mathbf C}}
\def\HH{{\mathbf H}}
\def\Hh{{\mathcal H}_\ell}
\def\kp{\varkappa}
\def\Sim{\operatorname{Sim}}
\def\Isom{\operatorname{Isom}}
\def\Mob{\operatorname{M\ddot ob}}
\def\Conf{\operatorname{Conf}}
\def\Symz{{\mathcal S}_\ell}
\def\im{\operatorname{im}}
\def\ren{{\mathrm{ren}}}
\def\bub{{\mathrm{bub}}}
\def\bas{{\mathrm{base}}}
\def\E{{\mathcal E}}
\def\Th{\Theta}
\def\cent{\operatorname{cent}}
\def\scal{\operatorname{scal}}
\def\Ric{\operatorname{Ric}}
\def\Area{\operatorname{Area}}
\def\ext{\mathbf{ext}}
\providecommand{\MopsL}{\overline{\Mo}^{\,\mathrm{ps}}_\ell(S^n)}
\providecommand{\Xss}{X^{\mathrm{ss}}}
\providecommand{\Xst}{X^{\mathrm{s}}}
\providecommand{\Gmod}{\mathcal G_{n,\ell}(s)}
\providecommand{\Polys}{\mathcal P_\ell(s)}
\providecommand{\pt}{\operatorname{pt}}
\providecommand{\pAB}{{\mathfrak p}_{A|B}}
\newcommand{\conj}{\operatorname{conj}}
\usetikzlibrary{calc,shapes.geometric,decorations.pathreplacing}
\tikzset{
  vtx/.style={circle,fill=black,inner sep=1.6pt},
  leaf/.style={circle,draw,fill=white,inner sep=1.4pt},
  blob/.style={draw,shape=ellipse,minimum width=20mm,minimum height=11mm,fill=black!5},
  br/.style={draw,shape=ellipse,fill=black!5}
}
\def\vp{\varpi}
\let\phi=\varphi
\let\epsilon=\varepsilon
\dedicatory{Dedicated to the memory of Misha Kapovich}
\begin{document}

\title[Canonical metrics on punctured spheres, II]
      {Canonical metrics on punctured spheres, II:
        \\
compactified moduli spaces of black holes}

\author{Boris Botvinnik}

\address{Department of Mathematics,
  University of Oregon, Eugene, OR 97403, USA}
\email{botvinn@uoregon.edu}


\begin{abstract}
Let $g_{\pl}$ denote the canonical scalar-flat metric on
$S^n\setminus\pl$ introduced in Part~I, which for $n=3$ corresponds to
the Brill--Lindquist initial data for $\ell$ black holes, and let
$\eta\in\Hh$ be its associated mass matrix. In this paper, we
investigate the asymptotic behavior of the metrics and the mass matrix
$\eta$ as subsets of punctures collide. Such collisions cause every
mass to diverge; if a cluster $A$ of black holes collapses at a
relative scale $t$, the mass matrix satisfies
$\eta=\E(A)\ln(1/t)+O(1)$, diverging along an explicit ray
$\E(A)\in\Hh$. For nested families of clusters, this degeneracy
structure naturally organizes into a space of rooted
trees. Subtracting the divergence gives a renormalized mass matrix
which extends smoothly to every stratum of the Axelrod--Singer
compactified moduli space $\overline{\mathcal
  M}{}^{\,\mathrm{AS}}_\ell=\overline C_{\ell-1}(\R^n)/\mathrm O(n)$
and splits naturally into one mass matrix per vertex of the bubble
tree.

Geometrically, at each vertex carrying punctures the suitably
rescaled metrics converge to a complete scalar-flat metric with the
nodes filled in, adjacent such pieces being joined by Schwarzschild
necks, and the renormalized matrix acting as its mass data. For $n=3$,
the necks contain minimal spheres, which are common apparent horizons
of the merging holes. At vertices bearing two holes we determine on
which side of the limiting horizon each node lies, and we formulate
as a conjecture the resulting criterion for a neck to be a component
of the outermost horizon. We also compare the mass lower bound of
Part~I with the Riemannian Penrose inequality. Because the mass matrix is invariant under rotations of
the bubbles, it lives on a Deligne--Mumford-type quotient of the
compactification.
\end{abstract}

\vspace*{-10mm}

\maketitle
\vspace*{-12mm}

\renewcommand{\baselinestretch}{.7}
\setcounter{tocdepth}{1}
\begin{small}

\tableofcontents
\end{small}
\vspace*{-10mm}

\renewcommand{\baselinestretch}{1}
\section{Introduction}\label{s0}

\subsection{Black holes at rest}
Brill--Lindquist's paper \cite{BL} describes $\ell$ ``black holes at
rest,'' i.e., a scalar-flat metric which is asymptotically flat
near each black hole. This metric depends on scalar parameters which
can be chosen so that the metric depends only on the configuration of
the punctures modulo M\"obius transformations $\Mob(S^3)$. Then the
ADM masses become functions on the moduli space
$\mathcal{M}_\ell(S^3)$ of $\ell$ points on the sphere modulo the
group $\Mob(S^3)$. The same construction gives, for every $n\ge3$, a
canonical complete scalar-flat metric on $S^n$ minus $\ell$
points. This normalization has an important property: every mass
diverges when the black holes collide, including the masses of black
holes far from the collision (Theorem~\ref{thm:mainA}). This paper
determines what the divergence is, and what survives it.

We show that the divergence is universal: logarithmic, with explicit
rational exponents that depend only on which black holes merge and on
how many are left as spectators (Theorem~\ref{thm:mainA}). Subtracting
the divergence leaves a finite part which extends smoothly to the
strata of the Axelrod--Singer compactification of the moduli space
(Theorem~\ref{thm:mainB}), and which splits into one mass matrix per
vertex of the bubble tree and one length per edge
(Theorem~\ref{thm:mainC}). This finite part is the mass data of the
limiting metrics: once suitably rescaled, the metrics converge at each
vertex of the tree that carries black holes to a complete scalar-flat
metric whose ends are the black holes attached to that vertex,
adjacent such pieces being joined by Schwarzschild necks
(Theorem~\ref{thm:mainD}).

\begin{figure}[ht]
\centering
\begin{tikzpicture}[scale=0.9,every node/.style={font=\scriptsize}]
\begin{scope}
\draw (0,0) circle (1.4);
\fill (0.2,0.9) circle (1.6pt) node[above right]{$4$};
\fill (-0.35,-1.05) circle (1.6pt);
\node at (-0.55,-1.00){$5$};
\fill (-0.85,0.4) circle (1.6pt) node[right]{$6$};
\fill (0.95,-0.12) circle (1.6pt) node[right]{$3$};
\fill (0.58,-0.42) circle (1.3pt);
\fill (0.66,-0.50) circle (1.3pt);
\draw[dashed] (0.72,-0.30) circle (0.36);
\draw[dashed] (0.62,-0.46) circle (0.12);
\node at (0.22,-0.78){$1,2$};
\node at (0,-1.95){(a)\ \ $A=\{1,2,3\}\supset B=\{1,2\}$};
\end{scope}
\begin{scope}[xshift=4.6cm]
\coordinate (r) at (0,0); \coordinate (v) at (1.2,0); \coordinate (w) at (2.4,0);
\draw (r)--(v)--(w);
\fill (r) circle (2pt); \fill (v) circle (2pt); \fill (w) circle (2pt);
\draw (r)--(-0.8,0.7) node[left]{$4$};
\draw (r)--(-0.9,0) node[left]{$5$};
\draw (r)--(-0.8,-0.7) node[left]{$6$};
\draw (v)--(1.2,0.8) node[above]{$3$};
\draw (w)--(3.2,0.6) node[right]{$1$};
\draw (w)--(3.2,-0.6) node[right]{$2$};
\node[above] at (0.6,0.05){$t_A$}; \node[above] at (1.8,0.05){$t_B$};
\node[below] at (r){$r$}; \node[below] at (v){$v$}; \node[below] at (w){$w$};
\node at (1.2,-1.95){(b)\ \ the bubble tree};
\end{scope}
\begin{scope}[xshift=9.6cm]
\draw[thick] (0,0) ellipse (1.0 and 0.6);
\draw[thick] (2.4,0) ellipse (0.75 and 0.45);
\draw[thick] (4.3,0) ellipse (0.75 and 0.45);
\draw[thick] (0.98,0.12) .. controls (1.3,0.06) and (1.4,0.06) .. (1.67,0.10);
\draw[thick] (0.98,-0.12) .. controls (1.3,-0.06) and (1.4,-0.06) .. (1.67,-0.10);
\draw[thick] (3.13,0.10) .. controls (3.35,0.05) and (3.4,0.05) .. (3.57,0.09);
\draw[thick] (3.13,-0.10) .. controls (3.35,-0.05) and (3.4,-0.05) .. (3.57,-0.09);
\node at (0,0){$g^{(r)}$}; \node at (2.4,0){$g^{(v)}$}; \node at (4.3,0){$g^{(w)}$};
\node[below] at (0,-0.62){ends $4,5,6$};
\node[below] at (2.4,-0.47){end $3$};
\node[below] at (4.3,-0.47){ends $1,2$};
\node[above] at (1.33,0.08){$S_{e_A}$}; \node[above] at (3.35,0.08){$S_{e_B}$};
\node at (2.2,-1.95){(c)\ \ the geometric limit};
\end{scope}
\end{tikzpicture}
\caption{{\small Here we see a degeneration of six punctures: cluster $A$
  collapses at relative scale $t_A$, and within it, cluster $B$
  collapses at relative scale $t_B$ (a). The nested family $\{A,B\}$
  forms a rooted tree with one internal edge per cluster (b). At each
  vertex bearing black holes, the rescaled metrics converge to a
  complete scalar-flat metric whose ends correspond precisely to those
  black holes, while the nodes act as regular points. Adjacent
  geometric components are joined by Schwarzschild necks whose minimal
  spheres $S_e$ shrink to points with respect to the normalization on
  either side (c); see Theorem~\ref{thm:mainD}. The renormalized mass
  matrix encodes the mass data of the three components, while the
  divergent part captures the relative scales $t_A$ and $t_B$
  (Theorems~\ref{thm:mainB} and~\ref{thm:mainC}).}}
\label{fig:intro}
\end{figure}
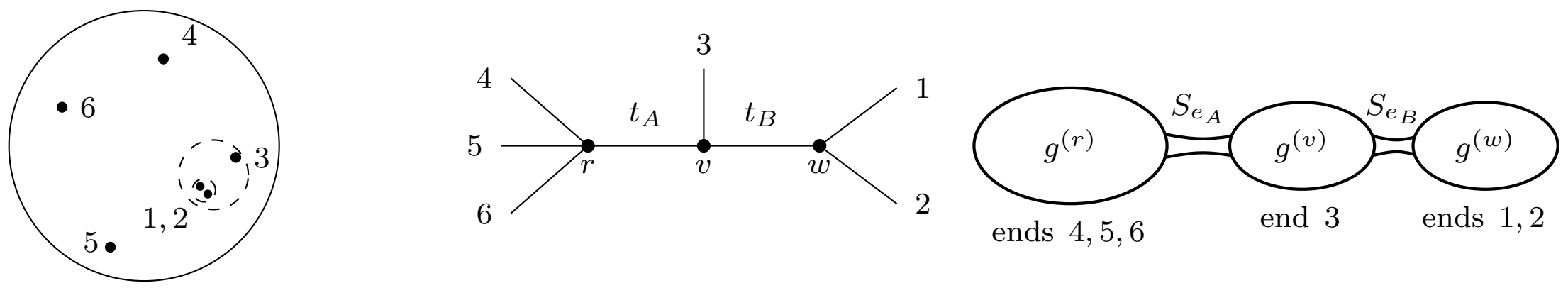
For $n=3$, the necks contain minimal spheres, which are common
apparent horizons for the merging black holes. At vertices bearing two
black holes, we determine on which side of the limiting horizon each
node lies (Theorem~\ref{thm:mainE}); that this decides whether a neck
is a component of the outermost horizon is stated as a conjecture.
Finally, because the mass matrices are invariant under the relative
rotations of the bubbles, they descend to a Deligne--Mumford-type
quotient of the Axelrod--Singer compactification. For $\ell\ge6$, on
the other hand, the polystable compactification introduced in Part~I
is too coarse to record them (Theorem~\ref{thm:mainF}).

\subsection{The setting}
Let $h$ be the round metric on $S^n$, $n\ge3$, let
$\pl=(p_1,\dots,p_\ell)$ be an ordered configuration of $\ell\ge4$
distinct points, and let $\Mo_\ell(S^n)=\Conf_\ell(S^n)/\Mob(S^n)$ be
the moduli space of such configurations modulo the M\"obius group.
The canonical metric of Part~I is
\[
g_{\pl}=\Phi_{\pl}^{4/\kp}h,\qquad
\Phi_{\pl}=\sum_{i=1}^\ell c_i\,|x-p_i|^{-\kp},\qquad \kp=n-2,
\]
complete, scalar-flat and asymptotically flat with one end at each
puncture, the constants $c_i=e^{f_i}$ being fixed by the \emph{mass
normalization}
\[
\eta_{ij}=-\kp\ln|p_i-p_j|+f_i+f_j\ \ (i\ne j),\qquad
\sum_{j\ne i}\eta_{ij}=0 .
\]
The \emph{mass matrix} $\eta=(\eta_{ij})$ is a complete conformal
invariant of $[\pl]$, lying in the space $\Hh$ of symmetric matrices
with zero diagonal and zero row sums, and it encodes the masses: if
$\m_i$ is the coefficient of $\rho^{2-n}$ in the expansion of
$g_{\pl}$ at the $i$-th end, then
\[
\hm_i:=\tfrac{\kp}{4}\,\m_i=\sum_{j\ne i}e^{\eta_{ij}} .
\]
The mass map $\Xi\colon\Mo_\ell(S^n)\to\Hh$, given by
$[\pl]\mapsto\eta$, is injective and proper, and is therefore a homeomorphism onto a
closed image \cite[Theorem~A]{PartI}. Part~I also described a second
normalization, in which a configuration is realized as the Gauss image
of an equilateral polygon in $\HH^{n+1}$. The space of such polygons
is compact and corresponds to the polystable quotient $\MopsL$ of
Kapovich, Leeb, and Millson, which serves as a compactification of
$\Mo_\ell(S^n)$ \cite[Theorem~B and Corollary~4.11]{PartI}. The mass
normalization blows up when black holes collide, whereas the polygon
normalization does not; establishing the precise relationship between
the two at the boundary is one of the main results of this paper
(Theorem~\ref{thm:mainF}).

\subsection{The case \texorpdfstring{$n=3$}{n=3}}
In units where $G=c=1$ and within a stereographic chart placing
$p_\ell$ at infinity, the metric $g_{\pl}$ assumes the
Brill--Lindquist form
\[
g=\Big(1+\sum_{i<\ell}\frac{m_i}{2\,|y-y_i|}\Big)^{4}|dy|^2
\]
(see Lemma~\ref{L:chart}). Here, the end at infinity corresponds to
the puncture $p_\ell$. Consequently, the Brill--Lindquist data for $N$
holes coincides with the canonical metric for $\ell=N+1$ punctures,
where all ends are treated symmetrically. The ADM mass of the $i$-th
end is given by $M_i=2\hm_i$. In terms of the Brill--Lindquist
parameters, this is expressed as $M_i=m_i\big(1+\sum_{j\ne
  i}m_j/2|y_i-y_j|\big)$ for $i<\ell$ and
$M_\ell=\sum_{i<\ell}m_i$. Translated into the pair-mass notation of
Part~I, both expressions take the unified form $M_i=2\sum_{j\ne
  i}\xi_{ij}$, where $\xi_{ij}=\tilde c_i\tilde c_j/d_{ij}$ with $d_{ij}=|p_i-p_j|$
the chordal distance on $S^3$. The
canonical gauge fixes the bare masses $m_i$ as explicit functions of
the positions \cite[Section~2]{PartI}; in particular, they are \emph{not}
held constant along a degeneration. The divergence of the masses $M_i$
asserted below is therefore a consequence of the canonical gauge,
while the primary content of Theorems~\ref{thm:mainB}--\ref{thm:mainD}
is the precise identification of the gauge-independent remainder.

\subsection{Main results}

We say that a cluster $A\subseteq\{1,\dots,\ell\}$ of $k$ points
\emph{collapses at scale $\epsilon\to0$} if, in a fixed stereographic
chart, the points of $A$ take the form $\epsilon a_i+o(\epsilon)$ for
distinct points $a_i$, while the remaining points converge to distinct
limits strictly bounded away from the cluster. We restrict our
attention to the range $2\le k\le\ell-2$, as these are the only cases
that yield non-trivial asymptotics.

To each such cluster $A$, we associate a \emph{ray} $\E(A)\in\Hh$,
which is a symmetric matrix depending only on $A$, $\ell$ and $n$. This matrix has positive entries on the diagonal blocks
$A\times A$ and $A^c\times A^c$, and negative entries on the
off-diagonal blocks, satisfying the symmetry $\E(A)=\E(A^c)$. The
explicit entries of $\E(A)$ are given in \eqref{eq:ray}, and they
completely determine the exponents $\theta_i$ discussed below. A
family $\mathcal{A}$ of such subsets is called \emph{nested} if any
two of its members are either disjoint or comparable under inclusion,
and \emph{reduced} if no two members are equal or complementary. It is
natural to identify a reduced nested family with a tree $T$ having
$\ell$ labelled leaves and exactly one internal edge $e_A$
corresponding to each set $A\in\mathcal{A}$. We denote by $\ell_v$ the
valence of an internal vertex $v$ of $T$.

\begin{theoremletter}[universal divergence]\label{thm:mainA}
Let $\ell\ge4$ and $n\ge3$.
\begin{enumerate}[label=\textup{(\roman*)},leftmargin=*]
\item If a cluster $A$ of $k$ points, $2\le k\le\ell-2$, collapses at
scale $\epsilon$, then
\[
\eta=\E(A)\ln(1/\epsilon)+O(1),\qquad
\hm_i\asymp\epsilon^{-\theta_i},\qquad
\theta_i=\begin{cases}
\dfrac{(n-2)(\ell-k)(\ell-k-1)}{(\ell-1)(\ell-2)}, & i\in A,\\[3mm]
\dfrac{(n-2)\,k(k-1)}{(\ell-1)(\ell-2)}, & i\in A^c .
\end{cases}
\]
In particular the mass of every black hole diverges, and the exponent
$\theta_i$ of a black hole inside the cluster is determined by the
number of spectators.
\item
Suppose the punctures collide according to the hierarchy of a nested
family $\mathcal{A}$. Let $\epsilon_A$ denote the spatial scale of a
cluster $A\in\mathcal{A}$, such that the distance between any two
punctures $p_i$ and $p_j$ satisfies
$\vert{}p_i-p_j\vert{}\asymp\epsilon_{A(i,j)}$, where $A(i,j)$ is the
smallest cluster containing both points $p_i$ and $p_j$. For each
cluster $A$, we define the relative scale
$t_A=\epsilon_A/\epsilon_{\hat{A}}$, where $\hat{A}$ denotes the
parent cluster of $A$. Under these conditions, the mass matrix admits
the asymptotic expansion
\[
\eta=\sum_{A\in\mathcal{A}}\ln(1/t_A)\,\E(A)+O(1).
\]
The $O(1)$ remainder is bounded in terms of the implied constants in
$|p_i-p_j|\asymp\epsilon_{A(i,j)}$ only.
\item For a reduced nested family $\mathcal A$ the rays $\E(A)$,
  $A\in\mathcal A$, are linearly independent, and the cones
  $\sigma_{\mathcal A}=\sum_{A\in\mathcal A}\R_{>0}\E(A)$ form a fan in
  $\Hh$ isomorphic to the space of phylogenetic trees of
  Billera--Holmes--Vogtmann \cite{BHV}, i.e.\ to the tropical
  Grassmannian of lines of Speyer--Sturmfels \cite{SS}.
\end{enumerate}
\end{theoremletter}
Thus no rescaling of the masses extends across the boundary, but the
mass map \emph{tropicalizes}: along every degeneration the normalized
mass matrix $\eta/\Vert{}\eta\Vert{}$ converges to a point of the fan
of (iii), and the exponents $\theta_i$ are the tropicalizations of the
rows of $\E(A)$ (Remark~\ref{R:tropic}). For $\ell=5$ the link of the
fan is the Petersen graph, and a merging pair diverges with three
times the exponent of the spectators (Section~\ref{ss:petersen}). This
is the real counterpart of the tropicalization of
$\mathcal{M}_{0,\ell}$ to the same fan \cite{SS,Te}: for every $n$, the
real moduli space $\mathcal{M}_\ell(S^n)$ realizes it through the mass
map.

The divergence we described can be subtracted. Let
$\overline{\mathcal{M}}{}^{\,\mathrm{AS}}_\ell(S^n)=\overline
C_{\ell-1}(\R^n)/\mathrm{O}(n)$ denote the Axelrod--Singer
compactification of $\mathcal{M}_\ell(S^n)$ \cite{AS,FM,Si}. This
space is compact and stratified by reduced nested families; away from
points with positive-dimensional stabilizers it is an orbifold with
corners. These corners admit charts $(\sigma,t)$ comprising one
\emph{screen} $\sigma_v$ per internal vertex of the tree (representing
a configuration of branches at $v$ modulo translation and dilation)
and one small relative scale $t_A\ge0$ on each edge, see
Section~\ref{s:AS}.
\begin{theoremletter}[the renormalized mass matrix]\label{thm:mainB}
Let $\mathcal{A}$ be a reduced nested family and let $(\sigma,t)$
denote the corner chart of its stratum in the moduli space
$\overline{\mathcal{M}}{}^{\,\mathrm{AS}}_\ell(S^n)$. Then the
matrix-valued function
\[
\eta-\sum_{A\in\mathcal{A}}\ln(1/t_A)\,\E(A),
\]
defined for $t\in(0,\delta)^{\mathcal{A}}$, extends smoothly to the
corner $t=0$. Its boundary value $\eta^\ren$ lies in the subspace
$W_{\mathcal{A}}\subseteq\Hh$ consisting of matrices for which the
block coupling each cluster $A\in\mathcal{A}$ to its complement is
additive (meaning $\zeta_{ij}=\mu_i+\mu_j$ for $i\in A$ and $j\notin
A$). Furthermore, this boundary value is canonical modulo the subspace
$V_{\mathcal{A}}=\operatorname{span}\{\E(A):A\in\mathcal{A}\}$, which
fixes the ambiguity in the boundary defining functions, and it remains
invariant under independent rotations of the screens.
\end{theoremletter}

This means, in particular, that the mass map is a smooth function on
$\overline\Mo^{\,\mathrm{AS}}_\ell$ with logarithmic singularities
along the boundary hypersurfaces, with the constant residue $\E(A)$
along the hypersurface of the split $\{A,A^c\}$. The boundary value
$\eta^\ren$ is the restriction to the corresponding stratum of a smooth
function on the chart, see Theorem~\ref{T:bv}.

\begin{theoremletter}[factorization]\label{thm:mainC}
For every reduced nested family $\mathcal{A}$ with associated tree
$T$, the gluing map $
\gamma_{\mathcal{A}}\colon\ \bigoplus_{v}\mathcal{H}_{\ell_v}\oplus\R^{\mathcal{A}}
\longrightarrow W_{\mathcal{A}}, $ where the direct sum ranges over
the internal vertices of $T$, is a linear isomorphism. This map
propagates a mass matrix at each vertex out to the leaves and
incorporates the corresponding rays; consequently, $\dim
W_{\mathcal{A}}=\sum_v\dim\mathcal{H}_{\ell_v}+\vert{}\mathcal{A}\vert{}$.
\end{theoremletter}
In geometric terms, once the universal divergence is subtracted, the
clusters decouple. The surviving data at a stratum of codimension $r$
consist of exactly one mass matrix per vertex of the bubble tree, with
the $r$ divergent coordinates corresponding to its edge lengths. This
reflects the mass-matrix analogue of the product structure
$\prod_v\overline{\mathcal{M}}_{0,\ell_v}$ on a boundary stratum of
the Deligne--Mumford space.  The explicit procedure for recovering the
tree, the edge lengths, and the vertex data from a given mass matrix
is detailed in Section~\ref{s7}, beginning with the neighbor-joining
algorithm from phylogenetics (Remark~\ref{R:NJ}).

The metrics converge accordingly. Denote by
$\Th:=\sum_A\ln(1/t_A)\E(A)$ the divergent matrix; its entries are
the table of rescalings needed to see the various pieces of the limit.

\begin{theoremletter}[the geometric limit]\label{thm:mainD}
Let $v$ be a vertex of the bubble tree carrying at least one
puncture. As $t\to0$, the metrics $g_{\pl}$, rescaled by an explicit
factor $\lambda_v$ \textup{(}see \eqref{eq:Nv}\textup{)}, converge in
$C^\infty_{loc}$ to a complete scalar-flat metric $g^{(v)}$. The
asymptotically flat ends of $g^{(v)}$ are exactly the punctures
attached to $v$, with the nodes acting as regular points; if $v$
carries a single puncture, $g^{(v)}$ is flat. If $v$ and an adjacent
vertex both carry punctures, the two components are joined by a
Schwarzschild neck, whose minimal sphere shrinks to a point in the
normalization of either side. If $v$ carries at least two punctures
$i\ne i'$, then $\lambda_v=e^{-2\Th_{ii'}/\kp}$, so that the black holes
at $v$ have unit size, and the masses converge after the subtraction:
\[
e^{-\Th_{ii'}}\,\hm_i(g_{\pl})\longrightarrow
\hm_i(g^{(v)})=\sum_{j}e^{\eta^\ren_{ij}},
\]
where the sum ranges
over the other punctures at $v$.
\end{theoremletter}
For $n=3$, expressed in terms of the Brill--Lindquist masses: if a
black hole $i$ is located at a vertex $v$ alongside at least one other
black hole, then
\[
M_i(g_{\pl})=e^{\Th_{ii'}}\,M_i\big(g^{(v)}\big)\,(1+o(1)),\qquad
e^{\Th_{ii'}}\to\infty,
\]
where $g^{(v)}$ is the near-zone data of the holes at $v$: the clusters
below $v$ collapsed to points, the holes above $v$ receded to infinity,
and the weights given by \eqref{eq:weights}. The factor $e^{\Th_{ii'}}$
depends only on the tree and the relative scales, and the necks are
Schwarzschild throats whose masses tend to zero in the normalization of
either side (Proposition~\ref{P:neck}, Remark~\ref{R:pinch}). At a
vertex without black holes the geometric limit depends on the path to
the corner, whereas $\eta^\ren$ does not (Remark~\ref{R:leafless});
this is why $\eta^\ren$, not the limit metric, is the object attached
to a boundary stratum.

For $n=3$, the metric $g_{\pl}$ is time-symmetric initial data, and
its apparent horizons are its minimal surfaces \cite{Gi}. Let
$\Sigma_i$ be the outermost minimal surface of the end $i$; by
Huisken--Ilmanen \cite{HI} it is a smooth, outer-minimizing union of
spheres enclosing every closed minimal surface, and we write $\ext_i$
for its exterior region.

\begin{theoremletter}[necks and horizons]\label{thm:mainE}
Let $n=3$.
\begin{enumerate}[label=\textup{(\roman*)},leftmargin=*]
\item Along any degeneration, for every internal edge $e$ of the
  bubble tree both of whose endpoints carry holes, separating a
  cluster $B$ from its complement, and at all sufficiently small
  scales, the metric $g_{\pl}$ admits a minimal
  sphere $S_e$. This sphere is obtained from the strictly stable
  horizon of the limiting Schwarzschild neck via the implicit function
  theorem. It serves as a common apparent horizon for the holes in
  $B$: it is enclosed by $\Sigma_i$ for every end $i\notin B$, and,
  from the opposite side, by $\Sigma_j$ for every $j\in B$. Its area
  tends to zero with respect to the normalization on either side.
\item Let $v$ be a vertex bearing exactly two holes, $i$ and
  $i'$. Then the limiting component $g^{(v)}$ is the spatial
  Schwarzschild metric, and its horizon is an Apollonius sphere. A
  node $a_c$ at $v$ lies in $\ext_i^{(v)}$, exactly on the horizon, or
  in $\ext_{i'}^{(v)}$, depending on whether the explicit ratio $R_c$
  from \eqref{eq:R} is strictly greater than, equal to, or strictly
  less than $1$.
\item If a single cluster of $\ell-2$ holes collapses, the root
  component is the Schwarzschild metric joining the two spectator
  holes, and the node of the cluster lies \emph{exactly} on its
  horizon. This holds universally for every shape of the cluster,
  every position of the node, and every value of $\ell$.
\end{enumerate}
\end{theoremletter}
In part (iii) the root carries one spectator on each side of the
Apollonius sphere, and the exponent that decides the side of the node
in \eqref{eq:R} vanishes exactly when $|A|=\ell-2$. Part (ii) is a
statement about the limit pieces. If, as we conjecture, it decides
whether a neck sphere is a \emph{component} of $\Sigma_i$ or is hidden
behind it, then this is a property of the screen rather than of the
tree, and both outcomes occur within a single boundary stratum. For
instance, when a pair of holes merges in the presence of three
spectators, the node of the pair lies inside the spectators' common
limiting horizon if the pair is closer than approximately eight
spectator separations, and outside it beyond that distance, see
Example~\ref{E:horizons}. We do not, however, establish the
convergence of the outermost minimal surfaces along the degeneration.
The two statements that depend on it are formulated as conjectures in Section~\ref{ss:horizons}, with their
supporting arguments: a visibility criterion for the necks in terms of
the limit pieces $g^{(v)}$ (Conjecture~\ref{C:vis}), and the behaviour
at the boundary of the \emph{Penrose ratio}
$P_i=|\Sigma_i|/64\pi\hm_i^{\,2}\in(0,1)$, the canonical gauge turning the
scale-invariant Riemannian Penrose inequality \cite{HI,Bray} into a
function on the moduli space (Conjecture~\ref{C:penrose}).
Finally, we compare three distinct compactifications of
$\Mo_\ell(S^n)$. The first is the Axelrod--Singer
compactification. The second is its quotient
$\overline\Mo^{\,\mathrm{DM}}_\ell$, obtained by taking the screens of
a boundary point modulo independent rotations---that is, modulo the
automorphisms of each bubble that fix its node. This space serves as
the analogue of the Deligne--Mumford compactification, with which it
shares its tree-indexed strata; whether it is Hausdorff remains an
open problem (see Question~\ref{Q:DM}). The third is the polystable
quotient $\MopsL$ introduced in Part~I. This corresponds to the
Kapovich--Leeb--Millson space of equilateral polygons, which for $n=2$
coincides with the Deligne--Mostow GIT quotient with equal weights
\cite{KLM,DM}. Finally, we define an internal vertex of a tree to be a
\emph{centroid} if every branch emanating from it carries at most
$\ell/2$ leaves.

\begin{theoremletter}[three compactifications]\label{thm:mainF}
  $\mbox{ \ }$
\begin{enumerate}[label=\textup{(\roman*)},leftmargin=*]
\item There are continuous surjections which restrict to the identity
  on the main stratum $\Mo_\ell(S^n)$:
\[
\overline\Mo^{\,\mathrm{AS}}_\ell(S^n)\longrightarrow
\overline\Mo^{\,\mathrm{DM}}_\ell(S^n)\longrightarrow\MopsL .
\]
The second map retains only the screen at the centroid vertex of a
bubble tree. Specifically, it maps the open stratum of a tree $T$ onto
the stratum of $\MopsL$ indexed by the partition of the leaves into
the branches at the centroid. If $T$ contains an edge $e_A$ that
splits the leaves into two equal halves, the stratum maps instead to
the point $\pt_{A\vert{}A^c}$.
\item The renormalized mass matrix $\eta^\ren$ is constant on the
  fibres of the first map. Consequently, the mass data described in
  Theorems~\ref{thm:mainB} and \ref{thm:mainC} descends to the
  quotient $\overline\Mo^{\,\mathrm{DM}}_\ell$.
\item For $\ell\ge6$ the mass data do not descend to $\MopsL$: there
  are points of one open boundary stratum with the same image in
  $\MopsL$ whose vertex mass matrices (the components of $\eta^\ren$ in
  $\bigoplus_v\mathcal H_{\ell_v}$, Theorem~\ref{thm:mainC}) differ. For
  $\ell=4,5$, on the other hand,
  every vertex of valence at least $4$ is a centroid, and the vertex
  mass matrices are determined by the image in $\MopsL$.
\end{enumerate}
\end{theoremletter}
In short, the compactification adapted to the metrics is the bubble-tree
compactification; the mass data descend to its Deligne--Mumford quotient,
and for $\ell\ge6$ not to the polystable quotient.

\subsection*{Relations to other work}
\emph{Geometric analysis.} For one point $p$, the metric $G_p^{4/(n-2)}h$, with
$G_p$ the Green's function of the conformal Laplacian
$\frac{4(n-1)}{n-2}\Delta_h+\scal_h$, is scalar-flat and asymptotically flat, and
the positivity of its mass completed the solution of the Yamabe problem
\cite{S84,LP}. The metric $g_{\pl}$ is the $\ell$-point version, with the weights
of the Green's functions fixed conformally. Complete metrics of constant
\emph{positive} scalar curvature on $S^n\setminus\{p_1,\dots,p_\ell\}$ in the round
conformal class were constructed by Schoen \cite{S88}, and their moduli spaces
were studied by Mazzeo--Pollack--Uhlenbeck \cite{MPU} and Mazzeo--Pacard
\cite{MPa}. The scalar-flat problem is the degenerate member of that family, with
Schwarzschild ends in place of Delaunay ends; here the metric is explicit, and so
is its degeneration. Habermann and Jost \cite{HJ} built canonical metrics on
locally conformally flat manifolds from the Green's function of the conformal
Laplacian and studied their behaviour under surgery-type degenerations.

\emph{Relativity.} For $n=3$ the metrics $g_{\pl}$ are the Brill--Lindquist data
\cite{BL}, on which the puncture method of numerical relativity is built
\cite{BB}. For time-symmetric data the apparent horizons are the minimal surfaces
\cite{Gi}; the outermost one is smooth and outer-minimizing, and its area is
bounded by the mass \cite{HI,Bray}. When two Brill--Lindquist holes share a
common apparent horizon was determined numerically \cite{Ca,Bi,Alc,Th}.
Chru\'sciel and Mazzeo \cite{CM} showed that for many-black-hole data with
sufficiently small holes the outermost apparent horizon has several components.
Theorem~\ref{thm:mainE} and Conjecture~\ref{C:vis} concern the opposite regime,
in which the holes merge. In astrophysical terms, Theorem~\ref{thm:mainD} says
that for black holes at rest the hierarchical description of a multiple system
becomes exact in the limit, with the bubble tree as the hierarchy.

\emph{Moduli spaces.} For $n=2$, $\Mo_\ell(S^2)$ is $\mathcal M_{0,\ell}$ modulo
complex conjugation, and the three compactifications of Theorem~\ref{thm:mainF}
are the analogues of three standard compactifications of $\mathcal M_{0,\ell}$:
the Fulton--MacPherson space in the real form of Axelrod--Singer and Kontsevich
\cite{FM,AS,Ko,Si}; the Deligne--Mumford--Knudsen space \cite{DeMu,Kn}, whose real
points were studied by Devadoss and by Etingof--Henriques--Kamnitzer--Rains
\cite{Dev,EHKR}; and the Deligne--Mostow quotient \cite{DM}, which for general $n$
is the Kapovich--Leeb--Millson space of polygons \cite{KLM}. The pattern of
Theorems~\ref{thm:mainB}--\ref{thm:mainD} (a divergence with universal exponents
when points collide, a finite part obtained by subtracting it, and a factorization
of that finite part over the vertices of a tree) is the pattern of the operator
product expansion and of the factorization axiom of two-dimensional conformal
field theory on the boundary of $\overline{\mathcal M}_{0,\ell}$; here the
factorization is exact and elementary.

\subsection*{Plan of the paper} Section~\ref{s:recall}
reviews the necessary background from Part~I and establishes the
stereographic chart in which all computations are
performed. Section~\ref{s4} proves Theorem~\ref{thm:mainA}, covering
the rays, the convergence rates, superposition along nested families,
linear independence, and the tree space, concluding with the Petersen
fan for $\ell=5$. Section~\ref{s:AS} details the Axelrod--Singer
compactification and its corner charts. Section~\ref{s10} focuses on
the underlying linear algebra, specifically the arrangement of the
subspaces $W_{\mathcal A}$ and the gluing isomorphism of
Theorem~\ref{thm:mainC}. Section~\ref{s5} proves
Theorems~\ref{thm:mainB} and~\ref{thm:mainD}, while
Section~\ref{ss:horizons} proves Theorem~\ref{thm:mainE} and
formulates Conjectures~\ref{C:vis} and \ref{C:penrose}.
Section~\ref{s7} shows how to read all of this off a given mass
matrix, with examples. Section~\ref{s8} proves
Theorem~\ref{thm:mainF}, and Section~\ref{s:questions} collects open
questions.

\subsection*{Acknowledgement}
The author is deeply grateful to Kazuo Akutagawa, who initiated this
project through his computations of the ADM mass matrix in this
setting. This work is dedicated to the memory of Misha Kapovich, whose
work on polygons and configurations at infinity, and whose generosity
in many conversations, shaped the moduli-theoretic side of these
papers.

\subsection*{AI disclosure}
Many of the constructions and first versions of the proofs in this paper
were developed in conversations with the AI model Claude Fable 5.1
(Anthropic), and the paper was revised with Claude Opus 5.5 (Anthropic).
The author substantially reworked all of this material into its present
form and takes responsibility for it.

\section{Preliminaries and the chart dictionary}\label{s:recall}

\subsection{The mass normalization as a projection}
We use the notation of Section~\ref{s0}. More generally, for any metric
$(\sum_i\tilde c_i|x-p_i|^{-\kp})^{4/\kp}h$ the mass of the $i$-th end is
$\sum_{j\ne i}\xi_{ij}$ with $\xi_{ij}=\tilde c_i\tilde
c_j|p_i-p_j|^{-\kp}$ \cite[Proposition~1.1]{PartI}. The mass map
$\Xi$ is proper \cite[Proposition~3.5]{PartI}, with closed image cut
out by an inertia condition on $M(\eta)_{ij}=e^{-2\eta_{ij}/\kp}$
\cite[Theorem~3.6]{PartI}. Let $\Symz$ be the symmetric $\ell\times\ell$
matrices with zero diagonal, with $\langle
S,S'\rangle=\sum_{i<j}S_{ij}S'_{ij}$, let $\vp(a)_{ij}=a_i+a_j$
$(i\ne j)$, and let $\Lambda_\ell=\operatorname{im}\vp$, the ambiguity
created by the conformal group. For the log-distance matrix
$L_{ij}=\ln|p_i-p_j|$ the mass normalization reads $\eta=-\kp L+\vp(f)$
with $f$ determined by the row-sum condition. Write
\begin{equation}\label{eq:cent}
\cent(L)_i=\frac{\kp}{\ell-2}\Big(\sum_{j\ne i}L_{ij}-\frac{1}{\ell-1}\sum_{a<b}L_{ab}\Big)
\end{equation}
for the centering vector of $L\in\Symz$, with row sums
$T_i=\sum_{j\ne i}L_{ij}$ and total $T=\sum_{a<b}L_{ab}$, so that $\eta=\Pi(L)=-\kp
L+\vp(\cent L)$.

\begin{Lemma}[{\cite[Lemma 2.4]{PartI}}]\label{L:proj0}
There is an isomorphism $\Symz=\Hh\oplus^\perp\Lambda_\ell$, and the
map $\Pi:L\mapsto\eta$ of the mass normalization equals $-\kp P$,
where $P$ is the orthogonal projection of $\Symz$ onto $\Hh$. In
particular $\Pi$ is linear with kernel $\Lambda_\ell$, and
$\|\Pi(S)\|\le\kp\|S\|$. Moreover $\cent$ is linear and
$\cent(\vp(a))=\kp a$.
\end{Lemma}

We also use the tropical bounds of \cite[Lemma~2.7]{PartI}: for every $\eta\in\Hh$,
\begin{equation}\label{eq:trop}
\eta_i^{\max}\le\ln\hm_i\le\eta_i^{\max}+\ln(\ell-1),\qquad
\frac{\|\eta_{i\cdot}\|_1}{2(\ell-1)}\le\eta_i^{\max}\le\frac{\|\eta_{i\cdot}\|_1}{2},
\end{equation}
where $\eta_i^{\max}=\max_{j\ne i}\eta_{ij}$; together with the properness of $\Xi$ they
show that $\max_i\hm_i\to\infty$ exactly when $\|\eta\|\to\infty$.

For $0\le k\le\ell$ put $D=(\ell-1)(\ell-2)$ and
\begin{equation}\label{eq:sigmas}
\sigma_k=\frac{\kp(k-1)(2\ell-2-k)}{2D},\qquad
\sigma'_k=\frac{\kp k(k-1)}{2D},\qquad
\omega_k=\sigma_k+\sigma'_k=\frac{\kp(k-1)}{\ell-2}.
\end{equation}

\subsection{Stereographic projection}
Let $\psi\colon\R^n\to S^n$ be stereographic projection with $\psi(\infty)=p_\ell$,
and $p_i=\psi(y_i)$ for $i<\ell$. Write $L^{\R}\in\Symz$ for the \emph{Euclidean}
log-distance matrix with the row of $\ell$ set to zero:
$L^\R_{ij}=\ln|y_i-y_j|$ for $i\ne j<\ell$, $L^\R_{i\ell}=0$.

\begin{Lemma}\label{L:chart}
Let $L$ be the chordal log-distance matrix of $\pl$. Then
$L=L^\R+\vp(\rho)$ with $\rho_i=\frac12\ln2-\frac12\ln(1+|y_i|^2)$ for $i<\ell$ and
$\rho_\ell=\frac12\ln2$;
hence $\eta(\pl)=\Pi(L^\R)$, and with $f^\R:=\cent(L^\R)$,
\begin{equation}\label{eq:chartmetric}
\psi^*g_{\pl}=\Big(e^{f^\R_\ell}+\sum_{i<\ell}e^{f^\R_i}\,|y-y_i|^{-\kp}\Big)^{4/\kp}\delta .
\end{equation}
More generally, for a metric $g=\big(C+\sum_{i<\ell}w_i|y-y_i|^{-\kp}\big)^{4/\kp}\delta$,
$C,w_i>0$, which is $\Phi^{4/\kp}h$ on $S^n$ with
$\Phi=\sum_i\tilde c_i\phi(\cdot,p_i)^{-\kp/2}$, the pair products are
\begin{equation}\label{eq:xi}
\xi_{ij}(g)=w_iw_j|y_i-y_j|^{-\kp}\ (i,j<\ell),\qquad \xi_{i\ell}(g)=w_iC,
\qquad \hm_i(g)=\sum_{j\ne i}\xi_{ij}(g).
\end{equation}
\end{Lemma}
\begin{proof}
The relations
\[\vert{}\psi(y)-\psi(z)\vert{}=2\vert{}y-z\vert{}\big((1+\vert{}y\vert{}^2)(1+\vert{}z\vert{}^2)\big)^{-1/2}
\ \ \mbox{and} \ \
\vert{}\psi(y)-p_\ell\vert{}=2(1+\vert{}y\vert{}^2)^{-1/2}
\]
yield the first identity. Since the projection $\Pi$ annihilates
$\Lambda_\ell$, we have $\Pi(L)=\Pi(L^\R)$. Furthermore,
$\cent(L)=f^\R+\kp\rho$ by Lemma~\ref{L:proj0}, which implies
$c_i=e^{f^\R_i}2^{\kp/2}(1+\vert{}y_i\vert{}^2)^{-\kp/2}$ and
$c_\ell=2^{\kp/2}e^{f^\R_\ell}$. The computation in the proof of
\cite[Proposition~1.1]{PartI} yields
\[
\psi^*g=\tfrac14\big(c_\ell+\sum_ic_i(1+\vert{}y_i\vert{}^2)^{\kp/2}\vert{}y-y_i\vert{}^{-\kp}\big)^{4/\kp}\delta,
\]
where the overall constant simplifies since
$\tfrac14(2^{\kp/2})^{4/\kp}=1$. To obtain \eqref{eq:xi}, we invert
this parameterization: the metric $g$ is $\Phi^{4/\kp}h$ with $\tilde
c_i=2^{\kp/2}w_i(1+\vert{}y_i\vert{}^2)^{-\kp/2}$ and $\tilde
c_\ell=2^{\kp/2}C$, and substituting these and the first two relations into
$\tilde c_i\tilde c_j\vert{}p_i-p_j\vert{}^{-\kp}$ gives \eqref{eq:xi}.
\end{proof}
Consequently, when $p_\ell$ is placed at infinity, every chart factor
has rank one. The mass matrix, the weights of the metric, and the pair
products can be extracted directly from the Euclidean configuration
$(y_i)_{i<\ell}$, with the point at infinity contributing the constant
term.

\section{Degeneration: proof of Theorem~\ref{thm:mainA}}\label{s4}
\subsection{The rays}
For $A\subseteq\{1,\dots,\ell\}$ with $k=|A|$ define $\E(A)$, symmetric with
zero diagonal, by
\begin{equation}\label{eq:ray}
\E(A)_{ij}=
\begin{cases}
\ \ \dfrac{\kp(\ell-k)(\ell-k-1)}{(\ell-1)(\ell-2)}, & i,j\in A,\\[3mm]
-\dfrac{\kp(k-1)(\ell-1-k)}{(\ell-1)(\ell-2)}, & i\in A,\ j\in A^c,\\[3mm]
\ \ \dfrac{\kp k(k-1)}{(\ell-1)(\ell-2)}, & i,j\in A^c .
\end{cases}
\end{equation}

For $A\subseteq\{1,\dots,\ell\}$ let $\chi_A$ be the symmetric $0$--$1$ matrix with
$(\chi_A)_{ij}=1$ iff $i\ne j$ and $i,j\in A$. The whole of the degeneration analysis
rests on one identity, which is a statement about the linear map $\Pi$ of
Lemma~\ref{L:proj0} and contains no asymptotics.

\begin{Lemma}\label{L:rays}
For every $A$ with $|A|=k$,
\begin{equation}\label{eq:centchi}
\cent(\chi_A)=\sigma_k\mathbf 1_A-\sigma'_k\mathbf 1_{A^c},\qquad
\Pi(\chi_A)=-\E(A);
\end{equation}
equivalently $\E(A)=\kp P(\chi_A)$, the orthogonal projection of $\kp\chi_A$ onto
$\Hh$. Moreover
\begin{equation}\label{eq:ids}
\begin{gathered}
\E(A)_{AA}=\kp-2\sigma_k,\qquad \E(A)_{A^cA^c}=2\sigma'_k,\\
\E(A)_{AA^c}-\E(A)_{A^cA^c}=-\omega_k,\qquad
\E(A)_{AA}-\E(A)_{AA^c}=\kp-\omega_k .
\end{gathered}
\end{equation}
\end{Lemma}

\begin{proof}
For $L=\chi_A$ one has $T_i=k-1$ for $i\in A$, $T_i=0$ for $i\notin A$,
and $T=\binom k2$, so \eqref{eq:cent} gives
$$
f_i=\frac{\kp(k-1)\big(2(\ell-1)-k\big)}{2(\ell-1)(\ell-2)}=\sigma_k\ (i\in A),\qquad
f_j=-\frac{\kp k(k-1)}{2(\ell-1)(\ell-2)}=-\sigma'_k\ (j\notin A).
$$
Substituting into $\eta_{ij}=-\kp(\chi_A)_{ij}+f_i+f_j$: for $i,j\in A$,
$$
\eta_{ij}=\kp\,\frac{(k-1)(2\ell-2-k)-(\ell-1)(\ell-2)}{(\ell-1)(\ell-2)}
=-\frac{\kp(\ell-k)(\ell-k-1)}{(\ell-1)(\ell-2)},
$$
by the polynomial identity $(k-1)(2\ell-2-k)-(\ell-1)(\ell-2)=-(\ell-k)(\ell-k-1)$,
both sides expanding to $-k^2+2\ell k-k+\ell-\ell^2$; for $i\in A$, $j\notin A$ the
entry of $\chi_A$ vanishes and
\[
\eta_{ij}=\frac{\kp(k-1)}{2(\ell-1)(\ell-2)}[(2\ell-2-k)-k]
=\frac{\kp(k-1)(\ell-1-k)}{(\ell-1)(\ell-2)};
\]
and for $i,j\notin A$, $\eta_{ij}=-\frac{\kp
  k(k-1)}{(\ell-1)(\ell-2)}$. These are $-\E(A)_{ij}$ in the three
cases of \eqref{eq:ray}. The computation is valid for every $0\le
k\le\ell$, and returns $\E(A)=0$ for $k\in\{0,1,\ell-1,\ell\}$. The
identities \eqref{eq:ids} are the same polynomial identity read three
more times.
\end{proof}
\begin{Theorem}[Theorem~\ref{thm:mainA}(i)]\label{T:degen}
Let $2\le k\le\ell-2$ and let $\pl(\epsilon)$ be a family such that, in a fixed
stereographic chart, $p_a(\epsilon)=\psi(\epsilon a_a+o(\epsilon))$ for $a\in A$ with
$a_a\in\R^n$ distinct, and $p_j(\epsilon)\to p^0_j$ for $j\notin A$, the $p^0_j$
distinct and distinct from $\psi(0)$. Then
\begin{equation}\label{eq:degen}
\eta(\pl(\epsilon))=\E(A)(- \ln\epsilon) +O(1),
\end{equation}
so $\E(A)\in\Hh$, and by \eqref{eq:trop}
\begin{equation}\label{eq:rates}
\hm_i\asymp\epsilon^{-\theta_i},\qquad
\theta_i=\begin{cases}
\dfrac{(n-2)(\ell-k)(\ell-k-1)}{(\ell-1)(\ell-2)},& i\in A,\\[3mm]
\dfrac{(n-2)k(k-1)}{(\ell-1)(\ell-2)},& i\notin A.
\end{cases}
\end{equation}
In particular every mass diverges.
\end{Theorem}

\begin{proof}
In the given chart $L_{ab}=\ln\epsilon+O(1)$ for $a\ne b$ in $A$, while
$L_{aj},L_{jk}=O(1)$ otherwise, i.e.,
$$
L=\ln\epsilon\cdot\chi_A+b,\qquad \|b\|_\infty=O(1).
$$ Use the projection $\Pi$. It is linear, $\Pi(\chi_A)=-\E(A)$ by
Lemma~\ref{L:rays}, and $\|\Pi(b)\|\le\kp\|b\|$ by
Lemma~\ref{L:proj0}; this gives \eqref{eq:degen}, and $\E(A)\in\Hh$
because $\Pi$ takes values in $\Hh$. Since $\theta_i=\max_{j\ne
  i}\E(A)_{ij}$, \eqref{eq:trop} gives \eqref{eq:rates}.
\end{proof}

\begin{Remark}\label{R:tropic}
Dividing \eqref{eq:trop} by $\ln(1/\epsilon)$ and using
\eqref{eq:degen} gives
\begin{equation}\label{eq:thetatrop}
\theta_i=\max_{j\ne i}\E(A)_{ij},
\end{equation}
the tropicalization of the $i$-th row of the ray $\E(A)$. The $O(1)$
term and the tropical error $\ln(\ell-1)$ both contribute
$O(1/\ln(1/\epsilon))$ to the normalized exponents.
\end{Remark}

\begin{Proposition}\label{P:rays}
For $2\le k\le\ell-2$: $\E(A)=\E(A^c)$, and $\E(A)_{ij}>0$ iff
$i,j$ lie on the same side of the partition. For arbitrary $0\le k\le\ell$,
$\E(A)=0$ iff $k\in\{0,1,\ell-1,\ell\}$, i.e.\ exactly for the degenerations undone
by a conformal transformation. Consequently $\{A,A^c\}\mapsto\R_{>0}\E(A)$ is
injective on splits with $2\le|A|\le\ell-2$.
\end{Proposition}

\begin{proof}
For the first identity, by Lemma~\ref{L:rays} it suffices that
$\chi_A-\chi_{A^c}\in\ker\Pi=\Lambda_\ell$, and indeed
$$
(\chi_A-\chi_{A^c})_{ij}=
\begin{cases}
+1,& i,j\in A,\\
-1,& i,j\in A^c,\\
\ \ 0,& \text{otherwise,}
\end{cases}
$$
which is $\varpi(a)$ for $a_i=\tfrac12$ on $A$ and $a_i=-\tfrac12$ on $A^c$. The signs
are read off \eqref{eq:ray}; $\E(A)_{AA}=0$ forces $k\in\{\ell-1,\ell\}$ and
$\E(A)_{A^cA^c}=0$ forces $k\in\{0,1\}$; and the partition is recovered from
the sign pattern.
\end{proof}
The identity $\E(A)=\E(A^c)$ reflects the fact that on $S^n$ the collapse
of $A$ and the collapse of $A^c$ are the same degeneration, the pinching of
a neck between the two clusters, described in two charts that differ by a
M\"obius transformation.

\subsection{Superposition}\label{ss:super}
We define a family $\mathcal A=\{A_1,\dots,A_r\}$ of subsets, each of
size between $2$ and $\ell-2$, to be \emph{nested} if any two sets in
$\mathcal A$ are either disjoint or comparable under inclusion. We
call it \emph{reduced} if, additionally, no two sets in the family are
equal or complementary. Adjoining a formal root $A_0=\{1,\dots,\ell\}$
with $\epsilon_0:=1$, the collection $\mathcal A\cup\{A_0\}$ ordered
by inclusion forms a rooted tree. Consequently, the \emph{parent}
$\hat v$, defined as the smallest set properly containing $A_v$,
exists for every $v$ (with $\hat v=0$ when $A_v$ is
maximal).

Reducedness is a normalization. A set of pairwise compatible splits
$\{A,A^c\}$ is the same as a tree with $\ell$ labelled leaves, and to
represent it by a family of subsets one chooses a side of each split.
Choosing the side disjoint from
a fixed leaf---which we take to be the leaf $\ell$, corresponding to the
end sent to $\infty$ in the shape-space model of \cite{PartI}---produces
representatives that are automatically nested and reduced. We refer to
this as the \emph{normal form} of $\mathcal A$, characterized by the
condition that $\ell\notin A$ for every set in the family. Conversely,
any nested family can be made reduced by replacing any set containing
$\ell$ with its complement. This substitution leaves the relevant
geometric quantities invariant, since $\E(A)=\E(A^c)$ by
Proposition~\ref{P:rays} and $W_A=W_{A^c}$, as is evident from
\eqref{eq:defect} below. Proposition~\ref{P:nested} holds for any nested
family; reducedness matters for the dimension counts that follow.
\begin{Proposition}[Theorem~\ref{thm:mainA}(ii)]\label{P:nested}
Let $\mathcal A$ be a nested family, assign scales
$\epsilon_v\in(0,1)$ with $\epsilon_v<\epsilon_w$ when $A_v\subsetneq
A_w$, and suppose $|p_i-p_j|\asymp\epsilon_{v(i,j)}$ where
$A_{v(i,j)}$ is the smallest member containing $i$ and $j$
($\epsilon:=1$ if none). Then
\begin{equation}\label{eq:super}
\eta=\sum_{v}\tau_v\,\E(A_v)+O(1),\qquad
\tau_v=\ln\frac{\epsilon_{\hat v}}{\epsilon_v}>0 .
\end{equation}
\end{Proposition}
The proof rests on a telescoping identity.
\begin{Lemma}\label{L:chain}
Fix $i\ne j$ and define $\mathcal C(i,j)=\{A_v\in\mathcal A : i,j\in
A_v\}$. Then $\mathcal C(i,j)$ forms a chain $A_{v_0}\subsetneq
A_{v_1}\subsetneq\dots\subsetneq A_{v_m}$ with
$A_{v_0}=A_{v(i,j)}$. Furthermore, consecutive sets in this chain are
related by the parent map: $\widehat{v}_t=v_{t+1}$ for $t<m$, and
$\widehat{v}_m=0$. Consequently,
\[
\sum_v\tau_v\,(\chi_{A_v})_{ij}=-\ln\epsilon_{v(i,j)}\qquad (i\ne j),
\]
under the convention $\epsilon_{v(i,j)}=1$ when $\mathcal
C(i,j)=\varnothing$.
\end{Lemma}
\begin{proof}
Two members of $\mathcal C(i,j)$ both contain $i$, so they are
comparable, and $\mathcal C(i,j)$ is a chain with least element
$A_{v(i,j)}$. A member properly containing $A_{v_t}$ contains $i,j$,
hence lies in $\mathcal C(i,j)$ and contains $A_{v_{t+1}}$; so
$\widehat{v_t}=v_{t+1}$, and $\widehat{v_m}=0$. Only members of
$\mathcal C(i,j)$ contribute to the sum, which telescopes to
$\sum_{t=0}^{m}\ln(\epsilon_{v_{t+1}}/\epsilon_{v_t})=-\ln\epsilon_{v(i,j)}$,
with $\epsilon_{v_{m+1}}:=\epsilon_0=1$.
\end{proof}
\begin{proof}[Proof of Proposition~\ref{P:nested}]
By hypothesis there is $B<\infty$, depending only on the implied constants in
$|p_i-p_j|\asymp\epsilon_{v(i,j)}$, with
\[
L_{ij}=\ln|p_i-p_j|=\ln\epsilon_{v(i,j)}+b_{ij},\qquad |b_{ij}|\le B ,
\]
so that $L=-\sum_v\tau_v\chi_{A_v}+b$ by Lemma~\ref{L:chain}. Apply the linear map
$\Pi$ and use Lemma~\ref{L:rays}:
\[
\eta=\Pi(L)=\sum_v\tau_v\,\E(A_v)+\Pi(b) .
\]
By Lemma~\ref{L:proj0},
$\|\Pi(b)\|\le\kp\|b\|\le\kp\binom\ell2^{1/2}B$; entrywise,
\eqref{eq:cent} gives $|T_i|\le(\ell-1)B$ and $|T|\le\binom\ell2 B$,
hence $|f_i|\le\kp B(3\ell-2)/\big(2(\ell-2)\big)$ and
$$
\max_{i\ne j}|\Pi(b)_{ij}|\ \le\ \frac{4\kp(\ell-1)}{\ell-2}\,B .
$$ Both bounds are uniform in the scales, which is the meaning of the
$O(1)$ in \eqref{eq:super}.  Finally we have
$\tau_v=\ln(\epsilon_{\hat v}/\epsilon_v)>0$ because
$\epsilon_v<\epsilon_{\hat v}$.
\end{proof}

\begin{Remark}\label{R:realize}
  The hypothesis $|p_i-p_j|\asymp\epsilon_{v(i,j)}$ constrains
distances, not charts; that families realizing it exist for every
nested $\mathcal A$ and every admissible system of scales is a short
induction on the forest. In a fixed stereographic chart, let the
ancestor chain of a leaf $i$ be $A_{u_1(i)}\subsetneq\dots\subsetneq
A_{u_m(i)}$ and set
$$
y_i=a^{(0)}_i+\epsilon_{u_m(i)}a^{(1)}_i+\dots+\epsilon_{u_1(i)}a^{(m)}_i ,
$$ where at each vertex of the forest the points assigned to distinct
children, and to the leaves attached there, are chosen distinct. Two
leaves agree in every term above $A_{v(i,j)}$ and differ at that
level, so $|y_i-y_j|\asymp\epsilon_{v(i,j)}$; all $y_i$ remain
bounded, so chordal and Euclidean distances are comparable and the
same holds for $p_i=\psi(y_i)$. The case $m=1$ is the family of
Theorem~\ref{T:degen}.  This is exactly the corner chart of the
Axelrod--Singer compactification, \eqref{eq:recon} below.
\end{Remark}
The rays of a reduced nested family span a simplicial cone. Without
reducedness this fails:
\begin{Example}\label{E:notreduced}
Let $\ell=6$ and $\mathcal A=\{\{1,2\},\{3,4,5,6\}\}$. The elements of
$\mathcal A$ are disjoint, so $\mathcal A$ is nested, and both sizes
lie in $[2,\ell-2]$; but they are complementary, so $\E(A_1)=\E(A_2)$
and \eqref{eq:super} reads
$\eta=(\tau_1+\tau_2)\E(\{1,2\})+O(1)$. Modulo $\Mob(S^n)$ this is not
a corner but the ordinary codimension-one degeneration with neck
length $\tau_1+\tau_2$: the tree has one internal edge, and the two
scales contribute additively to its length. The redundancy of the
family is visible as the linear dependence of its rays.
\end{Example}
\begin{Theorem}[Theorem~\ref{thm:mainA}(iii), first part]\label{T:indep}
Let $\mathcal A$ be a reduced nested family. Then the rays $\E(A_v)$
for all $A_v\in\mathcal A$ are linearly independent. Consequently, the
space $\sigma_{\mathcal A}=\sum_v\R_{>0}\E(A_v)$ forms a simplicial
cone of dimension $\vert{}\mathcal A\vert{}$.
\end{Theorem}
\begin{proof}
By Lemma~\ref{L:proj0} and Lemma~\ref{L:rays}, $\E(A)=\kp P(\chi_A)$,
so the assertion is that the $\chi_{A_v}$ are linearly independent
modulo $\Lambda_\ell=\ker P$. Suppose $S:=\sum_v
c_v\chi_{A_v}\in\Lambda_\ell$ with some $c_v\ne0$; discarding the
members with $c_v=0$ leaves a family that is still nested and reduced
and all of whose coefficients are nonzero.

For distinct $i,j,k,l$ consider the four-point functional
$$
D(ij|kl)(S)=S_{ij}+S_{kl}-S_{ik}-S_{jl} ,
$$
which annihilates $\Lambda_\ell$, since
$(a_i+a_j)+(a_k+a_l)-(a_i+a_k)-(a_j+a_l)=0$, and which on $\chi_A$ equals
$[i,j\in A]+[k,l\in A]-[i,k\in A]-[j,l\in A]$.

Choose $A_u$ minimal in the family and $i\ne j$ in $A_u$; since
$|A_u|\le\ell-2$ there are at least two indices outside $A_u$, from
which $k\ne l$ will be chosen. By nestedness and minimality every
remaining member either contains $A_u$ or is disjoint from it, and the
contributions are: $1$ for $A_v=A_u$; for $A_v\supsetneq A_u$,
$1+[k,l\in A_v]-[k\in A_v]-[l\in A_v]$, which vanishes as soon as
$k\in A_v$; and for $A_v\cap A_u=\varnothing$, where $i,j\notin A_v$,
simply $[k,l\in A_v]$.

If some member properly contains $A_u$, let $B$ be the smallest such
and pick $k\in B\setminus A_u$, which is nonempty; the members
properly containing $A_u$ form a chain with least element $B$, so all
of them contain $k$ and drop out. If no member properly contains
$A_u$, pick $k\notin A_u$ arbitrarily. The members containing $k$ form
a chain, so among those disjoint from $A_u$ there is a largest, say
$C$, if any exist; pick $l\notin A_u\cup C$ with $l\ne k$ (any
$l\notin A_u$, $l\ne k$, if $C$ does not exist). Then a member
disjoint from $A_u$ either misses $k$, or contains $k$ and hence lies
in $C$ and misses $l$; either way it drops out. With these choices
$0=D(ij|kl)(S)=c_u$, a contradiction.

It remains to check that such an $l$ exists. This fails only if
$A_u\cup C=\{1,\dots,\ell\}$, i.e.\ $C=A_u^c$, which reducedness excludes. (If
$A_u^c$ is a member then $A_u$ is automatically maximal: a member $B\supsetneq A_u$
meets $A_u^c$, hence is comparable with it, hence contains it, hence has $\ell$
elements, contradicting $|B|\le\ell-2$; so the two cases in the choice of $k$ do not
interfere.)
\end{proof}

\begin{Proposition}[tree space; Theorem~\ref{thm:mainA}(iii), second part]\label{R:treespace}
  Let $\delta_A:=\mathbf 1-\chi_A-\chi_{A^c}$, the split metric of
$\{A,A^c\}$ ($\mathbf 1=J-I$; entries $1$ across the split, $0$ within
its sides), and recall that a tree metric on the tree $T$ of a reduced
nested family $\mathcal A$ is $\sum_{A\in\mathcal A}\lambda_A\delta_A$
with $\lambda_A>0$; the space of tree metrics with $\ell$ labelled
leaves, modulo its lineality space, is the space of phylogenetic trees
of Billera--Holmes--Vogtmann \cite{BHV} and the tropical Grassmannian
of lines of Speyer--Sturmfels \cite{SS}, a fan of dimension $\ell-3$
whose maximal cones are indexed by trivalent trees. Then:
\begin{enumerate}[label=\textup{(\arabic*)},leftmargin=*]
\item $\delta_A\equiv-2\chi_A$ modulo $\Lambda_\ell$, and the trivial
  split metrics $\delta_{\{i\}}=\vp(e_i)$ span $\Lambda_\ell$, which is
  therefore the lineality space of the space of tree metrics;
\item the linear map $-\frac\kp2P\colon\Symz\to\Hh$ carries the cone of
  tree metrics of $T$ onto $\sigma_{\mathcal A}=\sum_v\R_{>0}\E(A_v)$,
  isomorphically, and induces a linear isomorphism
  $\Symz/\Lambda_\ell\to\Hh$ carrying the space of phylogenetic trees
  onto the union of the cones $\sigma_{\mathcal A}$. Hence the
  $\sigma_{\mathcal A}$ form a fan in $\Hh$ isomorphic to the space of
  phylogenetic trees, and Theorem~\ref{T:indep} is the linear
  independence of compatible split metrics modulo the lineality space.
\end{enumerate}
\end{Proposition}
\begin{proof}
(1) $\mathbf 1=\vp(\tfrac12,\dots,\tfrac12)\in\Lambda_\ell$, and
  $\chi_{A^c}\equiv\chi_A$ modulo $\Lambda_\ell$ by the proof of
  Proposition~\ref{P:rays}, so $\delta_A\equiv-2\chi_A$. The trivial
  split $\{i\}|\{i\}^c$ has $\delta_{\{i\}}=\vp(e_i)$, and
  $\Lambda_\ell=\im\vp$ is spanned by these. (2) For a tree metric,
  $-\frac\kp2P(\sum_A\lambda_A\delta_A)=\kp\sum_A\lambda_AP(\chi_A)=\sum_A\lambda_A\E(A)$
  by Lemma~\ref{L:rays}, which runs over $\sigma_{\mathcal A}$ as the
  $\lambda_A$ run over $\R_{>0}$; the map is injective on the cone by
  Theorem~\ref{T:indep}. Since $\ker P=\Lambda_\ell$ is the lineality
  space, $P$ induces an isomorphism $\Symz/\Lambda_\ell\cong\Hh$ which
  carries the fan of \cite{BHV,SS}, cone by cone, onto the
  $\sigma_{\mathcal A}$.
\end{proof}
\begin{Remark}
The principal addition to the framework of \cite{BHV,SS} is the
identification of this fan with the asymptotic directions of the mass
map. Specifically, by Theorem~\ref{T:degen} and
Proposition~\ref{P:nested}, the tropicalization of $\Xi$ corresponds
to the tropical Grassmannian, where the sign convention reflects that
$\eta$ is the negative of the centred log-distance. For a family that
is not reduced, Example~\ref{E:notreduced} demonstrates that the
resulting cone is simply determined by the underlying set of splits,
and thus has strictly smaller dimension.
\end{Remark}

\begin{Corollary}\label{T:compact}
Along a degeneration as in Proposition~\ref{P:nested},
$\eta/\Vert{}\eta\Vert{}\to\big[\sum_v\tau_v\E(A_v)\big]$ whenever
the ratios $\tau_v/\sum_w\tau_w$ converge. A single cluster gives
$[\E(A)]$, which depends only on the partition, not on the limiting
configuration in the bubble.
\end{Corollary}
By contrast, no rescaling of the masses extends across the boundary:
each $\hm_i$ diverges along every degeneration, at the rate
\eqref{eq:rates}, while $\hm_i\ge\ell-1$ everywhere
\cite[Corollary~2.9]{PartI}.
\subsection{The fan at $\ell=5$}\label{ss:petersen}
\begin{Proposition}\label{P:petersen5}
  For $\ell=5$ and $A=\{i,j\}$, \eqref{eq:ray} reads $\E(A)_{ij}=\kp/2$,
$\E(A)_{kl}=\kp/6$ for $k,l\in A^c$, $\E(A)_{ik}=-\kp/6$ for
$k\in A^c$. Hence $\|\E(A)\|=\kp/\sqrt2$, the ten rays sum to zero, and
$$
\langle\E(\{i,j\}),\E(\{k,l\})\rangle=
\begin{cases}+\kp^2/6,&\{i,j\}\cap\{k,l\}=\emptyset,\\ -\kp^2/6,&|\cap|=1,\end{cases}
$$
so distinct rays meet at $\arccos(\pm\tfrac13)$, i.e.\ $70.53^\circ$ and
$109.47^\circ$. Two rays span a $2$-cone exactly when the pairs are disjoint, which
happens $15$ times; the link of the fan is the Petersen graph.
\end{Proposition}

\begin{proof}
  $\|\E\|^2=\kp^2(\tfrac14+3\cdot\tfrac1{36}+6\cdot\tfrac1{36})=\kp^2/2$. The
  inner products are computed entrywise: for $A=\{1,2\}$, $B=\{3,4\}$
  the products are $\tfrac1{12}$ on the pairs $12,34$, $+\tfrac1{36}$
  on $13,14,23,24$ and $-\tfrac1{36}$ on $15,25,35,45$, summing to
  $\kp^2/6$; for $A=\{1,2\}$, $B=\{1,3\}$ they are $-\tfrac1{12}$ on
  $12,13$, $+\tfrac1{36}$ on $14,15,23,45$ and $-\tfrac1{36}$ on
  $24,25,34,35$, summing to $-\kp^2/6$. The sum of all ten is
  $S_5$-invariant, hence a multiple of the all-equal matrix, hence
  $0$. Two $2$-subsets are nested or disjoint iff disjoint, and there
  are $\tfrac12\binom52\binom32=15$ such pairs; the resulting graph,
  with an edge between disjoint pairs, is the Kneser graph $KG(5,2)$,
  i.e.\ the Petersen graph.
\end{proof}
The $15$ two-dimensional cones correspond precisely to the $15$
trivalent trees with five leaves. If $p_i$ and $p_j$ collide at scale
$\epsilon$, then $\hm_i,\hm_j\asymp\epsilon^{-(n-2)/2}$, whereas the
three spectator masses grow only as
$\epsilon^{-(n-2)/6}$. Consequently, the merging pair diverges with an
exponent three times as large, in sharp contrast to the $\ell=4$ case
where all rates coincide. The convergence of the projectivized vector
$\eta/\Vert{}\eta\Vert{}$ to the fan is logarithmically slow;
specifically, after normalization, the error term in \eqref{eq:degen}
is $O\big(1/\ln(1/\epsilon)\big)$.

\section{The Axelrod--Singer compactification}\label{s:AS}
\subsection{Trees, screens, and the corner chart}
Let $\mathcal A$ be a reduced nested family in normal form
($\ell\notin A$ for all $A\in\mathcal A$, Section~\ref{ss:super}). Its tree
$T$ has internal vertices $V=\{r\}\cup\mathcal A$, where for
$v\in\mathcal A$ we write $A_v=v$ and $A_r=\{1,\dots,\ell\}$; the
parent $\hat v$ is the smallest member properly containing $A_v$, or
$r$. The branches at $v$ are the children of $v$, the leaves
$\Lambda_v:=A_v\setminus\bigcup_{\text{children}}A_w$ attached
directly to $v$, and the parent branch. One branch at each vertex is
\emph{at infinity}: the parent branch if $v\ne r$, and the leaf
$\ell\in\Lambda_r$ if $v=r$. Write $\mathcal B^\circ_v$ for the
remaining (finite) branches, so that $\ell_v=|\mathcal B^\circ_v|+1$
is the valence; $b_v(i)$ for the branch at $v$ containing a leaf $i$;
and $B_b$ for the leaf set of a branch $b$. For $i\ne j$ let $v(i,j)$
be the vertex at which $i$ and $j$ lie in different branches; it is
the smallest member of $\mathcal A\cup\{A_r\}$ containing both, as in
Lemma~\ref{L:chain}.
A \emph{screen} at $v$ is a point of
$C_{\ell_v-1}(\R^n):=\Conf_{\ell_v-1}(\R^n)/(\R^n\rtimes\R_{>0})$, a configuration
indexed by $\mathcal B^\circ_v$ modulo translation and dilation; the branch at
infinity is at $\infty$. We fix the slice
\[
\mathcal N_v=\Big\{(a_b)_{b\in\mathcal B^\circ_v}\in\Conf_{\ell_v-1}(\R^n):\ \sum_ba_b=0,\ \sum_b|a_b|^2=1\Big\},
\]
which maps diffeomorphically onto $C_{\ell_v-1}(\R^n)$ and is $O(n)$-invariant.
For $\sigma_v=(a_b)\in\mathcal N_v$ put
\begin{equation}\label{eq:alpha}
\alpha^{(v)}\in\mathcal S_{\ell_v}:\qquad
\alpha^{(v)}_{bc}=\ln|a_b-a_c|\ (b\ne c\in\mathcal B^\circ_v),\qquad
\alpha^{(v)}_{b\,\infty_v}=0 ,
\end{equation}
the Euclidean log-distance matrix of the screen with the row of the branch at
infinity set to zero, exactly as $L^\R$ in Section~\ref{s:recall}. Pullback along the branch
map is $(\pi_v^*\theta)_{ij}=\theta_{b_v(i)b_v(j)}$ if $b_v(i)\ne b_v(j)$, and $0$
otherwise.

For $t=(t_A)_{A\in\mathcal A}\in(0,\delta)^{\mathcal A}$ put
\begin{equation}\label{eq:tau}
\epsilon_{\le v}:=\prod_{A\in\mathcal A,\ A\supseteq A_v}t_A\quad(\epsilon_{\le r}=1),
\qquad \tau_A:=\ln(1/t_A),
\qquad \Th:=\sum_{A\in\mathcal A}\tau_A\E(A)\in\Hh ,
\end{equation}
and define the configuration $y(\sigma,t)$ by
\begin{equation}\label{eq:recon}
y_i=\sum_{s=0}^{m}\epsilon_{\le v_s}\,a^{(v_s)}_{b_{v_s}(i)}\qquad(i<\ell),
\end{equation}
where $r=v_0\supsetneq v_1\supsetneq\dots\supsetneq v_m$ is the chain of vertices
with $i\in A_{v_s}$, so that $i\in\Lambda_{v_m}$. The \emph{node position} of $v_m$
is $y_{v_m}:=\sum_{s<m}\epsilon_{\le v_s}a^{(v_s)}_{b_{v_s}(i)}$ (the same for all
$i\in A_{v_m}$), and $y_i=y_{v}+\epsilon_{\le v}a_i$ exactly for $i\in\Lambda_v$.
Note that $\epsilon_{\le\hat v}\,t_{A_v}=\epsilon_{\le v}$: $t_{A_v}$ is the scale of
$v$ relative to its parent, and $\tau_{A_v}$ is the $\tau_v$ of \eqref{eq:super}.
\begin{Definition}\label{D:AS}
Let $\overline C_m(\R^n)$ denote the compactification of $C_m(\R^n)$
obtained by applying the Fulton--MacPherson construction \cite{FM}, in
its real, codimension-one formulation \cite[Section~5]{AS}, to the
configuration space $\Conf_m(\R^n)$ modulo translations and
dilations. The corner charts utilized below are those introduced by
Sinha \cite{Si}. This space is a compact manifold with corners. Its
closed strata are indexed by nested families of proper subsets of
$\{1,\dots,m\}$ having cardinality at least $2$. The open stratum
associated with such a family is given by the product $\prod_v
C_{\ell_v-1}(\R^n)$ over the internal vertices of its corresponding
tree. Near the corner corresponding to a family $\mathcal A$ there is a
continuous function $\delta\colon\prod_v\mathcal N_v\to(0,1]$ such that the map
$(\sigma,t)\mapsto[y(\sigma,t)]$, defined in \eqref{eq:recon} for $t_A>0$,
extends to a diffeomorphism from
\[
\mathcal U_{\mathcal A}:=\Big\{(\sigma,t)\in\prod_v\mathcal N_v\times[0,1)^{\mathcal A}:\
t_A<\delta(\sigma)\text{ for all }A\in\mathcal A\Big\}
\]
onto an open neighborhood of the open corner stratum, where the locus
$\{t_A=0\}$ defines the boundary face $\partial_A$. The function $\delta$ is
needed because $\prod_v\mathcal N_v$ is not compact: as a screen approaches a
collision, the scales below it must be smaller for the points $y_i(\sigma,t)$ to
remain distinct and for the map to remain injective. On a compact subset of
$\prod_v\mathcal N_v$ a constant $\delta$ suffices. We define the quotient space
  \[
  \overline\Mo^{\,\mathrm{AS}}_\ell(S^n):=\overline
  C_{\ell-1}(\R^n)/O(n).
  \]
  We often omit $(S^n)$ from the notation, here and for the quotient
  $\overline\Mo^{\,\mathrm{DM}}_\ell$ of Section~\ref{s8}.
  Here, the orthogonal group $O(n)$ acts via its standard action on
  $\Conf_{\ell-1}(\R^n)$. Because this action commutes with both
  translations and dilations, it extends naturally to the
  compactification by functoriality.
\end{Definition}
Its interior is
$C_{\ell-1}(\R^n)/O(n)=\Conf_{\ell-1}(\R^n)/\Sim(\R^n)=\Mo_\ell(S^n)$,
\cite[(5.2)]{PartI}. With $m=\ell-1$ and
$\{1,\dots,\ell-1\}=\{1,\dots,\ell\}\setminus\{\ell\}$, the indexing
families are exactly the reduced nested families in normal
form. $O(n)$ is compact, so the quotient is compact Hausdorff and
stratified by the same trees, with open strata
$\prod_vC_{\ell_v-1}(\R^n)$ modulo the diagonal $O(n)$. It is not a
manifold with corners: configurations with symmetry have nontrivial
stabilizers, and these are positive-dimensional for configurations lying
on a line (stabilizer $\mathrm O(n-1)$). Over the locus with finite
stabilizers it is an orbifold with corners. For instance, for $\ell=n+2$
the configurations lying in a hyperplane form a hypersurface fixed by a
reflection; for $\ell\le n+1$ every configuration lies in a hyperplane
and the quotient is $C_{\ell-1}(\R^{\ell-2})/\mathrm O(\ell-2)$.
\begin{Remark}\label{R:bdf}
On $\mathcal U_{\mathcal A}$, $t_A$ is a boundary defining function
for $\partial_A$; any other is $u_At_A$ with $u_A>0$ smooth on
$\mathcal U_{\mathcal A}$, and changing the slices $\mathcal N_v$ has
the same effect. So $\tau_A$ is canonical up to a smooth function, and
its restriction to the corner up to a smooth function on the corner.
This is why the boundary value of Section~\ref{s5} is canonical modulo
$V_{\mathcal A}$, and not merely modulo $O(1)$.
\end{Remark}

\begin{Remark}\label{R:leaf}
  The construction singles out $p_\ell$. The identity of $\Mo_\ell(S^n)$
extends to a homeomorphism between the spaces built with $p_\ell$ and
with $p_m$ at infinity, taking $\partial_A$ to $\partial_A$ if
$m\notin A$ and to $\partial_{A^c}$ if $m\in A$, and relating the
scale coordinates by smooth positive factors. Indeed, let $\gamma$ be
a M\"obius map with $\gamma(y_m)=\infty$, $\gamma(\infty)$ finite. On
a cluster $A\not\ni m$, $\gamma$ is a diffeomorphism near the cluster,
so the cluster maps to a cluster with the same screen (up to
similarity) and relative scale multiplied by a smooth positive factor,
the ratio of the conformal factors of $\gamma$ at the node and at its
parent. On a cluster $A\ni m$ at scale $\epsilon$ around $y_m$,
$\gamma$ is to leading order an inversion centred at $y_m$: the points
of $A$ go to distance $\asymp\epsilon^{-1}$, i.e.\ in the new chart
the complement $A^c$ is a cluster of relative scale $\asymp\epsilon$,
its screen is the old root screen, and the new root screen is the
inverted old $A$-screen. All transition functions are smooth in
$(\sigma,t)$ up to $t=0$. Since $\E(A)=\E(A^c)$ and $\eta$ is M\"obius
invariant, everything below is insensitive to the choice, except for
the $V_{\mathcal A}$-ambiguity already present. We work in normal form
throughout.
\end{Remark}

\section{The arrangement of subspaces and the gluing isomorphism}\label{s10}
This section proves Theorem~\ref{thm:mainC}. The arguments are linear
algebra, organized by the trees of Section~\ref{s:AS}.

\subsection{The subspaces}
Since $|A|,|A^c|\ge2$ for every member of a reduced nested family, put
\begin{equation}\label{eq:defect}
W_A:=\{\zeta\in\Hh:\ \zeta_{ij}+\zeta_{kl}-\zeta_{il}-\zeta_{kj}=0
\ \text{for } i,k\in A,\ j,l\in A^c\},
\end{equation}
the matrices whose block coupling $A$ to $A^c$ is additive,
$\zeta_{ij}=\mu_i+\mu_j$ for $i\in A$, $j\notin A$; clearly
$W_A=W_{A^c}$. For a reduced nested family put $W_{\mathcal
  A}=\bigcap_{A\in\mathcal A}W_A$ and $V_{\mathcal
  A}=\operatorname{span}\{\E(A):A\in\mathcal A\}$. We write
$D(ij|kl)(S)=S_{ij}+S_{kl}-S_{ik}-S_{jl}$ for the four-point
functional, which kills $\Lambda_\ell$ and, for $\ell\ge4$, detects it
(a symmetric zero-diagonal matrix all of whose four-point defects
vanish is of the form $\vp(a)$, a routine check); membership in $W_A$
is the vanishing of the $2\times2$ defects \eqref{eq:defect}, and
since $P$ changes a matrix by an element of $\Lambda_\ell$, it changes
no defect.
\begin{Remark}\label{R:notcompact}
  The codimension of $W_A$ is $\tfrac{\ell(\ell-3)}2-\binom
k2-\binom{\ell-k}2+1$ (Corollary~\ref{C:dimW}), which equals $1$ for
$(\ell,k)=(4,2)$ but $2$ for $(5,2)$, $3$ for $(6,2)$ and $4$ for
$(6,3)$. So for $\ell\ge5$ the boundary values on a face of codimension one
lie in a subspace of $\Hh$ of codimension at least two, and the spaces
$\R^E_{>0}\times\prod_v\mathcal H_{\ell_v}$ do not attach to $\Hh$ as the faces of a
manifold with corners. We regard $\bigsqcup_T\big(\R^E_{>0}\times\prod_v\mathcal
H_{\ell_v}\big)$ as the associated graded of the degeneration: the target of the
boundary values, not a closure of the domain.
\end{Remark}

\subsection{The gluing map}\label{ss:gluing}
Write $T$ for the tree of $\mathcal A$ as in Section~\ref{s:AS}: the $\ell$
leaves, one internal vertex per member together with a root, and one
internal edge $e_A$ per member. For an internal vertex $v$ let
$\mathcal B_v$ be its set of branches --- the components of
$T\setminus\{v\}$, so $|\mathcal B_v|=\ell_v$ --- and let $\pi_v$ send
a leaf to the branch containing it. Each $B_b$ is a singleton, a
member of $\mathcal A$, or the complement of one.
\begin{figure}[h!]
\centering
\begin{tikzpicture}[scale=0.95]
\node[vtx,label=below:{root}] (r) at (0,0) {};
\node[vtx,label=below:{$\{1,2,3\}$}] (v1) at (2.2,0) {};
\node[vtx,label=below:{$\{1,2\}$}] (v2) at (4.4,0) {};
\node[leaf,label=left:{$4$}] (l4) at (-1.3,0.9) {};
\node[leaf,label=left:{$5$}] (l5) at (-1.6,0) {};
\node[leaf,label=left:{$6$}] (l6) at (-1.3,-0.9) {};
\node[leaf,label=above:{$3$}] (l3) at (2.2,1.25) {};
\node[leaf,label=right:{$1$}] (l1) at (5.7,0.75) {};
\node[leaf,label=right:{$2$}] (l2) at (5.7,-0.75) {};
\draw (r)--(l4) (r)--(l5) (r)--(l6) (v1)--(l3) (v2)--(l1) (v2)--(l2);
\draw[very thick] (r)--node[above]{$\tau_{A_2}$}(v1)
                  --node[above]{$\tau_{A_1}$}(v2);
\end{tikzpicture}
\caption{The tree of $\mathcal A=\{A_1,A_2\}$, $A_1=\{1,2\}\subset
  A_2=\{1,2,3\}$, at $\ell=6$: this is tree (d) of
  Section \ref{ss:trees}. The valences are $\ell_v=4,3,3$, so
  Theorem~\ref{T:gluing} gives $\dim W_{\mathcal A}=\dim\mathcal
  H_4+0+0+2=4$.}
\label{F:tree}
\end{figure}
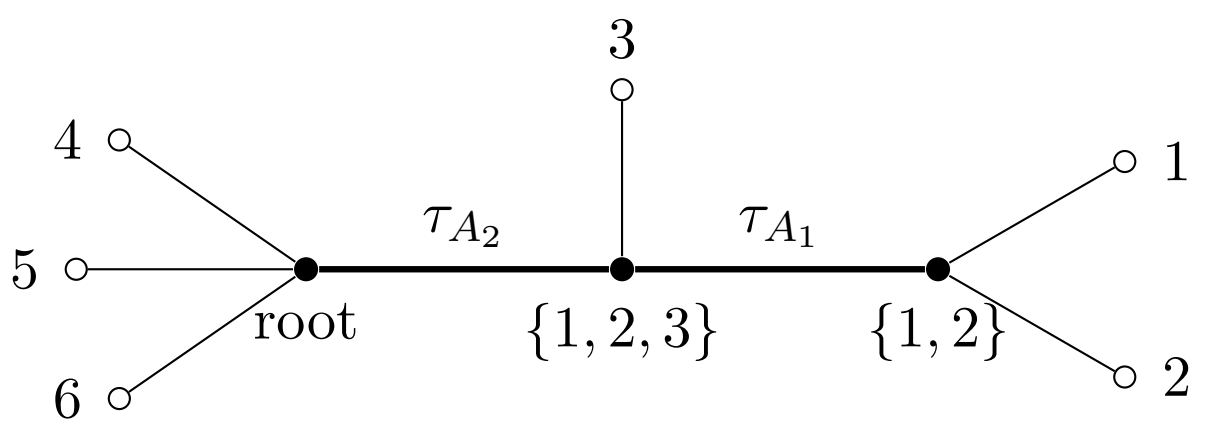
The gluing map is
\begin{equation}\label{eq:gamma}
\gamma_{\mathcal A}\big((\eta^{(v)}),(\tau_A)\big)=\sum_v P\pi_v^*\eta^{(v)}
+\sum_{A\in\mathcal A}\tau_A\,\E(A) ,
\qquad \eta^{(v)}\in\mathcal H_{\ell_v},\ \tau\in\R^{\mathcal A}:
\end{equation}
carry a mass matrix at each vertex out to the leaves along the branch
map, project to $\Hh$, and add the rays. One remark before the proofs:
because $\pi_v^*\vp(a)$ agrees with $\vp(\tilde a)$, $\tilde
a_i=a_{\pi_v(i)}$, off the diagonal blocks and vanishes on them,
\begin{equation}\label{eq:pullbacklambda}
\pi_v^*\vp(a)=\vp(\tilde a)-2\sum_ba_b\chi_{B_b},
\qquad\text{so}\qquad P\pi_v^*(\Lambda_{\ell_v})\subseteq V_{\mathcal A},
\end{equation}
each $P\chi_{B_b}$ being $0$ or $\kp^{-1}\E(A)$ by Lemma~\ref{L:rays} and
$\E(A)=\E(A^c)$.

\begin{figure}[h]
\centering
\begin{tikzpicture}[scale=0.9]
\begin{scope}
\node[vtx,label=below:{$u$}] (u) at (0,0) {};
\node[leaf,label=left:{$i$}] (i) at (-1.4,1.1) {};
\node[leaf,label=left:{$j$}] (j) at (-1.4,-1.1) {};
\node[leaf,label=right:{$k$}] (k) at (1.4,1.1) {};
\node[leaf,label=right:{$l$}] (l) at (1.4,-1.1) {};
\draw (u)--(i) (u)--(j) (u)--(k) (u)--(l);
\node at (0,-2.1) {(a) four distinct branches at $u$};
\end{scope}
\begin{scope}[xshift=7.4cm]
\node[vtx,label=below:{$v$}] (v) at (0,0) {};
\node[br,minimum width=20mm,minimum height=13mm] (bu) at (2.4,0.4) {};
\node at (2.4,0.4) {$u$};
\node[leaf] (i2) at (3.35,1.15) {};
\node[leaf] (j2) at (3.5,0.45) {};
\node[leaf] (k2) at (3.35,-0.25) {};
\node at (4.05,1.15) {$i$};\node at (4.2,0.45) {$j$};\node at (4.05,-0.25) {$k$};
\node[leaf,label=below:{$l$}] (l2) at (-1.5,-1.0) {};
\node[br,minimum width=11mm,minimum height=7mm] at (-1.9,0.8) {};
\draw (v)--(bu) (v)--(l2) (v)--(-1.9,0.8);
\node at (1.0,-2.1) {(b) the same four, seen from $v\ne u$};
\end{scope}
\end{tikzpicture}
\caption{Lemma~\ref{L:branch}(a) and Proposition~\ref{P:inj}: for
  $v\ne u$, all branches at $u$ except the one containing $v$ lie in a
  single branch at $v$, so at least three of the four leaves have the
  same image under $\pi_v$ and the $v$-term contributes nothing to the
  four-point defect.}
\label{F:inj}
\end{figure}
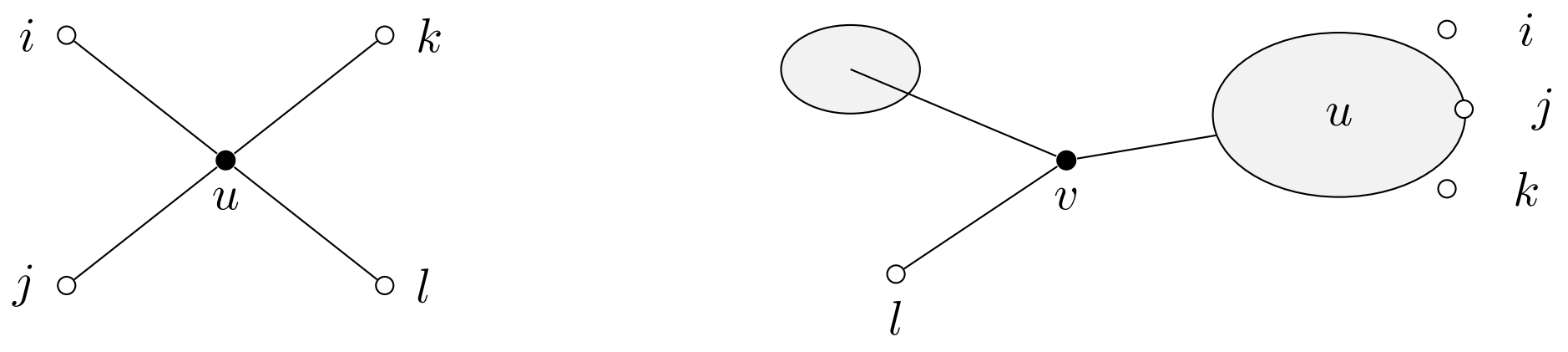

\begin{Lemma}\label{L:branch}
Let $u$ be an internal vertex of $T$ and $X\subset T$ a connected
subgraph not containing $u$. Then all leaves of $X$ lie in a single
branch at $u$. In particular:
\textup{(a)} if $v\ne u$ is another internal vertex, all branches at
$v$ except the one containing $u$ lie in a single branch at $u$
\textup{(}Figure~\ref{F:inj}(b)\textup{)};
\textup{(b)} if $A\in\mathcal A$, the leaves on the side of $e_A$ not
containing $u$ lie in a single branch at $u$.
\end{Lemma}
\begin{proof}
The branches at $u$ are the components of $T\setminus\{u\}$. For (a)
take $X$ to be $v$ together with its branches not containing $u$; for
(b), the component of $T\setminus e_A$ not containing $u$.
\end{proof}

\begin{Proposition}\label{P:into}
For any internal vertex $u$ and any symmetric, zero-diagonal matrix
$\theta$ indexed by the branches at $u$, we have $P\pi_u^*\theta\in
W_{\mathcal A}$. In particular, it follows that
$\operatorname{im}\gamma_{\mathcal A}\subseteq W_{\mathcal A}$.
\end{Proposition}

\begin{proof}
Fix $A\in\mathcal A$ and $i,k\in A$, $j,l\in A^c$; as $P$ preserves
defects it is enough to check $\pi_u^*\theta$ and $\chi_{A'}$. For
$\pi_u^*\theta$: if $u$ is on the $A$ side of $e_A$ then
$\pi_u(j)=\pi_u(l)=:b_0$ by Lemma~\ref{L:branch}(b), and the defect is
$\theta_{\pi_u(i)b_0}+\theta_{\pi_u(k)b_0}-\theta_{\pi_u(i)b_0}-\theta_{\pi_u(k)b_0}=0$;
if $u$ is on the $A^c$ side then $\pi_u(i)=\pi_u(k)$ and the same
cancellation occurs. For $\chi_{A'}$, nestedness leaves three cases: if
$A'\subseteq A$ then $j,l\notin A'$ and all four terms vanish; if
$A'\subseteq A^c$ then $i,k\notin A'$ and again all vanish; if
$A'\supseteq A$ then $i,k\in A'$ and the defect is $[j\in A']+[l\in
  A']-[l\in A']-[j\in A']=0$.
\end{proof}

\begin{Proposition}\label{P:inj}
The homomorphism $\gamma_{\mathcal A}$ is injective.
\end{Proposition}

\begin{proof}
  Suppose $\gamma_{\mathcal A}((\eta^{(v)}),(\tau_A))=0$, that is
$S:=\sum_v\pi_v^*\eta^{(v)}+\kp\sum_A\tau_A\chi_A\in\Lambda_\ell$, so
every defect of $S$ vanishes. Fix an internal vertex $u$; if
$\ell_u=3$ then $\mathcal H_{\ell_u}=0$ and there is nothing to prove,
so let $\ell_u\ge4$ and choose leaves $i,j,k,l$ in four distinct
branches at $u$.

Let $v\ne u$ be another internal vertex. By Lemma~\ref{L:branch}(a) at least three of
$\pi_v(i)$, $\pi_v(j)$, $\pi_v(k)$, $\pi_v(l)$ coincide; say three equal $b$ and the fourth is
$c$ (possibly $c=b$). Using $\theta_{bb}=0$, each position of the odd one out gives
$$
D(ij|kl)(\pi_v^*\eta^{(v)})=
\begin{cases}
0+\theta_{bc}-0-\theta_{bc}, & \pi_v(l)=c,\\
\theta_{bb}+\theta_{cb}-\theta_{bc}-\theta_{bb}, & \pi_v(k)=c,\\
\theta_{cb}+\theta_{bb}-\theta_{cb}-\theta_{bb}, & \pi_v(i)=c,
\end{cases}
$$
all zero, the case $\pi_v(j)=c$ being the second with $i,j$ exchanged. By
Lemma~\ref{L:branch}(b) all leaves on the far side of $e_A$ from $u$ lie in one
branch at $u$, so at least three
of the four are on the same side of $A$; with three inside and one outside the defect
of $\chi_A$ is $1+0-1-0$, $1+0-0-1$, $0+1-1-0$ or $0+1-0-1$ according to which is
outside, with three outside and one inside it is $0$, and with all four on one side it
is $1+1-1-1$ or $0$.

Only the $u$-term is left:
$$
0=D(ij|kl)(S)=D\big(\pi_u(i)\pi_u(j)\,\big|\,\pi_u(k)\pi_u(l)\big)(\eta^{(u)}),
$$
and every $4$-tuple of distinct branches at $u$ arises this way, so all defects of
$\eta^{(u)}$ vanish and
$\eta^{(u)}\in\Lambda_{\ell_u}\cap\mathcal H_{\ell_u}=0$. Then
$\kp\sum_A\tau_A\chi_A\in\Lambda_\ell$, so $\tau=0$ by Theorem~\ref{T:indep}.
\end{proof}
The inverse is restriction to one leaf per branch, followed by recentring.
\begin{Definition}\label{D:rho}
  For any $\zeta\in W_{\mathcal A}$ and any internal vertex $v$, we
select a representative leaf $r_b\in B_b$ from each branch $b$ and
define $\rho_v(\zeta):=P_{\ell_v}\big[(\zeta_{r_br_c})_{b\ne
    c}\big]\in\mathcal H_{\ell_v}.$
\end{Definition}

\begin{Lemma}\label{L:welldef}
The homomorphism $\rho_v$ is well-defined. For any internal vertex $u$
and any symmetric, zero-diagonal matrix $\theta$ indexed by the
branches at $u$, we have
$\rho_v(P\pi_u^*\theta)=\delta_{uv}\,P_{\ell_v}\theta$. Furthermore,
$\rho_v(\E(A))=0$ for all $A\in\mathcal A$. In particular, the
composition $\rho_v\circ\gamma_{\mathcal A}$ acts as the projection
onto the $v$-th summand.
\end{Lemma}

\begin{proof}

  If $r_b,r_b'\in B_b$ and $|B_b|\ge2$ then $B_b$ or $B_b^c$ lies in
$\mathcal A$ and $W_{B_b}=W_{B_b^c}$, so that block of $\zeta$ is
additive and $\zeta_{r_br_c}-\zeta_{r_b'r_c}=\mu_{r_b}-\mu_{r_b'}$ is
independent of $r_c$; changing representatives one at a time alters
the matrix by an element of $\Lambda_{\ell_v}$, which $P_{\ell_v}$
kills.

Next, $P\pi_u^*\theta=\pi_u^*\theta+\vp(a)$ for some $a\in\R^\ell$,
and the restriction of $\vp(a)$ to representatives lies in
$\Lambda_{\ell_v}$. For $u=v$, $(\pi_v^*\theta)_{r_br_c}=\theta_{bc}$,
so $\rho_v(P\pi_v^*\theta)=P_{\ell_v}\theta$.  For $u\ne v$, by
Lemma~\ref{L:branch}(a) all branches at $v$ but one have the same
$\pi_u$-image, so $(\pi_u^*\theta)_{r_br_c}$ vanishes unless $b$ or
$c$ is the exceptional branch $b_0$ and equals a constant $\alpha$
when exactly one of them is --- that is $\vp(a)$ with $a_{b_0}=\alpha$
and $a_b=0$ otherwise, which projects to $0$. Only $\theta_{bb}=0$ was
used. Finally, for $A\in\mathcal A$: if $v$ is on the $A^c$ side of
$e_A$ then $(\chi_A)_{r_br_c}=0$; if on the $A$ side then, with $b_0$
the branch containing $e_A$, $(\chi_A)_{r_br_c}=1-[b=b_0]-[c=b_0]$
when $r_{b_0}\notin A$, and $\equiv1$ when $r_{b_0}\in A$; both are in
$\Lambda_{\ell_v}$. Either way $\rho_v(\E(A))=\kp\rho_v(P\chi_A)=0$.
\end{proof}
Now we prove surjectivity.
\begin{Lemma}\label{L:rigid}
If $\xi\in W_{\mathcal A}$ and $\rho_v(\xi)=0$ for every internal
vertex $v$, then $\xi\in V_{\mathcal A}$.
\end{Lemma}
The induction collapses a minimal cluster, which is always a cherry.
\begin{Lemma}\label{L:cherry}
Let $A$ be minimal in $\mathcal A$ and $u$ the endpoint of $e_A$ on
the $A$ side. Then $A$ is the set of leaves attached to $u$, and
$\ell_u=|A|+1$.
\end{Lemma}
\begin{proof}
Let $C\subsetneq A$ denote the leaf set of another branch at $u$ such
that $\vert{}C\vert{}\ge2$. Then the partition $\{C,C^c\}$ forms an
internal split, which implies that either $C\in\mathcal A$ or
$C^c\in\mathcal A$. The former condition directly contradicts the
minimality of $A$. In the latter case, the intersection of $C^c$ and
$A$ is precisely $A\setminus C\ne\varnothing$. However, because
$C^c\not\subseteq A$ and $A\not\subseteq C^c$, this configuration
contradicts the nested property of $\mathcal A$.
\end{proof}

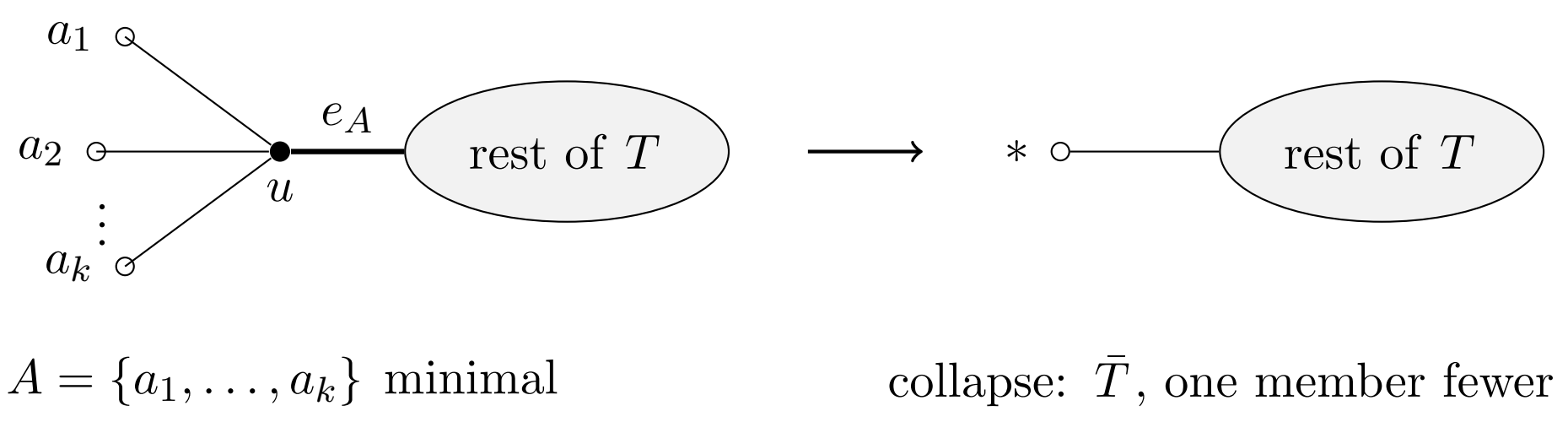
\begin{figure}[h]
\centering
\begin{tikzpicture}[scale=0.9]
\begin{scope}
\node[vtx,label=below:{$u$}] (u) at (0,0) {};
\node[leaf,label=left:{$a_1$}] at (-1.35,1.0) {};
\node[leaf,label=left:{$a_2$}] at (-1.6,0.0) {};
\node[leaf,label=left:{$a_k$}] at (-1.35,-1.0) {};
\node at (-1.55,-0.52) {$\vdots$};
\node[blob,minimum width=22mm] (rest) at (2.5,0) {rest of $T$};
\draw (u)--(-1.35,1.0) (u)--(-1.6,0) (u)--(-1.35,-1.0);
\draw[very thick] (u)--node[above]{$e_A$}(rest);
\node at (0,-2.0) {$A=\{a_1,\dots,a_k\}$ minimal};
\end{scope}
\begin{scope}[xshift=7.6cm]
\node[leaf,label=left:{$*$}] (st) at (-0.8,0) {};
\node[blob,minimum width=22mm] (rest2) at (2.0,0) {rest of $T$};
\draw (st)--(rest2);
\node at (0.6,-2.0) {collapse: $\bar T$, one member fewer};
\end{scope}
\draw[->,thick] (4.6,0)--(5.6,0);
\end{tikzpicture}
\caption{Lemmas~\ref{L:cherry} and \ref{L:rigid}. Vanishing local data
  at $u$ normalizes the $A$-block of $\xi$ to zero and makes the
  $A\times A^c$ block constant along $A$, which says exactly that
  $\xi$ is pulled back from the collapsed tree.}
\label{F:cherry}
\end{figure}

\begin{proof}[Proof of Lemma~\ref{L:rigid}]
  Induction on $|\mathcal A|$, working modulo $\Lambda_\ell$ with a
representative $S\in\Symz$ of $\xi$; we must show
$S\in\Lambda_\ell+\operatorname{span}\{\chi_A:A\in\mathcal A\}$.

If $\mathcal A=\varnothing$ then $T$ is a star, $\pi_u$ is a bijection
on leaves, and $\rho_u(\xi)=P\xi=\xi$, so $\xi=0$.

Let $A\in\mathcal A$ be minimal, $k=|A|$, and $u$ as in
Lemma~\ref{L:cherry}; fix $j_0\in A^c$. That $\rho_u(\xi)=0$ says the
matrix with entries $S_{ab}$ $(a\ne b\in A)$ and $S_{aj_0}$ is
additive: $S_{ab}=\alpha_a+\alpha_b$ and
$S_{aj_0}=\alpha_a+\alpha_*$. Since $\xi\in W_A$ we have
$S_{aj}=\mu_a+\mu_j$, so $\mu_a=\alpha_a+\text{const}$. Subtracting
$\vp(\beta)$ with $\beta_a=\alpha_a$ on $A$ and $0$ elsewhere, we may
assume
$$
S_{ab}=0\ (a,b\in A),\qquad S_{aj}=\nu_j\ \text{ independent of } a\in A,
$$
that is $S=\pi^*\bar S$ for the collapse $\pi$ of $A$ to a new leaf $*$, with
$\bar S_{jl}=S_{jl}$ and $\bar S_{*j}=\nu_j$ (Figure~\ref{F:cherry}).

Let $\bar{\mathcal A}$ be the induced family on $A^c\cup\{*\}$: a member
$B\supsetneq A$ becomes $(B\setminus A)\cup\{*\}$, a member disjoint from $A$ is
unchanged. It is nested and reduced --- reducedness of $\mathcal A$ rules out
$B=A^c$, which would give a member of size $\bar\ell-1$ --- with
$|\bar{\mathcal A}|=|\mathcal A|-1$ and tree $\bar T$. A defect of $\bar S$ not
involving $*$ is one of $S$; one involving $*$ becomes, on replacing $*$ by any
$a\in A$, a defect of $S$, since $a$ lies on the same side of $B$ as $*$ and
$\bar S_{*j}=S_{aj}$. So $\bar S\in W_{\bar{\mathcal A}}$, and
$\rho_v(\bar S)=\rho_v(S)=0$ for $v\ne u$, choosing $*$ as the representative of the
branch that contained $A$.

By induction $\bar S\in\Lambda_{\bar\ell}+\operatorname{span}\{\chi_{\bar B}\}$.
Pulling back, $\pi^*\chi_{\bar B}=\chi_B-\chi_A$ for $B\supsetneq A$ (the block
$A\times A$, which $\chi_B$ fills and $\pi^*\chi_{\bar B}$ leaves empty) and
$\pi^*\chi_{\bar B}=\chi_B$ for $B$ disjoint from $A$; and \eqref{eq:pullbacklambda}
applied to $\pi$ --- whose only non-singleton fibre is $A$ --- gives
$\pi^*\Lambda_{\bar\ell}\subseteq\Lambda_\ell+\R\chi_A$. Hence
$S\in\Lambda_\ell+\operatorname{span}\{\chi_B:B\in\mathcal A\}$ and $\xi\in V_{\mathcal A}$.
\end{proof}

\begin{Theorem}[Theorem~\ref{thm:mainC}, first part]\label{T:gluing}
For every reduced nested family $\mathcal A$ the map \eqref{eq:gamma}
is a linear isomorphism
$$
\gamma_{\mathcal A}:\ \bigoplus_v\mathcal H_{\ell_v}\oplus\R^{\mathcal A}
\ \xrightarrow{\ \sim\ }\ W_{\mathcal A},
$$
with inverse $\zeta\mapsto\big((\rho_v(\zeta)),(\tau_A)\big)$, where $\tau$ is
determined by
$\zeta-\sum_vP\pi_v^*\rho_v(\zeta)=\sum_{A\in\mathcal A}\tau_A\E(A)$. Hence
\begin{equation}\label{eq:gluingdim}
\dim W_{\mathcal A}=\sum_v\dim\mathcal H_{\ell_v}+|\mathcal A|,
\qquad
W_{\mathcal A}/V_{\mathcal A}\cong\bigoplus_v\mathcal H_{\ell_v}.
\end{equation}
\end{Theorem}

\begin{proof}
The image lies in $W_{\mathcal A}$ by Proposition~\ref{P:into}, and
$\gamma_{\mathcal A}$ is injective by Proposition~\ref{P:inj}. Given
$\zeta\in W_{\mathcal A}$, set $\eta^{(v)}=\rho_v(\zeta)$ and
$\xi=\zeta-\sum_vP\pi_v^*\eta^{(v)}$. Then $\xi\in W_{\mathcal A}$, and
$\rho_w(\xi)=0$ for all $w$ by Lemma~\ref{L:welldef}, so $\xi\in V_{\mathcal A}$ by
Lemma~\ref{L:rigid}, that is, $\xi=\sum_A\tau_A\E(A)$ with $\tau$ unique by
Theorem~\ref{T:indep}, and
$\zeta=\gamma_{\mathcal A}\big((\rho_v(\zeta)),\tau\big)$. The displayed statements
follow, the second because
$V_{\mathcal A}=\gamma_{\mathcal A}(0\oplus\R^{\mathcal A})$.
\end{proof}

\begin{Corollary}\label{C:dimW}
For a single split, $\dim W_A=\binom k2+\binom{\ell-k}2-1$ and
$\dim(W_A/\R\E(A))=\dim\mathcal H_{k+1}+\dim\mathcal H_{\ell-k+1}$. In
general, \eqref{eq:gluingdim} holds with $\vert{}\mathcal A\vert{}$
denoting the number of internal edges; for a family that is not
reduced, $\vert{}\mathcal A\vert{}$ represents the number of distinct
splits.
\end{Corollary}

\begin{proof}
It is enough to use $\dim\mathcal H_{m+1}=\binom m2-1$, applied twice.
\end{proof}

\begin{Remark}\label{R:numgluing}
For the family of Figure~\ref{F:tree} the codimensions of $W_{A_1}$,
$W_{A_2}$ and $W_{\mathcal A}$ are $3$, $4$ and $5$, not $3+4$: the
defects of different members are not independent, and no
transversality count replaces Lemma~\ref{L:rigid}.
\end{Remark}

\section{Renormalization}\label{s5}
This section proves Theorems~\ref{thm:mainB} and \ref{thm:mainD}.

\subsection{The expansion at a corner}

\begin{Lemma}\label{L:exp}
On $\mathcal U_{\mathcal A}$,
\begin{equation}\label{eq:exp}
L^\R(y(\sigma,t))=-\sum_{A\in\mathcal A}\tau_A\,\chi_A+\sum_{v\in V}\pi_v^*\alpha^{(v)}(\sigma_v)+E(\sigma,t),
\end{equation}
where $E$ is smooth on $\mathcal U_{\mathcal A}$ including $t=0$, $E(\sigma,0)=0$,
and the row of $\ell$ of every term is zero.
\end{Lemma}

\begin{proof}
Let $i\ne j<\ell$ and $v=v(i,j)$. By \eqref{eq:recon} the chains of $i$ and $j$
agree down to $v$, so
\[
y_i-y_j=\epsilon_{\le v}\Big[\big(a^{(v)}_{b_v(i)}-a^{(v)}_{b_v(j)}\big)+r_{ij}(\sigma,t)\Big],
\]
where $r_{ij}$ collects the deeper terms of the two chains divided by
$\epsilon_{\le v}$; each is a product of at least one $t_A$ with a screen point, and
screens are bounded by the normalization. The bracket does not vanish on
$\mathcal U_{\mathcal A}$: for $t=0$ the screen points are distinct, and for $t\ne0$
the points $y_i$ are distinct by the choice of $\delta$ in Definition~\ref{D:AS}. Hence
$L^\R_{ij}=\ln\epsilon_{\le v}+\alpha^{(v)}_{b_v(i)b_v(j)}+E_{ij}$ with $E_{ij}$
smooth and vanishing at $t=0$. Now $\ln\epsilon_{\le v}=-\sum_{A\supseteq A_v}\tau_A=-\sum_{A\ni i,j}\tau_A$
is $-\sum_A\tau_A(\chi_A)_{ij}$ (Lemma~\ref{L:chain}), and
$\alpha^{(v)}_{b_v(i)b_v(j)}=\sum_u(\pi_u^*\alpha^{(u)})_{ij}$, because for
$u\ne v$ on the path from $i$ to $j$ one of $i,j$ lies in the parent branch of $u$,
whose row of $\alpha^{(u)}$ is zero, and for $u$ off the path $\pi_u(i)=\pi_u(j)$.
For $j=\ell$ all terms vanish: $\ell\notin A$ for every member, $\pi_r(\ell)=\infty_r$,
and $\ell$ lies in the parent branch of every $v\ne r$.
\end{proof}

\subsection{The boundary value}

\begin{Theorem}[Theorem~\ref{thm:mainB}]\label{T:bv}
On $\mathcal U_{\mathcal A}$ the function $\eta-\Th=\eta-\sum_A\tau_A\E(A)$, defined
for $t\in(0,\delta)^{\mathcal A}$, extends smoothly to $t=0$. Its value on the corner is
\begin{equation}\label{eq:etaren}
\eta^\ren(\sigma)=\Pi\Big(\sum_{v\in V}\pi_v^*\alpha^{(v)}(\sigma_v)\Big)
=-\kp\sum_{v}P\big(\pi_v^*\alpha^{(v)}\big), \ \ \mbox{and}
\end{equation}
\begin{enumerate}[label=\textup{(\alph*)},leftmargin=*]
\item $\eta^\ren(\sigma)\in W_{\mathcal A}$ for every $\sigma$;
\item replacing the slices or the boundary defining functions, $t_A\mapsto u_At_A$ with
$u_A>0$ smooth, changes $\eta^\ren$ by $\sum_A(\ln u_A)|_{t=0}\,\E(A)$, an element of
$V_{\mathcal A}$ depending continuously on the point of the corner. So $\eta^\ren$
defines a continuous map $\partial^\circ_{\mathcal A}\to W_{\mathcal A}/V_{\mathcal A}$
on the open stratum $\partial^\circ_{\mathcal A}$ of $\mathcal A$ in
$\overline\Mo^{\,\mathrm{AS}}_\ell$, and \eqref{eq:etaren} is a continuous lift of it
to $W_{\mathcal A}$;
\item $\eta^\ren$ is invariant under the diagonal $O(n)$ and, more, under
$O(n)^{V}$ acting on the screens vertex by vertex.
\end{enumerate}
\end{Theorem}

\begin{proof}
By Lemma~\ref{L:chart}, $\eta=\Pi(L^\R)$; by Lemma~\ref{L:exp} and the linearity of
$\Pi$, $\eta=-\sum_A\tau_A\Pi(\chi_A)+\Pi(\sum_v\pi_v^*\alpha^{(v)})+\Pi(E)$, and
$\Pi(\chi_A)=-\E(A)$ by Lemma~\ref{L:rays}. So $\eta-\Th=\Pi(\sum_v\pi_v^*\alpha^{(v)})+\Pi(E)$,
smooth up to $t=0$ with the stated value there. (a) is Proposition~\ref{P:into}.
(b): $\tau_A\mapsto\tau_A-\ln u_A$. (c): $\alpha^{(v)}$ depends only on the distances
within the screen at $v$.
\end{proof}

\begin{Definition}\label{D:renorm}
The \emph{renormalized mass matrix} at a point of the open stratum of $\mathcal A$ is
the class $\eta^\ren\in W_{\mathcal A}/V_{\mathcal A}$ of Theorem~\ref{T:bv}(b); the
lift \eqref{eq:etaren} is its value for the normal-form corner chart with
$\tau_A=\ln(1/t_A)$. No limit along a path is involved: $\eta^\ren$ is the restriction to
the corner of a smooth function on the chart.
\end{Definition}

\subsection{The recentred reading}\label{ss:recentred}

\begin{Theorem}[Theorem~\ref{thm:mainC}, second part]\label{T:recentred}
For every internal vertex $v$,
\[
\rho_v\big(\eta^\ren\big)=\Xi_{\ell_v}\big(\hat\sigma_v\big)\in\mathcal H_{\ell_v},
\qquad
\hat\sigma_v:=\{\psi(a_b):b\in\mathcal B^\circ_v\}\cup\{\psi(\infty)\}\subset S^n ,
\]
the mass matrix of the canonical metric on the $\ell_v$-punctured sphere whose
punctures are the screen points of $v$ together with the point at infinity.
Consequently
\[
\eta^\ren=\gamma_{\mathcal A}\Big(\big(\Xi_{\ell_v}(\hat\sigma_v)\big)_{v\in V},\ \tau^\circ\Big)
\]
for a unique $\tau^\circ\in\R^{\mathcal A}$ (which depends on the slices), and the
surviving datum at the corner modulo $V_{\mathcal A}$ is exactly one mass matrix per
vertex of the bubble tree.
\end{Theorem}

\begin{proof}
By \eqref{eq:etaren} and Lemma~\ref{L:welldef},
$\rho_v(\eta^\ren)=-\kp P_{\ell_v}(\alpha^{(v)})=\Pi_{\ell_v}(\alpha^{(v)})$.
By \eqref{eq:alpha}, $\alpha^{(v)}$ is the Euclidean log-distance matrix of the
configuration $(a_b)$ with the row of the point at infinity set to zero, so by
Lemma~\ref{L:chart} applied with $\ell_v$ points, $\Pi_{\ell_v}(\alpha^{(v)})$ is the
mass matrix of $\hat\sigma_v$. The last statement is Theorem~\ref{T:gluing} applied
to $\eta^\ren\in W_{\mathcal A}$.
\end{proof}

In words: after the universal subtraction the clusters decouple, and what survives is
one canonical configuration per vertex --- the screen with its node adjoined as a
puncture --- read off by restricting to one leaf per branch and recentring. For a
single cluster the same can be written without the node row, at the price of a
correction term.

\begin{Corollary}[single cluster: the splitting formulas]\label{C:split}
Let $\mathcal A=\{A\}$, $|A|=k$. Then $\eta^\ren\in W_A$, so
$\eta^\ren_{ij}=\mu_i+\mu_j$ for $i\in A$, $j\notin A$, with $\mu$ unique up to
$\mu|_A\mapsto\mu|_A+c$, $\mu|_{A^c}\mapsto\mu|_{A^c}-c$. The mass matrix
$\eta^\bub:=\rho_{v_A}(\eta^\ren)\in\mathcal H_{k+1}$ of the bubble --- the
$(k+1)$-punctured sphere $\{\psi(a_a)\}_{a\in A}\cup\{\infty\}$ --- is
\begin{equation}\label{eq:splitbub}
\eta^\bub_{ij}=\eta^\ren_{ij}+\frac{\ell-k-1}{k-1}(\mu_i+\mu_j)+c_A,\qquad i,j\in A,
\end{equation}
$c_A$ being fixed by $\sum_{i\ne j\in A}\eta^\bub_{ij}=0$ and the node row by the
vanishing of row sums; symmetrically the mass matrix $\eta^\bas:=\rho_r(\eta^\ren)$ of
the base, the $(\ell-k+1)$-punctured sphere formed by the root screen and the node, is
\begin{equation}\label{eq:splitbase}
\eta^\bas_{ij}=\eta^\ren_{ij}+\frac{k-1}{\ell-k-1}(\mu_i+\mu_j)+c_{A^c},\qquad
i,j\notin A .
\end{equation}
The two coefficients are reciprocal, and \eqref{eq:splitbub}--\eqref{eq:splitbase} are
unchanged by adding a multiple of $\E(A)$ to $\eta^\ren$.
\end{Corollary}

\begin{proof}
Write $S=\pi_{v_A}^*\alpha^{\bub}+\pi_r^*\alpha^{\bas}$ for the matrix of
\eqref{eq:etaren}, $\hat f=\cent(S)$, so $\eta^\ren=-\kp S+\vp(\hat f)$. For
$a\in A$ and $j\notin A$ we have $S_{aj}=\alpha^\bas_{Aj}$, the entry of the node row
of the base, so $\eta^\ren_{aj}=\hat f_a+(\hat f_j-\kp\alpha^\bas_{Aj})$; we may therefore
take $\mu_a=\hat f_a$ for $a\in A$ and $\mu_j=\hat f_j-\kp\alpha^\bas_{Aj}$ for $j\notin A$.
Put $S_i=\sum_{b\in A,b\ne i}\alpha^\bub_{ib}$ and
$\Sigma=\sum_{a<b\in A}\alpha^\bub_{ab}$. For $i\in A$ the row sum of $S$ is
$S_i+R$, where $R=\sum_{j\notin A}S_{ij}=\sum_{j\notin A}\alpha^\bas_{Aj}$ is the
node row of the base and does not depend on $i\in A$, and the total $T$ of
\eqref{eq:cent} is $\Sigma+(k-1)R+\Sigma'$ with $\Sigma'$ the sum over the pairs of
root branches. Hence \eqref{eq:cent} with $\ell$ indices gives, for $i,j\in A$,
$$
\eta^\ren_{ij}=-\kp\alpha^\bub_{ij}+\frac{\kp}{\ell-2}(S_i+S_j)+c_1,\qquad
\mu_i=\frac{\kp}{\ell-2}S_i+c_2 ,
$$
with constants $c_1,c_2$ independent of $i,j\in A$ (they collect $\Sigma$, $\Sigma'$
and $R$), while by Theorem~\ref{T:recentred} and \eqref{eq:cent} with $k+1$ indices
(the node row of $\alpha^\bub$ being zero),
$$
\eta^\bub_{ij}=-\kp\alpha^\bub_{ij}+\frac{\kp}{k-1}(S_i+S_j)-\frac{2\kp\Sigma}{k(k-1)} .
$$
Subtracting, $\eta^\bub_{ij}-\eta^\ren_{ij}
=\kp\big(\tfrac1{k-1}-\tfrac1{\ell-2}\big)(S_i+S_j)+\mathrm{const}$, and
$\kp S_i=(\ell-2)\mu_i+\mathrm{const}$, giving \eqref{eq:splitbub} with coefficient
$\frac{\ell-2}{k-1}-1=\frac{\ell-k-1}{k-1}$. The constant is fixed because the
$A\times A$ block of an element of $\mathcal H_{k+1}$ has total sum zero.

For the base the computation is the same except for one point: the node appears once
in the base but once for each of the $k$ leaves of $A$ in $S$. For $i\notin A$ put
$S'_i=\sum_{b\notin A,\,b\ne i}\alpha^\bas_{ib}+\alpha^\bas_{iA}$, the row sum of
$\alpha^\bas$ in the $(\ell-k+1)$-point base; then the row sum of $S$ at $i$ is
$S'_i+(k-1)\alpha^\bas_{Ai}$. Hence, for $i,j\notin A$, with constants $c_3,c_4,c_5$,
\begin{gather*}
\eta^\ren_{ij}=-\kp\alpha^\bas_{ij}+\frac{\kp}{\ell-2}\big(S'_i+S'_j+(k-1)(\alpha^\bas_{Ai}+\alpha^\bas_{Aj})\big)+c_3,\\
\eta^\bas_{ij}=-\kp\alpha^\bas_{ij}+\frac{\kp}{\ell-k-1}(S'_i+S'_j)+c_4,\\
\mu_i=\frac{\kp}{\ell-2}\big(S'_i+(k-1)\alpha^\bas_{Ai}\big)-\kp\alpha^\bas_{Ai}+c_5 .
\end{gather*}
In $\eta^\bas_{ij}-\eta^\ren_{ij}$ the coefficient of $S'_i$ is
$\kp\big(\frac1{\ell-k-1}-\frac1{\ell-2}\big)=\frac{\kp(k-1)}{(\ell-k-1)(\ell-2)}$ and the
coefficient of $\alpha^\bas_{Ai}$ is $-\frac{\kp(k-1)}{\ell-2}$. In
$\frac{k-1}{\ell-k-1}\mu_i$ they are $\frac{\kp(k-1)}{(\ell-k-1)(\ell-2)}$ and
$\frac{\kp(k-1)}{\ell-k-1}\big(\frac{k-1}{\ell-2}-1\big)=-\frac{\kp(k-1)}{\ell-2}$, the same.
This gives \eqref{eq:splitbase}, the constant being fixed as before. Finally, adding
$t\E(A)$ changes the $A$-block and $\mu|_A$ each by a constant, hence changes
the right side of \eqref{eq:splitbub} by a constant, which the normalization removes.
\end{proof}

The correction term $\mu$ records how the $\ell$-point centering
differs from the $(k+1)$- and $(\ell-k+1)$-point centerings of the pieces; recentring
the blocks without it gives the wrong answer. In the language of
Definition~\ref{D:rho} the same is said by keeping the node row.

\subsection{The geometric reading}\label{ss:geom}

Let $S:=\sum_v\pi_v^*\alpha^{(v)}\in\Symz$ and $\hat f:=\cent(S)$, so that
$\eta^\ren=-\kp S+\vp(\hat f)$, i.e.
\begin{equation}\label{eq:etaren2}
\eta^\ren_{ij}=\hat f_i+\hat f_j-\kp\,\alpha^{(v(i,j))}_{b(i)b(j)} .
\end{equation}
For a vertex $v$ with $\Lambda_v\ne\varnothing$ define, on
$\R^n\setminus\{a_i:i\in\Lambda_v\}$,
\begin{equation}\label{eq:gv}
U^{(v)}(z)=\sum_{i\in\Lambda_v}e^{\hat f_i}\,G_{a_i}(z),\qquad
g^{(v)}:=\big(U^{(v)}\big)^{4/\kp}|dz|^2,
\qquad G_a(z)=|z-a|^{-\kp},\ \ G_\infty\equiv1,
\end{equation}
the constant term occurring only at the root, for the leaf $\ell$ at infinity.
Also put, for $i\in\Lambda_v$ (the members containing a leaf of $v$ are exactly the
$A\supseteq A_v$),
\begin{equation}\label{eq:Nv}
\ln N_v:=-\sum_{A\supseteq A_v}\tau_A\sigma_{|A|}+\sum_{A\not\supseteq A_v}\tau_A\sigma'_{|A|},
\qquad \lambda_v:=\epsilon_{\le v}^{\,2}\,N_v^{-4/\kp} .
\end{equation}

\begin{Lemma}[dominance]\label{L:dom}
Let $i\ne i'$ be leaves of $v$ and $m\notin\Lambda_v$. Then
\[
\Th_{mi'}-\Th_{ii'}=
\begin{cases}
-\displaystyle\sum_{A\ni m,\ A\subseteq B_b}\tau_A\,\omega_{|A|}, & m\in B_b,\ b\text{ a child of }v,\\[4mm]
-\displaystyle\sum_{A_v\subseteq A\subsetneq A_u}\tau_A(\kp-\omega_{|A|})
-\sum_{A\ni m,\ A\cap A_v=\varnothing}\tau_A\,\omega_{|A|}, & m\notin A_v,\ u=v(m,i),
\end{cases}
\]
and in both cases this tends to $-\infty$ as $t\to0$: in the first the sum contains
$B_b$, in the second the chain contains $A_v$, and $\omega_k<\kp$ for $k\le\ell-2$.
The same formulas give $\ln(e^{f^\R_m}/e^{f^\R_i})$ up to $\hat f_m-\hat f_i+O(t)$ in the
first case and $\ln\big(e^{f^\R_m}\epsilon_{\le u}^{-\kp}/e^{f^\R_i}\epsilon_{\le v}^{-\kp}\big)$ in the
second.
\end{Lemma}

\begin{proof}
A member containing $i,i'$ but not $m$ contributes $\E(A)_{AA^c}-\E(A)_{AA}=-(\kp-\omega_{|A|})$
to $\Th_{mi'}-\Th_{ii'}$; one containing $m$ but not $i,i'$ contributes
$\E(A)_{AA^c}-\E(A)_{A^cA^c}=-\omega_{|A|}$ (both by \eqref{eq:ids}); others contribute
$0$. Members containing $i$ are the $A\supseteq A_v$. For $m\in B_b$ none of these
misses $m$, and the members containing $m$ but not $i$ are those $m\in A\subseteq B_b$.
For $m\notin A_v$ the members containing $i$ but not $m$ form the chain from $A_v$
up to, but excluding, $A_u$. The last sentence is the same bookkeeping with
$\cent(\chi_A)_m-\cent(\chi_A)_i=\pm\omega_{|A|}$ from \eqref{eq:centchi} and
$\ln(\epsilon_{\le v}/\epsilon_{\le u})=-\sum_{A_v\subseteq A\subsetneq A_u}\tau_A$.
\end{proof}

\begin{Theorem}[Theorem~\ref{thm:mainD}, first part]\label{T:geom}
Let $v$ be a vertex with $\Lambda_v\ne\varnothing$, and $z$ the coordinate
$y=y_v+\epsilon_{\le v}z$ (with $y_r=0$). Then, as $t\to0$, uniformly for $\sigma$ in
compact sets and in $C^\infty_{loc}$ of $\R^n\setminus\{a_i:i\in\Lambda_v\}$,
\begin{enumerate}[label=\textup{(\roman*)},leftmargin=*]
\item $\lambda_v\,\psi^*g_{\pl(\sigma,t)}\longrightarrow g^{(v)}$; if $|\Lambda_v|\ge2$
then $\lambda_v=e^{-2\Th_{ii'}/\kp}$ for any $i\ne i'\in\Lambda_v$.
\item $g^{(v)}$ is scalar-flat, extends smoothly across every node (the points
$a_b$ for children $b$, and $\infty$ if $v\ne r$), and has exactly $|\Lambda_v|$
asymptotically flat ends, one at each leaf of $v$; it is flat $\R^n$ if $|\Lambda_v|=1$.
\item For $i,j\in\Lambda_v$: $\eta^\ren_{ij}=\ln\xi_{ij}(g^{(v)})$. For $i\in A_v$ and
$j\notin A_v$: $\eta^\ren_{ij}=\mu_i+\mu_j$ with $\mu_i=\hat f_i$, so the cross
entries are sums of log-weights.
\item If $|\Lambda_v|\ge2$ then for $i\in\Lambda_v$,
\[
e^{-\Th_{ii'}}\,\hm_i\big(g_{\pl(\sigma,t)}\big)\longrightarrow\hm_i(g^{(v)})=\sum_{j\in\Lambda_v\setminus\{i\}}e^{\eta^\ren_{ij}} .
\]
\end{enumerate}
\end{Theorem}

\begin{proof}
(i) By \eqref{eq:chartmetric} and \eqref{eq:xi}, $U:=e^{f^\R_\ell}+\sum_{m<\ell}e^{f^\R_m}|y-y_m|^{-\kp}$
and $\xi_{mi'}=e^{f^\R_m}e^{f^\R_{i'}}|y_m-y_{i'}|^{-\kp}$ (resp.\ $e^{f^\R_\ell}e^{f^\R_{i'}}$).
By Lemma~\ref{L:exp}, $f^\R=\cent(L^\R)=-\sum_A\tau_A\cent(\chi_A)+\hat f+O(t)$, so by
\eqref{eq:centchi} $f^\R_i=\ln N_v+\hat f_i+O(t)$ for $i\in\Lambda_v$. The leaf terms of
$U$ at $y=y_v+\epsilon_{\le v}z$ are $N_v\epsilon_{\le v}^{-\kp}e^{\hat f_i}|z-a_i|^{-\kp}(1+O(t))$,
since $y_i=y_v+\epsilon_{\le v}a_i$ exactly. For $m\in B_b$, $b$ a child, the term is
$e^{f^\R_m}\epsilon_{\le v}^{-\kp}|z-a_b+O(t)|^{-\kp}$; for $m\notin A_v$ it is
$e^{f^\R_m}|y_v-y_m|^{-\kp}(1+O(\epsilon_{\le v}|z|/\epsilon_{\le u}))$ with
$|y_v-y_m|\asymp\epsilon_{\le u}$, $u=v(m,i)$; and the constant term is $e^{f^\R_\ell}$,
the case $u=r$. By Lemma~\ref{L:dom} each of these is $o(N_v\epsilon_{\le v}^{-\kp})$
on compact sets. Hence $N_v^{-1}\epsilon_{\le v}^{\kp}U\to U^{(v)}$ with all derivatives,
and $\psi^*g=U^{4/\kp}\epsilon_{\le v}^2|dz|^2$ gives $\lambda_v$. For the formula
$\lambda_v=e^{-2\Th_{ii'}/\kp}$: $\Th_{ii'}=\sum_{A\supseteq A_v}\tau_A\E(A)_{AA}+\sum_{A\not\supseteq A_v}\tau_A\E(A)_{A^cA^c}$,
and by \eqref{eq:ids}
\[
-\tfrac2\kp\Th_{ii'}=-2\sum_{A\supseteq A_v}\tau_A+\tfrac4\kp\Big(\sum_{A\supseteq A_v}\tau_A\sigma_{|A|}-\sum_{A\not\supseteq A_v}\tau_A\sigma'_{|A|}\Big)=2\ln\epsilon_{\le v}-\tfrac4\kp\ln N_v .
\]

(ii) $U^{(v)}$ is harmonic off its poles, so $g^{(v)}$ is scalar-flat. Near a leaf
$a_i$ the metric is asymptotically flat in the inverted coordinate, exactly as in the
proof of \cite[Proposition~1.1]{PartI}. At $z=\infty$ (no constant term when $v\ne r$):
with $w=z/|z|^2$ one has $|dz|^2=|w|^{-4}|dw|^2$ and
$|z|^2|z-a|^{-2}=(1-2a\cdot w+|a|^2|w|^2)^{-1}$, so
$g^{(v)}=\big(|z|^{\kp}U^{(v)}(z)\big)^{4/\kp}|dw|^2$ with
$|z|^{\kp}U^{(v)}=\sum_ie^{\hat f_i}(1-2a_i\cdot w+|a_i|^2|w|^2)^{-\kp/2}$ smooth and
positive near $w=0$. The points $a_b$ of child branches are not poles of $U^{(v)}$
at all.

(iii) By \eqref{eq:etaren2} and \eqref{eq:xi} applied to $g^{(v)}$, whose weights are
$w_i=e^{\hat f_i}$ and whose constant is $e^{\hat f_\ell}$ (root only):
$\xi_{ij}(g^{(v)})=e^{\hat f_i+\hat f_j}|a_i-a_j|^{-\kp}$, resp.\ $e^{\hat f_i+\hat f_\ell}$,
which is $e^{\eta^\ren_{ij}}$. For $i\in A_v$, $j\notin A_v$ the vertex $v(i,j)=u$ and
the branch $b_u(i)$ are the same for all $i\in A_v$ --- write $b_u(v)$ for this branch
at $u$, the one containing the vertex $v$ --- so \eqref{eq:etaren2} is
$\hat f_i+[\hat f_j-\kp\alpha^{(u)}_{b_u(v)b_u(j)}]$.

(iv) $\hm_i=\sum_j\xi_{ij}$ with $\xi_{ij}=e^{\eta_{ij}}=e^{\Th_{ij}+\eta^\ren_{ij}+O(t)}$.
For $j\in\Lambda_v$, $\Th_{ij}=\Th_{ii'}$; for $j\notin\Lambda_v$, Lemma~\ref{L:dom}.
\end{proof}

\begin{Proposition}[necks; Theorem~\ref{thm:mainD}, second part]\label{P:neck}
Let $b$ be a child of $v$ with $\Lambda_v\ne\varnothing\ne\Lambda_b$, and put
$W_e:=\sum_{m\in B_b}e^{f^\R_m}$, $C_e:=U(y_b)-\sum_{m\in B_b}e^{f^\R_m}|y_b-y_m|^{-\kp}$
(the rest of $U$ evaluated at the node), $r_e:=(W_e/C_e)^{1/\kp}$. Then in the
coordinate $y=y_b+r_ex$,
\[
C_e^{-1}\,U(y_b+r_ex)\longrightarrow1+|x|^{-\kp},\qquad
r_e^{-2}C_e^{-4/\kp}\,\psi^*g_{\pl}\longrightarrow\big(1+|x|^{-\kp}\big)^{4/\kp}|dx|^2
\]
in $C^\infty_{loc}(\R^n\setminus0)$: the spatial Schwarzschild metric, unique up to
scale, with minimal sphere $\{|x|=1\}$. Moreover
\[
r_e\asymp\epsilon_{\le v}\,t_{B_b}^{\frac{|B_b|-1}{\ell-2}},\qquad
\epsilon_{\le b}\ll r_e\ll\epsilon_{\le v},\qquad
r_e^{-2}C_e^{-4/\kp}\asymp e^{-2\Th_{ij}/\kp}\ \ (i\in\Lambda_v,\ j\in\Lambda_b).
\]
\end{Proposition}

\begin{proof}
For $x$ in a compact subset of $\R^n\setminus0$ the terms outside $B_b$ vary on the
scale $\epsilon_{\le v}$ of the distances from $y_b$ to the other branches at $v$,
giving $C_e(1+O(r_e/\epsilon_{\le v}))$, and the terms in $B_b$ give
$W_er_e^{-\kp}|x|^{-\kp}(1+O(\epsilon_{\le b}/r_e))$ since $|y_m-y_b|\lesssim\epsilon_{\le b}$;
$W_er_e^{-\kp}=C_e$ by definition. By Lemma~\ref{L:dom}, $C_e\asymp N_v\epsilon_{\le v}^{-\kp}$
(dominated by the leaves of $v$) and $W_e\asymp N_b$ (dominated by the leaves of $b$),
so $r_e^{\kp}\asymp(N_b/N_v)\epsilon_{\le v}^{\kp}$, and $\ln(N_b/N_v)=-\tau_{B_b}\omega_{|B_b|}$
because $B_b$ is the only member containing a leaf of $b$ but not a leaf of $v$.
Finally $\Th_{ij}-\Th_{ii'}=-\tau_{B_b}\omega_{|B_b|}$ by Lemma~\ref{L:dom} with $m=j$,
which turns $2\ln\epsilon_{\le v}-\frac4\kp\ln N_v+\frac2\kp\tau_{B_b}\omega_{|B_b|}$ into
$-\frac2\kp\Th_{ij}$. The inversion $x\mapsto x/|x|^2$ is an isometry of the limit,
since $1+|x|^{-\kp}=|x|^{-\kp}(1+|x|^{\kp})$, so $\{|x|=1\}$ is minimal.
\end{proof}

\begin{Remark}\label{R:table}
Theorem~\ref{T:geom}(i) and Proposition~\ref{P:neck} say that, up to constants, the
rescaling needed to see the piece of the limit containing the ends $i$ and $j$ is
$e^{-2\Th_{ij}/\kp}$: for $i,j$ leaves of one vertex this is that vertex's metric; for
$i,j$ leaves of adjacent vertices it is the neck between them. For a single cluster,
$\Th=\tau\E(A)$ has three distinct entries, $\E(A)_{AA},\E(A)_{A^cA^c}>0>\E(A)_{AA^c}$:
the bubble and the base are large in $g_{\pl}$ and are scaled down, the neck is small
and is scaled up. Masses scale like $(\text{length})^{\kp}$, which is why the exponents
of \eqref{eq:rates} are the entries of $\E(A)$.
\end{Remark}

\begin{Remark}\label{R:pinch}
By Lemma~\ref{L:stable} and Proposition~\ref{P:necksphere} below, for $t$ small the
neck on the edge $(v,b)$ contains a minimal sphere $S_e$ of $g_{\pl}$, close to
$\{|y-y_b|=r_e\}$ in the coordinate $x$. There $U=2C_e(1+o(1))$, so its radius in
$g_{\pl}$ is $(2C_e)^{2/\kp}r_e(1+o(1))\asymp e^{\Th_{ij}/\kp}$
\textup{(}$i\in\Lambda_v$, $j\in\Lambda_b$\textup{)}. This need not tend to zero: for
$\ell=6$ and $A_1=\{1,2\}\subset A_2=\{1,2,3\}$, on the edge between the vertices of
$A_2$ and $A_1$ one has $\Th_{31}=\kp\,(0.3\,\tau_{A_2}-0.15\,\tau_{A_1})$, which tends to
$+\infty$ when $\tau_{A_2}>\tau_{A_1}/2$. What tends to zero is its size relative to the
adjacent pieces: in the normalization in which $g^{(v)}$ converges it is
$\asymp r_e/\epsilon_{\le v}$, and in that of $g^{(b)}$ it is
$\asymp t_{B_b}^{(\ell-|B_b|-1)/(\ell-2)}$, and both tend to zero. Seen from either side the node closes up, which is the geometric form of
Theorem~\ref{T:geom}(ii). For $n=3$ the minimal sphere $S_e$ is an apparent horizon
of the time-symmetric data $g_{\pl}$ \cite{Gi}, enclosing the whole cluster $B_b$:
a common horizon of the merging holes. Whether it is a component of the outermost
apparent horizon of an end outside the cluster is the subject of Section~\ref{ss:horizons}.
\end{Remark}

\begin{Remark}\label{R:leafless}
If $\Lambda_v=\varnothing$, the terms of $U$ at scale $\epsilon_{\le v}$ are the
child terms, of sizes $\asymp N_b\epsilon_{\le v}^{-\kp}$ for the children $b$, and
which of them dominate depends on the ratios $N_b/N_{b'}$, i.e.\ on the relative rates
at which the children's scales go to zero. If one child $b$ dominates, the limit at
scale $\epsilon_{\le v}$ is $\asymp(N_b|z-a_b|^{-\kp})^{4/\kp}|dz|^2$, flat space with
$a_b$ at infinity: $v$ is absorbed into the bubble of $b$ as a flat neighbourhood of
that bubble's node, and there is no neck on the edge $(v,b)$. If several children
have comparable $N_b$, the piece at $v$ has ends at those nodes, with weights
depending on the ratios. In either case the geometric limit at $v$ depends on the
path to the corner, whereas $\eta^\ren$ does not (Theorem~\ref{T:bv}). This is why
the geometric reading is stated only at vertices with leaves --- which is where the
masses are --- and why $\eta^\ren$, not the limit metric, is the object attached to
a boundary point.
\end{Remark}

\begin{Remark}\label{R:two}
The block of $\eta^\ren$ on $\Lambda_v$ can be read raw, as $\ln\xi(g^{(v)})$,
whose node-excluded row sums are the limiting masses (Theorem~\ref{T:geom}); or
recentred, as $\rho_v(\eta^\ren)=\Xi_{\ell_v}(\hat\sigma_v)$, the mass matrix of the
canonical metric on the screen with the node adjoined (Theorem~\ref{T:recentred}).
The two carry the same information, and the second is what the gluing isomorphism is
built from; but the canonical $\ell_v$-punctured metric is not a limit of the
$g_{\pl}$: it has $\ell_v$ ends, the limit $g^{(v)}$ has $|\Lambda_v|$. Explicitly,
since $\hat f_i-\hat f_j=\frac{\kp}{\ell-2}\big(\sum_{c\ne i}S_{ic}-\sum_{c\ne j}S_{jc}\big)$,
the weights of $g^{(v)}$ are
\begin{equation}\label{eq:weights}
e^{\hat f_i}=K_v\prod_{b\in\mathcal B^\circ_v\setminus\{i\}}|a_i-a_b|^{\frac{\kp|B_b|}{\ell-2}}\qquad(i\in\Lambda_v),
\end{equation}
each screen point weighted by the number of leaves it carries, with $K_v$ independent
of $i$. For a single cluster this is $\prod_{c\in A\setminus\{i\}}|a_i-a_c|^{\kp/(\ell-2)}$,
whereas the canonical $(k+1)$-point metric with the node at $\infty$ has weights
$\propto\prod_{c\ne i}|a_i-a_c|^{\kp/(k-1)}$ plus a positive constant term and the
canonical $k$-point metric ($k\ge3$) has weights $\propto\prod_{c\ne i}|a_i-a_c|^{\kp/(k-2)}$:
the limit bubble remembers $\ell$ and is canonical for neither. The $\R\E(A)$
ambiguity is a single parameter rescaling $g^\bub$ and $g^\bas$ together, by
$e^{2t\E(A)_{AA}/\kp}$ and $e^{2t\E(A)_{A^cA^c}/\kp}$: the neck couples the units of
length on the two sides.
\end{Remark}

\begin{Corollary}\label{C:floors}
The recentred data satisfy the bounds of \cite[Corollary~2.9]{PartI}:
every end of the canonical metric on $\hat\sigma_v$ has
$\hm\ge\ell_v-1$, in particular $\hm^\bub_i\ge k$ and
$\hm^\bas_j\ge\ell-k$ for a single cluster. The geometric limits
$g^{(v)}$ satisfy no universal floor: their weights \eqref{eq:weights}
are not row-normalized, and the arithmetic--geometric mean inequality
gives $\hm_i(g^{(v)})\ge(|\Lambda_v|-1)\big(\prod_{j\ne
  i}\xi_{ij}(g^{(v)})\big)^{1/(|\Lambda_v|-1)}$, a bound that moves
with the $V_{\mathcal A}$ scale.
\end{Corollary}

\subsection{Necks, horizons, and the Penrose ratio}\label{ss:horizons}
For $n=3$ the metric $g_{\pl}$ is time-symmetric initial data, and its apparent
horizons are its minimal surfaces \cite{Gi}. This subsection examines them along a
degeneration; the algebra is written for general $n$, and $n=3$ is assumed wherever
horizons or the Penrose inequality are mentioned. Two of the statements, the
visibility criterion and the behaviour of the Penrose ratio at the boundary, rest on
the convergence of outermost minimal surfaces along the degeneration, which we do not
prove; they are stated as conjectures, with the supporting argument, and what is
proved about the limit geometries is separated out.

Let $(M^3,g)$ be complete, scalar-flat, with $\ell$ asymptotically flat ends, and
fix an end $i$. The \emph{trapped region} of end $i$ is the union of the compact
regions bounded by closed minimal surfaces (together with the other ends); its
boundary $\Sigma_i$ is the \emph{outermost minimal surface} of end $i$. By
Huisken--Ilmanen \cite{HI}, $\Sigma_i$ is a smooth, embedded, outer-minimizing
minimal surface whose components are spheres, and every closed minimal surface in
$M$ is enclosed by $\Sigma_i$. For $K=0$ data, $\Sigma_i$ is the apparent horizon of
end $i$. The \emph{exterior region} $\ext_i\subset M$ is the component of
$M\setminus\Sigma_i$ containing end $i$. We write $|\Sigma_i|=\Area_g(\Sigma_i)$.

For $n=3$ (Huisken--Ilmanen for connected $\Sigma_i$, Bray \cite{Bray} in general),
$M^{\rm ADM}_i\ge\sqrt{|\Sigma_i|/16\pi}$, with equality iff $\ext_i$ is isometric to the
exterior of a Schwarzschild horizon. In the normalization of \cite{PartI},
\begin{equation}\label{eq:penrose}
|\Sigma_i|\le 64\pi\,\hm_i^{\,2},\qquad\text{equality iff $\ext_i$ is Schwarzschild.}
\end{equation}
Both sides scale like length$^2$: the inequality is invariant under homothety.

\begin{Lemma}\label{L:stable}
The horizon $\{|x|=1\}$ of $(1+|x|^{-\kp})^{4/\kp}|dx|^2$ is strictly stable: its
Jacobi operator $L=-\Delta_\Sigma-(\Ric(\nu,\nu)+|A|^2)$ is positive.
\end{Lemma}

\begin{proof}
The horizon is the fixed set of the inversion $x\mapsto x/|x|^2$, an isometry, so it
is totally geodesic, $A=0$. The Gauss equation with $\scal_g=0$ and $H=0$ gives
$\scal_\Sigma=-2\Ric(\nu,\nu)$, and $\Sigma$ is a round $S^{n-1}$, so
$\Ric(\nu,\nu)=-\tfrac12\scal_\Sigma<0$ and $L=-\Delta_\Sigma+\tfrac12\scal_\Sigma>0$.
\end{proof}

\begin{Proposition}[neck spheres]\label{P:necksphere}
Along the degeneration, for every edge $e=(v,b)$ with
$\Lambda_v\ne\varnothing\ne\Lambda_b$ and all sufficiently small scales
there is a minimal sphere $S_e$ of $g_{\pl}$ which, in the coordinate $x$ of
Proposition~\ref{P:neck}, is $C^\infty$-close to $\{|x|=1\}$. It is enclosed by the outermost
minimal surface $\Sigma_i$ of every end $i\notin B_b$, and from the other side by
$\Sigma_j$ for every $j\in B_b$.
\end{Proposition}

\begin{proof}
The rescaled metrics converge in $C^\infty_{loc}$ to Schwarzschild on a
neighbourhood of $\{|x|=1\}$ (Proposition~\ref{P:neck}), and the limit surface is
strictly stable (Lemma~\ref{L:stable}), so the implicit function theorem gives a
unique nearby minimal sphere. The enclosure statements are the defining property of
the outermost minimal surface.
\end{proof}

The minimal sphere of the \emph{limit} is $\{|x|=1\}$; for finite scales there is a
minimal sphere $C^\infty$-close to it, with the radii of Remark~\ref{R:pinch}. The
question is whether $S_e$ is a component of $\Sigma_i$, or hidden behind one.

\subsubsection*{Visibility}
Fix a vertex $v$ with $|\Lambda_v|\ge1$, a leaf $i\in\Lambda_v$, and a child $b$
with $\Lambda_b\ne\varnothing$, so that the neck sphere $S_e$, $e=(v,b)$, of
Proposition~\ref{P:necksphere} exists.
In the limit piece $g^{(v)}$, which has $|\Lambda_v|$ ends and in which the node
$a_b$ is a regular point, let $\Sigma^{(v)}_i$ be the outermost minimal surface of
end $i$ and $\ext^{(v)}_i$ its exterior region (if $|\Lambda_v|=1$, $g^{(v)}$ is
flat, $\Sigma^{(v)}_i=\emptyset$ and $\ext^{(v)}_i=\R^n$).

\begin{Conjecture}[visibility criterion]\label{C:vis}
For sufficiently small scales, the neck sphere $S_e$, $e=(v,b)$, is a component of
$\Sigma_i$ if $a_b\in\ext^{(v)}_i$, and is not a component of $\Sigma_i$ if $a_b$
lies in the interior of the trapped region of end $i$ in $g^{(v)}$.
\end{Conjecture}

The argument. In the units of $v$ the neck sphere shrinks to the regular point
$a_b$, and $\lambda_vg_{\pl}\to g^{(v)}$ in $C^\infty_{loc}$ away from the leaves
and nodes. If $\Sigma_i$ converges, away from the nodes, to $\Sigma^{(v)}_i$, then:
when $a_b$ is enclosed by $\Sigma^{(v)}_i$, the sphere $S_e$ lies in the interior
of the trapped region of $g_{\pl}$ for small scales and cannot meet its boundary;
when $a_b\in\ext^{(v)}_i$, $\Sigma_i$ must still enclose $S_e$
(Proposition~\ref{P:necksphere}), and in a neighbourhood of $a_b$ of size
$\ll1$ in $v$-units and $\gg r_e$ the geometry is Schwarzschild, whose only closed
minimal surface is its horizon; so the component of $\Sigma_i$ enclosing $S_e$ is
$S_e$ itself. If $a_b\in\Sigma^{(v)}_i$ one expects $S_e$ to be absorbed into a
single sphere with a small bump; this borderline case occurs, exactly, in
Corollary~\ref{C:border}. What is not proved is the convergence of the outermost
minimal surfaces; a strictly stable $\Sigma^{(v)}_i$ (the generic case) perturbs to
a minimal surface of $g_{\pl}$ by the implicit function theorem, which settles the
hidden direction, but the identification of that perturbation with a component of
$\Sigma_i$ needs an outer-minimizing argument that we do not carry out.

It would be interesting to describe, in terms of the tree and the screens, which
necks are visible from which ends.

Independently of the conjecture, which nodes lie in which exterior regions of the
limit pieces is an exact question about the explicit metrics \eqref{eq:gv}. The
answer depends on the end $i$, and on the screen, not only on the tree
(Example~\ref{E:horizons}). If the conjecture holds, $S_e$ is at most one component
of $\Sigma_i$, and $\Sigma_i$ consists of neck spheres alone only when $v$ has the
single leaf $i$.

\begin{Proposition}[vertices with two leaves]\label{P:twoleaf}
Let $v$ have exactly two leaves $i,i'$, and let $\mathcal C_v$ be its set of children.
Then $g^{(v)}$ is the
spatial Schwarzschild metric, its horizon is the Apollonius sphere
\[
\Sigma^{(v)}_i=\Sigma^{(v)}_{i'}=\big\{z:\ e^{\hat f_i}|z-a_i|^{-\kp}=e^{\hat f_{i'}}|z-a_{i'}|^{-\kp}\big\},
\]
and $\ext^{(v)}_i=\{e^{\hat f_i}|z-a_i|^{-\kp}>e^{\hat f_{i'}}|z-a_{i'}|^{-\kp}\}$.
A node $a_c$, $c\in\mathcal C_v$, lies in $\ext^{(v)}_i$, on the horizon, or in
$\ext^{(v)}_{i'}$ according as
\begin{equation}\label{eq:R}
R_c:=\Big(\frac{|a_i-a_c|}{|a_{i'}-a_c|}\Big)^{\kp(|B_c|-\ell+2)/(\ell-2)}
\prod_{b\in\mathcal C_v\setminus\{c\}}\Big(\frac{|a_i-a_b|}{|a_{i'}-a_b|}\Big)^{\kp|B_b|/(\ell-2)}
\end{equation}
is $>1$, $=1$ or $<1$. If $v\ne r$, the parent node $\infty$ lies in
$\ext^{(v)}_i$, on the horizon, or in $\ext^{(v)}_{i'}$ according as
$e^{\hat f_i}/e^{\hat f_{i'}}=\prod_{b\in\mathcal C_v}(|a_i-a_b|/|a_{i'}-a_b|)^{\kp|B_b|/(\ell-2)}$
is $>1$, $=1$ or $<1$. If $v=r$, one of the two leaves is the leaf $\ell$ at
infinity, say $i'=\ell$, and the same holds with every factor $|a_{i'}-a_b|$ in
\eqref{eq:R} replaced by $1$, and the horizon is a round sphere about $a_i$.
\end{Proposition}

\begin{proof}
A two-ended metric of the form \eqref{eq:gv} is Schwarzschild: a M\"obius map
sending $a_{i'}\mapsto\infty$, $a_i\mapsto0$ puts $U^{(v)}$ in the form
$w|x|^{-\kp}+w'$ (the constant term, if $v=r$, is the end at $\infty$). Its unique
closed minimal surface is the fixed set of the inversion exchanging the two ends,
which is where the two terms of $U^{(v)}$ agree, and the exterior region of end $i$
is where the $i$-term dominates. Evaluating the ratio of the two terms at $a_c$ with
the weights \eqref{eq:weights}, the common factor $|a_i-a_{i'}|^{\kp/(\ell-2)}$
cancels, the factor from $c$ itself is $(|a_i-a_c|/|a_{i'}-a_c|)^{\kp|B_c|/(\ell-2)}$
from the weights times $(|a_i-a_c|/|a_{i'}-a_c|)^{-\kp}$ from the Green's functions,
and the other children contribute the product in \eqref{eq:R}. At $\infty$ the
Green's functions contribute equally and only the weights remain. If $i'=\ell$ is the
leaf at infinity, its term is the constant $e^{\hat f_\ell}$, and since the row of
$\ell$ in $S$ is zero, \eqref{eq:cent} gives
$e^{\hat f_i}/e^{\hat f_\ell}=\prod_{b}|a_i-a_b|^{\kp|B_b|/(\ell-2)}$, the product over
the finite branches $b\ne i$ at $r$, which are the children of $r$. The ratio of the two
terms at $a_c$ is then $R_c$ with the factors $|a_{i'}-a_b|$ omitted, and the horizon is
the sphere $|z-a_i|=(e^{\hat f_i}/e^{\hat f_\ell})^{1/\kp}$.
\end{proof}

\begin{Corollary}[a single cluster of $\ell-2$ holes]\label{C:border}
Let $A$ be a single cluster with $|A|=\ell-2$, so that the root has two leaves
$i,i'$ and the one child $A$. Then the root piece is the Schwarzschild metric
joining the two spectators, and the node of the cluster lies \emph{exactly on its
horizon}, for every position of the node and every $\ell$: in the chart with
$i'=\infty$, the horizon is the sphere $|z-a_i|=|a_i-a_A|$ through the node.
\end{Corollary}

\begin{proof}
In \eqref{eq:R} with a single child $c=A$ the exponent is $\kp(|A|-\ell+2)/(\ell-2)=0$,
so $R_A=1$. In the chart with $i'$ at infinity, $U^{(r)}=e^{\hat f_i}|z-a_i|^{-\kp}+C$
with $e^{\hat f_i}/C=|a_i-a_A|^{\kp}$ by \eqref{eq:weights} (the row of the leaf at
infinity being zero in the chart dictionary of Lemma~\ref{L:chart}), and the
horizon is $|z-a_i|=(e^{\hat f_i}/C)^{1/\kp}$.
\end{proof}

In Corollary~\ref{C:border} the root carries one spectator on each side of the
Apollonius sphere, and the exponent of the child's factor in \eqref{eq:R} vanishes
because $|A|=\ell-2$; so the node lies on the limiting horizon for every screen. In
this case the neck is not expected to be
a component of $\Sigma_i$ for finite scales but to be absorbed into the deformed
horizon of the root piece.

\begin{Remark}\label{R:nonroot}
If $v\ne r$ has two leaves and a single child $c$, then
$|B_c|=|A_v|-2\le\ell-4$, the exponent in \eqref{eq:R} is negative,
and the node $c$ is visible from the \emph{nearer} leaf:
$a_c\in\ext^{(v)}_i$ iff $|a_i-a_c|<|a_{i'}-a_c|$. The parent node is
visible from the \emph{heavier} leaf, which by \eqref{eq:weights} is
the one farther from the children, weighted by the numbers of holes
they hide. With several children the conditions couple through the
product in \eqref{eq:R}.
\end{Remark}

\begin{Example}\label{E:horizons}
Let $n=3$, $\ell=5$, $A=\{1,2\}$ a single cluster at the root. In the chart with
$5=\infty$ the root screen consists of the spectators $a_3=(0,0,\tfrac12)$,
$a_4=(0,0,-\tfrac12)$ and the node $a_A=(h,0,0)$ on their perpendicular bisector,
$h\ge0$; the screen is taken modulo similarity, so $h$ is the ratio of the pair's
distance from the spectators to the spectators' separation. The root piece is, after
dividing by its constant,
\begin{equation}\label{eq:ex}
U^{(r)}=1+\rho\,|z-a_3|^{-1}+\rho\,|z-a_4|^{-1},\qquad \rho=\big(\tfrac14+h^2\big)^{1/3}.
\end{equation}
It is Brill--Lindquist data for two equal holes of bare mass $M=2\rho$ at
separation $d=1$, with $d/M=1/2\rho\le0.794$, so for every $h$ the end at infinity
has a common apparent horizon enclosing both spectators. Indeed, by
\eqref{eq:weights}, $e^{\hat f_3}/C=|a_3-a_4|^{1/3}|a_3-a_A|^{2/3}=(\tfrac14+h^2)^{1/3}$
and the same for $a_4$ (checked against the full canonical metric at
$\epsilon=10^{-6}$ for $h=0,1,8$ to six digits); the critical separation for a common
horizon in equal-mass Brill--Lindquist data is $d/M\approx1.53$ \cite{BL,Alc,Th}, and
$\rho\ge(1/4)^{1/3}=0.630$.

The common horizon $\Sigma^{(r)}_\infty$ of \eqref{eq:ex} is axisymmetric about the
axis of the spectators and reflection-symmetric; by uniqueness of the outermost
minimal surface it is the surface found by an axisymmetric horizon finder (ancillary
files; it returns the Schwarzschild horizon to six digits and the equal-mass critical
separation $1.530$--$1.535$ of \cite{Alc,Th}). Its polar height $z_0$ and equatorial radius $r_{\rm eq}$, and
whether the node $(h,0,0)$ lies inside it:

\begin{center}
\begin{tabular}{@{}r rrr l@{}}
\toprule
$h$ & $\rho$ & $z_0$ & $r_{\rm eq}$ & node \\
\midrule
0 & 0.630 & 1.388 & 1.179 & inside\\
1 & 1.077 & 2.234 & 2.111 & inside\\
2 & 1.620 & 3.294 & 3.212 & inside\\
4 & 2.533 & 5.101 & 5.048 & inside\\
6 & 3.310 & 6.646 & 6.606 & inside\\
8 & 4.005 & 8.033 & 7.999 & outside\\
10 & 4.645 & 9.310 & 9.281 & outside\\
\bottomrule
\end{tabular}
\end{center}

The threshold is $h_c=7.998$. So, granting Conjecture~\ref{C:vis}: for $h<h_c$ the
neck of the merging pair is hidden inside the spectators' common horizon, and for
$h>h_c$ it is a separate component of the apparent horizon of the end at infinity.
Both outcomes occur in one boundary stratum. The value is essentially $2\rho=h$,
i.e.\ $h=8$: for large $h$ the two close spectators look, from far away, like one
Schwarzschild hole of radius $2\rho\approx2h^{2/3}$, and the pair's node is outside
it iff $h>2h^{2/3}$.
\end{Example}

\subsubsection*{The floor and the Penrose ratio}
The floor $\hm_i\ge\ell-1$ is an arithmetic--geometric mean inequality in the
canonical gauge; its gauge-free form is
$\hm_i\ge(\ell-1)\big(\prod_{j\ne i}\xi_{ij}\big)^{1/(\ell-1)}$. It fixes a scale.
The Penrose inequality \eqref{eq:penrose} is scale-invariant and gauge-free. There is
nothing to reconcile between them; what the canonical gauge does is turn the
scale-invariant quantity into a function on the moduli space.

\begin{Definition}\label{D:P}
For $n=3$ and an end $i$, the \emph{Penrose ratio} of the canonical metric is
\[
P_i:=\frac{|\Sigma_i|}{64\pi\,\hm_i^{\,2}},\qquad P_i:\Mo_\ell(S^3)\to(0,1).
\]
By \eqref{eq:penrose}, $P_i\le1$ with equality iff $\ext_i$ is isometric to a
Schwarzschild exterior, and this never happens for $\ell\ge3$. Indeed, both metrics
are conformal to the round metric, so by Liouville's theorem such an isometry is the
restriction of a M\"obius transformation. Then on the open set $\ext_i$ the conformal
factor $\Phi_{\pl}$ would equal a positive combination of the Green's functions
of the conformal Laplacian at two points. By unique continuation this identity would
hold on all of $S^3$ minus the poles, whereas $\Phi_{\pl}$ has $\ell\ge3$ poles. For $\ell=3$, $P=0.99978$
\cite[Question~6.1]{PartI}.
\end{Definition}

\begin{Proposition}[scaling along a degeneration]\label{P:scaling}
Let $i\in\Lambda_v$ with $|\Lambda_v|\ge2$, $i'\ne i$ another leaf of $v$. Then
$\hm_i(g_{\pl})=e^{\Th_{ii'}}\hm_i(g^{(v)})(1+o(1))$ with $\Th_{ii'}\to+\infty$,
and the area of any family of surfaces converging in $v$-units to a surface of area
$a$ in $g^{(v)}$ is $e^{(n-1)\Th_{ii'}/\kp}a\,(1+o(1))$ in $g_{\pl}$. In particular
the neck spheres have area $o(e^{(n-1)\Th_{ii'}/\kp})$.
\end{Proposition}

\begin{proof}
The mass statement is Theorem~\ref{T:geom}(iv); $\Th_{ii'}\to+\infty$ because
$\Th_{ii'}$ dominates its row Lemma~\ref{L:dom} and the row sums of
$\Th\in\Hh$ vanish. Lengths scale by $\lambda_v^{-1/2}=e^{\Th_{ii'}/\kp}$, areas by
its $(n-1)$-st power. The neck spheres shrink to points in $v$-units.
\end{proof}

\begin{Conjecture}[the Penrose ratio at the boundary]\label{C:penrose}
\textup{(a)} On the closure of the strata in which hole $i$ shares its vertex $v$
with another hole, $P_i$ extends continuously, with boundary value
$P_i(g^{(v)})$, the Penrose ratio of end $i$ in the limit piece \eqref{eq:gv}.
\textup{(b)} At a corner where $i$ is the only leaf of its vertex $v$ and every
vertex adjacent to $v$ carries a hole, with necks $e$ at $v$ of Schwarzschild masses
$m_e$ in $v$-units,
\[
P_i=\frac{\sum_e m_e^2}{\big(\sum_e m_e\big)^2}+o(1);
\]
since the ratios $m_e/m_{e'}$ are powers of the relative scales, $P_i$ does not
extend continuously to such corners, its limit along a path being $1$ whenever one
neck dominates and $1/\#\{e\}$ when all are balanced.
\end{Conjecture}

For (a): by Proposition~\ref{P:scaling} the factor $e^{2\Th_{ii'}}$ cancels in $P_i$
($n=3$), the necks contribute nothing to the area in the limit, and
Conjecture~\ref{C:vis} identifies the limit of $\Sigma_i$ with $\Sigma^{(v)}_i$
plus necks. Note that $g^{(v)}$ is not the canonical metric of the screen
Remark~\ref{R:two}: the boundary value is the Penrose ratio of a
non-canonical datum, with the weights \eqref{eq:weights}. For (b): $g^{(v)}$ is flat,
$\Sigma^{(v)}_i=\emptyset$, every node is in $\ext^{(v)}_i$, so by the criterion
$\Sigma_i$ is the union of the neck spheres, of areas $16\pi m_e^2$, while the mass
of the flat end with small Schwarzschild necks is $\sum_em_e$ to leading order.
Under the hypothesis of (b) every edge at $v$ carries a neck
(Proposition~\ref{P:necksphere}), and there are at least two of them (a non-root
vertex has its parent edge and at least one child; the root with one leaf has at
least two children, since $k=\ell-1$ is excluded).

\begin{Remark}
For $3\le n\le7$ the Penrose inequality holds in the form of Bray--Lee
\cite{BrayLee}, and $P_i$ is defined by normalizing
$|\Sigma_i|^{(n-2)/(n-1)}/\hm_i$ to be $1$ on Schwarzschild; it is again scale-invariant,
and Proposition~\ref{P:scaling} and the argument for Conjecture~\ref{C:penrose} go
through verbatim.
\end{Remark}

\section{Reading the data off the mass matrix}\label{s7}

\subsection{Trees}\label{ss:trees}
We draw the tree with the root (the surviving configuration on $S^n$) on the left, one
internal vertex per collapsing cluster, and the $\ell$ labelled leaves. A vertex of
valence $\ell_v$ carries a configuration of $\ell_v$ punctures and contributes
$\mathcal H_{\ell_v}$, of dimension $\ell_v(\ell_v-3)/2$; each edge carries a length
$\tau$.

\medskip
\begin{center}
\begin{tabular}{@{}c@{\qquad}c@{\qquad}c@{}}
\begin{tikzpicture}[scale=0.55,every node/.style={font=\scriptsize}]
\draw (0,0)--(2,0); \fill (0,0) circle (2.2pt); \fill (2,0) circle (2.2pt);
\draw (0,0)--(-0.9,0.7) node[left]{$3$}; \draw (0,0)--(-0.9,-0.7) node[left]{$4$};
\draw (2,0)--(2.9,0.7) node[right]{$1$}; \draw (2,0)--(2.9,-0.7) node[right]{$2$};
\node at (1,0.35) {$\tau$};
\end{tikzpicture}
&
\begin{tikzpicture}[scale=0.55,every node/.style={font=\scriptsize}]
\draw (0,0)--(2,0); \fill (0,0) circle (2.2pt); \fill (2,0) circle (2.2pt);
\draw (0,0)--(-0.9,0.9) node[left]{$3$}; \draw (0,0)--(-1.1,0) node[left]{$4$};
\draw (0,0)--(-0.9,-0.9) node[left]{$5$};
\draw (2,0)--(2.9,0.7) node[right]{$1$}; \draw (2,0)--(2.9,-0.7) node[right]{$2$};
\node at (1,0.35) {$\tau$};
\end{tikzpicture}
&
\begin{tikzpicture}[scale=0.55,every node/.style={font=\scriptsize}]
\draw (0,0)--(2,0); \fill (0,0) circle (2.2pt); \fill (2,0) circle (2.2pt);
\draw (0,0)--(-0.9,0.9) node[left]{$4$}; \draw (0,0)--(-1.1,0) node[left]{$5$};
\draw (0,0)--(-0.9,-0.9) node[left]{$6$};
\draw (2,0)--(2.9,0.9) node[right]{$1$}; \draw (2,0)--(3.1,0) node[right]{$2$};
\draw (2,0)--(2.9,-0.9) node[right]{$3$};
\node at (1,0.35) {$\tau$};
\end{tikzpicture}
\\[2mm]
(a) $\ell=4$, $2{+}2$ & (b) $\ell=5$, $2{+}3$ & (c) $\ell=6$, $3{+}3$\\[1mm]
$\mathcal H_3\oplus\mathcal H_3=0$ &
$\mathcal H_3\oplus\mathcal H_4=\R^2$ &
$\mathcal H_4\oplus\mathcal H_4=\R^4$
\end{tabular}

\medskip
\begin{tabular}{@{}c@{\qquad\qquad}c@{}}
\begin{tikzpicture}[scale=0.55,every node/.style={font=\scriptsize}]
\draw (0,0)--(2,0)--(4,0); \fill (0,0) circle (2.2pt); \fill (2,0) circle (2.2pt);
\fill (4,0) circle (2.2pt);
\draw (0,0)--(-0.9,0.9) node[left]{$4$}; \draw (0,0)--(-1.1,0) node[left]{$5$};
\draw (0,0)--(-0.9,-0.9) node[left]{$6$};
\draw (2,0)--(2,-1.1) node[below]{$3$};
\draw (4,0)--(4.9,0.7) node[right]{$1$}; \draw (4,0)--(4.9,-0.7) node[right]{$2$};
\node at (1,0.35) {$\tau_{A_2}$}; \node at (3,0.35) {$\tau_{A_1}$};
\end{tikzpicture}
&
\begin{tikzpicture}[scale=0.55,every node/.style={font=\scriptsize}]
\draw (0,0)--(2,1); \draw (0,0)--(2,-1);
\fill (0,0) circle (2.2pt); \fill (2,1) circle (2.2pt); \fill (2,-1) circle (2.2pt);
\draw (0,0)--(-0.9,0.7) node[left]{$5$}; \draw (0,0)--(-0.9,-0.7) node[left]{$6$};
\draw (2,1)--(3.0,1.6) node[right]{$1$}; \draw (2,1)--(3.1,0.7) node[right]{$2$};
\draw (2,-1)--(3.1,-0.7) node[right]{$3$}; \draw (2,-1)--(3.0,-1.6) node[right]{$4$};
\node at (1.0,0.85) {$\tau_{A_1}$}; \node at (1.0,-0.85) {$\tau_{A_2}$};
\end{tikzpicture}
\\[2mm]
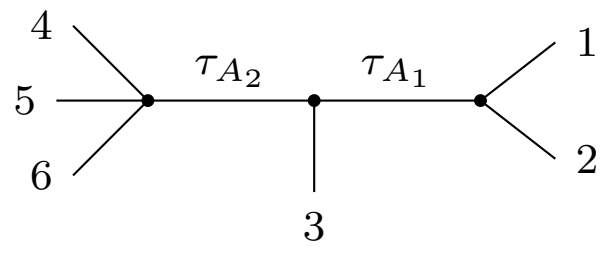
(d) $\ell=6$, nested $\{1,2\}\subset\{1,2,3\}$ &
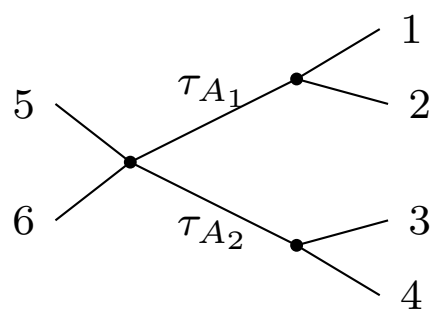
(e) $\ell=6$, disjoint $\{1,2\},\{3,4\}$\\[1mm]
$\mathcal H_4\oplus\mathcal H_3\oplus\mathcal H_3=\R^2$ &
$\mathcal H_4\oplus\mathcal H_3\oplus\mathcal H_3=\R^2$
\end{tabular}
\end{center}

\medskip
Two features are visible already. A vertex of valence $3$ carries
$\mathcal H_3=0$: three-ended pieces are rigid, their canonical mass matrix
vanishes and all three masses equal $2$. A vertex of valence $4$ carries
$\mathcal H_4\cong\R^2$ and, by \cite[Proposition~5.1]{PartI}, has all four masses equal. Note also
that the surviving dimension is much smaller than $\dim(\Hh/V_{\mathcal A})$: in (d)
and (e) the latter is $9-2=7$ while the former is $2$; the difference is the
constraint $\eta^\ren\in W_{\mathcal A}$. Geometrically, in (d) the middle vertex has a
single leaf and its limit metric is flat (Theorem~\ref{T:geom}(ii)); the two necks
on either side of it are Schwarzschild (Proposition~\ref{P:neck}).

\subsection{The procedure}

\begin{Proposition}[reading the data]\label{Alg}
Let $\eta=\Xi(\pl)$ for a configuration $\pl$ close to the corner of a reduced nested
family $\mathcal A$ in $\overline\Mo^{\,\mathrm{AS}}_\ell$. The following steps recover
the data of Theorems~\ref{thm:mainB}--\ref{thm:mainD} from $\eta$ alone.
\begin{enumerate}[label=\textup{(\roman*)},leftmargin=*]
\item {\em Detect the tree.} By \eqref{eq:ray} the entries $\eta_{ij}$ are large and
positive for $i,j$ on the same side of a split and large and negative across;
equivalently $M_{ij}\to0$ for $i,j$ on the same side. For a single split this sign
pattern determines the split. In general the splits, hence $\mathcal A$, are found by
neighbor joining, which returns the tree of $\mathcal A$ once the edge lengths are large
compared with the bounded part of $L$ (Remark~\ref{R:NJ}). By $\E(A)=\E(A^c)$ it does not matter which side is called the cluster,
and we take the normal form.
\item {\em Measure the necks.} Put
$\tau=G^{-1}\big(\langle\eta,\E(A_v)\rangle\big)_v$ with
$G_{vw}=\langle\E(A_v),\E(A_w)\rangle$, which is invertible by
Theorem~\ref{T:indep}; for a single split $\tau_A=\langle\eta,\E(A)\rangle/\|\E(A)\|^2$,
and for $\ell=5$ the entries of $G$ are those of Proposition~\ref{P:petersen5}. This
is the \emph{orthogonal lift}: $\eta-\sum_A\tau_A\E(A)\perp V_{\mathcal A}$. It differs
from the chart lift $\eta-\sum_A\ln(1/t_A)\E(A)$ of Theorem~\ref{T:bv} by an element
of $V_{\mathcal A}$ which stays bounded as $t\to0$, i.e.\ by the ambiguity of
Theorem~\ref{T:bv}(b).
\item {\em Subtract.} $\eta^\ren:=\eta-\sum_A\tau_A\E(A)$.
\item {\em Check the corner.} The defects \eqref{eq:defect} of $\eta^\ren$ are $O(t)$
by Theorem~\ref{T:bv}; their size measures the distance to the corner.
\item {\em Split.} For each vertex $v$ output $\rho_v(\eta^\ren)$, the mass matrix of
the canonical metric on the screen at $v$ with its node (Theorem~\ref{T:recentred});
for a single split this is \eqref{eq:splitbub}--\eqref{eq:splitbase}. For each vertex
with at least two leaves output also $\sum_{j\in\Lambda_v\setminus i}e^{\eta^\ren_{ij}}$,
which by Theorem~\ref{T:geom}(iv) is the mass of the $i$-th end of the limit metric
$g^{(v)}$ up to the constant rescaling of $g^{(v)}$ caused by the choice of lift
in~\textup{(ii)}.
\end{enumerate}
The output --- one canonical $\ell_v$-punctured sphere per vertex with its mass
matrix, the masses of the limit metrics up to those constants, and the neck lengths up to
bounded additive constants --- is independent of whether the clusters are split off
one at a time or all at once.
\end{Proposition}

\begin{proof}
(i) is Theorem~\ref{T:degen}, Proposition~\ref{P:nested} and Remark~\ref{R:NJ}. For (ii), both lifts of
$\eta$ to $W_{\mathcal A}$ modulo $V_{\mathcal A}$ differ by an element of
$V_{\mathcal A}$; the chart lift converges by Theorem~\ref{T:bv}, and its
$V_{\mathcal A}$-component is bounded because $\eta^\ren$ is (the chart lift is not
orthogonal to the rays in general), so the orthogonal lift
converges to $\eta^\ren$ minus its orthogonal projection to $V_{\mathcal A}$. The
projections $\rho_v$ kill $V_{\mathcal A}$ (Lemma~\ref{L:welldef}), which gives (v)
from Theorems~\ref{T:gluing} and \ref{T:recentred}, and adding $c\E(A)$ to
$\eta^\ren$ multiplies the block of $e^{\eta^\ren}$ on $\Lambda_v$ by a constant,
which is the rescaling of $g^{(v)}$ by Remark~\ref{R:two}. (iii)--(iv) are
Theorem~\ref{T:bv}, and the last sentence is Theorem~\ref{T:gluing}.
\end{proof}

\begin{Remark}\label{R:NJ}
Writing out \eqref{eq:cent}, with $T_i=\sum_{k\ne i}L_{ik}$ and $T=\sum_{a<b}L_{ab}$,
\[
\eta_{ij}=-\kp L_{ij}+\frac{\kp}{\ell-2}(T_i+T_j)-\frac{2\kp T}{(\ell-1)(\ell-2)}
=-\frac{\kp}{\ell-2}\,Q_{ij}(L)-\frac{2\kp T}{(\ell-1)(\ell-2)}\qquad(i\ne j),
\]
where $Q_{ij}(d)=(\ell-2)\,d_{ij}-r_i-r_j$, $r_i=\sum_{k\ne i}d_{ik}$, is the
neighbor-joining matrix of Saitou--Nei in the form of Studier--Keppler \cite{SN,SK}.
So the mass matrix is, up to the factor $-\kp/(\ell-2)$ and a constant common to all
off-diagonal entries, $Q(L)$, and step~(i) of
Proposition~\ref{Alg} --- take the pair with the largest entry of $\eta$, collapse it
to a node, repeat --- is the neighbor-joining algorithm run on $L$. The algorithm is
invariant under $L\mapsto L+(a_i+a_j)$, and by Proposition~\ref{R:treespace}(1) the
divergent part of $L$ in Proposition~\ref{P:nested} is, modulo such matrices, the
tree metric $\frac12\sum_A\ln(1/t_A)\,\delta_A$ of the tree of $\mathcal A$ with
internal edge lengths $\frac12\ln(1/t_A)$. Atteson's theorem \cite{At} --- neighbor
joining recovers the tree from any matrix within $\ell_\infty$-distance less than half
the shortest internal edge length of a tree metric --- therefore makes step~(i)
quantitative: with $B$ the bound of Proposition~\ref{P:nested} on the bounded part of
$L$, step~(i) returns the tree of $\mathcal A$ whenever
$\min_{A\in\mathcal A}\ln(1/t_A)>4B$. Buneman's four-point condition \cite{Bu}, which
characterizes tree metrics, is the vanishing of the defects \eqref{eq:defect} for every
split of the tree, and the subspace $W_{\mathcal A}$ of Section~\ref{s10} is the space of mass
matrices satisfying it for the splits of $\mathcal A$ alone.
\end{Remark}

The procedure is the inverse of gluing. Applying the contraction to the polystable
quotient (Section~\ref{s8}) amounts to running it and keeping only the screen at the
centroid vertex of the tree; applying the tropical map amounts to keeping only $\tau$
and the combinatorial type.

\subsection{Examples}
In the examples $\hm^\bub$, $\hm^\bas$ denote the masses of the recentred data, i.e.\
of the canonical metrics on the screens with their nodes; $\tau_A$ is the orthogonal
lift of Proposition~\ref{Alg}(ii). All numbers are for $n=3$ and were computed from
the mass matrix alone.

\begin{Example}[a completely symmetric merger]\label{E:sym}
$n=3$, $\ell=6$, $A=\{1,2,3\}$, screen $\{a_1,a_2,a_3\}$ an equilateral triangle and
base $\{b_4,b_5,b_6\}$ an equilateral triangle on the unit sphere of the chart, so
that inversion at the node fixes it. Both pieces are then regular simplices with four
vertices, and the procedure returns, for every $\epsilon$,
$$
\eta^\ren=\tau_A\E(A)\ \text{exactly},\qquad \mu\equiv0,\qquad
\eta^\bub=\eta^\bas=0,
$$
so all four masses of each canonical piece equal $3$, the minimum allowed by
the floor of \cite[Corollary~2.9]{PartI}, while the six masses of the original system
diverge like $\epsilon^{-3/10}$ ($7.21$, $23.39$, $91.00$ at $\epsilon=10^{-2},10^{-4},
10^{-6}$). The orthogonal lift is $\tau_A=\ln(1/\epsilon)-\ln3$ exactly: the constant
is the choice of scale in the screen, i.e.\ exactly the $\R\E(A)$ ambiguity of
Definition~\ref{D:renorm}; see Proposition~\ref{Alg}(ii).
\end{Example}

\begin{Example}[$\ell=5$: the bubble is rigid]\label{E:l5}
$n=3$, $\ell=5$, $A=\{1,2\}$, $\epsilon=10^{-6}$. The five masses are
$(1139.12,1139.12,19.89,21.45,22.15)$: the merging pair and the three spectators
diverge at the different rates $\epsilon^{-1/2}$ and $\epsilon^{-1/6}$ of \eqref{eq:rates},
the ratio $10^3$ against $10^1$ being visible in the raw numbers. The procedure
returns $\eta^\bub=0$, $\hm^\bub=(2,2,2)$ (the three-punctured sphere is unique, so
the merged pair, seen at its own scale, is always the same object) and a base
with $\hm^\bas=(3.0068,\dots)$. All surviving moduli are in the base; this is tree
(b). Geometrically the same rigidity appears as follows: the limit metric $g^{(v)}$ at
the cluster has two ends and is conformally flat and scalar-flat, hence Schwarzschild,
so the merged pair always leaves a Schwarzschild bubble, joined to the base by a
Schwarzschild neck.
\end{Example}

\begin{Example}[two-step merging]\label{E:nested}
$n=3$, $\ell=6$, nested $A_1=\{1,2\}\subset A_2=\{1,2,3\}$. In the chart with $p_6$
at infinity take
\[
y_1=(\epsilon_1,0,0),\quad y_2=(-\epsilon_1,0,0),\quad y_3=(\epsilon_2,\epsilon_2,0),\quad
y_4=(1,0,0),\quad y_5=(0,2,0),
\]
with $\epsilon_2=10^{-2}$ and $\epsilon_1=10^{-6}$, so that, up to bounded constants,
$\tau_{A_2}=\ln(1/\epsilon_2)=4.605$ and $\tau_{A_1}=\ln(\epsilon_2/\epsilon_1)=9.210$.
The screens are $\{0,y_4,y_5\}$ at the root, $\{0,(1,1,0)\}$ at the vertex of $A_2$
(the node of $A_1$ at $0$), and $\{(\pm1,0,0)\}$ at the vertex of $A_1$. The six masses
are
\[
(778.37,\ 778.37,\ 5.01,\ 16.67,\ 21.90,\ 22.51).
\]
For $\Th=\tau_{A_1}\E(A_1)+\tau_{A_2}\E(A_2)$ the row maxima are
$e^{\Th_{12}}=\epsilon_1^{-3/5}\epsilon_2^{3/10}=10^3$ for the merging pair,
$e^{\Th_{45}}=\epsilon_1^{-1/10}\epsilon_2^{-1/5}=10$ for the spectators $4,5,6$, and
$e^{\Th_{31}}=\epsilon_1^{3/20}\epsilon_2^{-9/20}$ and
$e^{\Th_{34}}=\epsilon_1^{-1/10}\epsilon_2^{3/10}$ for hole~$3$, both equal to $1$ at
these scales, in agreement with Theorem~\ref{thm:mainA}(ii).

\emph{Splitting off $A_2$ alone.} The orthogonal lift is $\tau_{A_2}=7.404$, not
$4.605$. The single-split lift picks up $\tfrac13\tau_{A_1}$, because
$\langle\E(A_1),\E(A_2)\rangle=\tfrac13\|\E(A_2)\|^2$, and $\ln(1/\epsilon_2)+\frac13\ln(\epsilon_2/\epsilon_1)=7.675$.
The cross-block defects of $\eta-\tau_{A_2}\E(A_2)$ are $1.0\cdot10^{-2}$, of order
$\epsilon_2$. The two four-ended canonical pieces have masses $3.0496$ (base, all four
equal) and $19.65$ (middle, all four equal), to within $O(\epsilon_2)$ depending on the
representative leaves in Definition~\ref{D:rho}. These are the masses of the canonical
metrics on $\{0,y_4,y_5\}\cup\{\infty\}$ and on
$\{y_1,y_2,y_3\}/\epsilon_2\cup\{\infty\}$, as Theorem~\ref{T:recentred} predicts. The
middle piece is itself close to a degeneration, with $1,2$ at relative distance
$2\cdot10^{-4}$, which is why its masses are large.

\emph{Splitting off $A_1$ inside the middle piece.} The orthogonal lift is
$\tau_{A_1}=8.874$, and the cross-block defect is $9.8\cdot10^{-5}$, of order
$\epsilon_1/\epsilon_2=10^{-4}$ rather than $\epsilon_1$ or $\epsilon_2$: the defect
reports the small parameter relevant at that step. The middle piece splits into two
three-ended pieces with all masses $2$.

\emph{Both clusters at once.} The lift of Proposition~\ref{Alg}(ii) with the Gram
matrix of $\E(A_1),\E(A_2)$ gives $(\tau_{A_1},\tau_{A_2})=(8.869,\,4.448)$, within
bounded constants of $(9.210,\,4.605)$, and the same base piece with masses $3.0496$,
as Theorem~\ref{T:gluing} requires. The final tree is (d): the only surviving mass
data is the two-dimensional $\mathcal H_4$ of the root. These numbers are reproduced
by the ancillary script \texttt{example75.py}.
\end{Example}

\section{Compactifications}\label{s8}

\subsection{The polystable quotient and its boundary}\label{ss:psb}
We recall the setting of \cite[Section~4.2]{PartI}. Let $X:=(S^n)^\ell$, let
$\Xst\subset\Xss\subset X$ be the stable and semistable loci of the equally weighted
configurations ($k_\zeta<\ell/2$, resp.\ $\le\ell/2$, for every $\zeta\in S^n$), and let
$q\colon\Xss\to\MopsL$ be the closed-orbit quotient; the closed $\Mob(S^n)$-orbits in
$\Xss$ are those of the polystable configurations \cite[Lemma~4.6]{PartI}, the quotient
is compact and metrizable and contains $\Mo_\ell(S^n)$ as an open dense subset
\cite[Corollary~4.11]{PartI}, and the Gauss map identifies it with the space
$\Po_\ell(s)=\Polys/\Isom(\HH^{n+1})$ of $s$-equilateral $\ell$-gons
\cite[Theorem~4.10]{PartI}, which in turn is homeomorphic via the Gram matrix of the
sides to the compact set $\Gmod\subset\mathrm{Sym}_\ell$ \cite[Theorem~4.4]{PartI}. For
$\ell$ even and a splitting $[\ell]=A\sqcup B$ with $|A|=|B|=\ell/2$, $X_{\{A,B\}}$ is
the single orbit of the configurations $\pAB$ with the points of $A$ at one point and
those of $B$ at another, and we put $\pt_{A|B}:=q(X_{\{A,B\}})$.

For $\pl\in X$ the \emph{coincidence pattern} $\pi(\pl)$ is the partition of $[\ell]$
into the fibres of $i\mapsto p_i$, and $X_\pi$ is the $\Mob(S^n)$-invariant set of
configurations with pattern $\pi$; we call $\pi$ \emph{stable} if every block has
fewer than $\ell/2$ elements. By \cite[Proposition~4.14]{PartI}, as a set
\[
\MopsL\;=\;\bigsqcup_{\pi\ \mathrm{stable}}X_\pi/\Mob(S^n)\ \ \sqcup
\bigsqcup_{\{A,B\}}\{\pt_{A|B}\},
\]
the second union over the splittings with $|A|=|B|=\ell/2$; each $X_\pi/\Mob(S^n)$,
with the quotient topology of $X_\pi$, is homeomorphic to $\Mo_{|\pi|}(S^n)$ via the
configuration of the distinct values, and $q^{-1}(\pt_{A|B})$ consists of the
semistable configurations in which the points of $A$, or those of $B$, coincide. The
polystable quotient thus records which punctures coincided and where the distinct
values are, and nothing else.

\subsection{Three compactifications}\label{ss:three}
Three compactifications of $\Mo_\ell(S^n)$ appear in this paper.
\begin{itemize}
\item[(AS)] The bubble-tree compactification in its real form, Definition~\ref{D:AS}:
for each cluster one records the {\em screen}, the limit of the cluster rescaled,
an element of $C_{|A|}(\R^n)$, i.e.\ modulo translations and dilations --- the
rotation relative to the ambient tangent space being retained. Its boundary
hypersurfaces are indexed by splits $\{A,A^c\}$ with $2\le|A|\le\ell-2$; they have
codimension one, and $\overline\Mo^{\,\mathrm{AS}}_\ell$ is compact, and an
orbifold with corners away from points with positive-dimensional stabilizers.
\item[(DM)] The same, with each screen taken modulo the full group of similarities,
i.e.\ modulo the automorphisms of the bubble fixing its node. We define it as a
topological quotient of (AS) in Proposition~\ref{P:three}(1), and do not claim that it
is Hausdorff. The name is by analogy: for $n=2$ its strata are those of the
Deligne--Mumford--Knudsen space $\overline{\mathcal M}_{0,\ell}$ \cite{DeMu,Kn} up to one
$\Z/2$ per bubble, acting independently; the resulting object is a further quotient of
$\Mbar\ell(\C)/\conj$, and differs from it from $\ell=6$ on, where a $3{+}3$ stratum is
$M_{0,4}\times M_{0,4}$ modulo $\Z/2\times\Z/2$ rather than modulo the diagonal.
\item[(PS)] The polystable quotient $\MopsL=\Po_\ell(s)$ of Section~\ref{ss:psb}, the
Kapovich--Leeb--Millson semistable quotient \cite{KLM} for equal weights, which for
$n=2$ is the Deligne--Mostow quotient $(\mathbf P^1)^\ell/\!\!/\mathrm{SL}_2$ with equal
weights \cite{DM}; we also call it the GIT quotient.
\end{itemize}

For a boundary point of $\overline\Mo^{\,\mathrm{AS}}_\ell$ with tree $T$, call an
internal vertex $v$ of $T$ a \emph{centroid} if each branch at $v$ carries at most
$\ell/2$ leaves.

\begin{Lemma}\label{L:centroid}
Every tree with $\ell$ labelled leaves has exactly one centroid, or exactly two, joined
by an edge $e_A$ with $|A|=\ell/2$. The root is a centroid if and only if every cluster
of the family has at most $\ell/2$ points. Under the collapse of an edge, a centroid
maps to a centroid.
\end{Lemma}

\begin{proof}
Since the branch sizes at a vertex sum to $\ell$, at most one branch at $v$ carries more
than $\ell/2$ leaves. If there is one, $b$ say, let $w$ be the neighbour of $v$ across
$b$: at $w$ the branch containing $v$ has $\ell-|B_b|<\ell/2$ leaves and the other
branches partition $B_b$, so the largest branch size at $w$ is smaller than at $v$.
Walking along large branches therefore ends at a centroid. If $v\ne v'$ are centroids,
the branch at $v$ containing $v'$ and the branch at $v'$ containing $v$ have at most
$\ell/2$ leaves each and together contain every leaf, so each has exactly $\ell/2$ and
no leaf hangs off an interior vertex of the path from $v$ to $v'$; as interior vertices
have valence $\ge3$ and every branch carries a leaf, the path is a single edge $e_A$
with $|A|=\ell/2$. The second statement is the definition, the branches at the root
being the maximal clusters of the family and the leaves attached to it. For the last, collapsing
an edge at a centroid $v$ merges $v$ with a neighbour $w$; the branches at the merged
vertex are branches at $v$ or branches at $w$ not containing $v$, and the latter are
contained in the branch at $v$ containing $w$, hence carry at most $\ell/2$ leaves.
\end{proof}

\begin{Proposition}\label{P:three}
\begin{enumerate}[label=\textup{(\arabic*)},leftmargin=*]
\item Let $\overline\Mo^{\,\mathrm{DM}}_\ell$ be the quotient of
$\overline\Mo^{\,\mathrm{AS}}_\ell$ by the equivalence relation whose classes are the
points of $\Mo_\ell(S^n)$ and the orbits of $\mathrm O(n)^V$ on each open boundary
stratum, the screens of a boundary point rotated independently. With the quotient
topology it is compact, the quotient map is continuous and restricts over each open
boundary stratum to the quotient by $\mathrm O(n)^E$, one factor per internal edge
(a principal bundle over the locus of screens without symmetry), and the boundary
value $\eta^\ren$ of Theorem~\ref{T:bv} is constant on its fibres.
\item For a boundary point $x$ with tree $T$ and a centroid $v_*$ of $T$, let
$\sigma(x,v_*)\in X$ be the configuration obtained from the screen at $v_*$ by placing
the leaves of each finite branch at the screen point of that branch and the leaves of
the branch at infinity at $\psi(\infty)$. It is semistable, and $c(x):=q(\sigma(x,v_*))$
does not depend on the choice of centroid. Together with the identity on
$\Mo_\ell(S^n)$ this defines a continuous surjection
\[
c\colon\overline\Mo^{\,\mathrm{AS}}_\ell\longrightarrow\MopsL=\Po_\ell(s)
\]
which factors through $\overline\Mo^{\,\mathrm{DM}}_\ell$. It maps the open stratum of
$T$ onto the stratum $X_\pi/\Mob$ of Section~\ref{ss:psb}, where $\pi$ is the
partition of $[\ell]$ into the branches at the centroid, if $\pi$ is stable, and to the
point $\pt_{A|A^c}$ if $T$ has an edge $e_A$ with $|A|=\ell/2$.
\end{enumerate}
\end{Proposition}

\begin{proof}
(1) The quotient of a compact space is compact and the quotient map is continuous. Over
an open boundary stratum $\prod_vC_{\ell_v-1}(\R^n)/\mathrm O(n)$ the classes are the
orbits of $\mathrm O(n)^V/\mathrm O(n)\cong\mathrm O(n)^E$, and the diagonal $\mathrm O(n)$
already divided out in Definition~\ref{D:AS} is one of the factors. Constancy of
$\eta^\ren$ on the fibres is Theorem~\ref{T:bv}(c).

(2) The multiplicities of $\sigma(x,v_*)$ are the branch sizes at $v_*$, which are at
most $\ell/2$, so it is semistable. If $T$ has two centroids they are joined by an edge
$e_A$ with $|A|=\ell/2$, and the two configurations are strictly semistable with
cluster $A$, resp.\ $A^c$; both are sent to $\pt_{A|A^c}$ by \cite[Lemma~4.6(3)]{PartI}.
Rotating the screen at $v_*$ replaces $\sigma(x,v_*)$ by a M\"obius-equivalent
configuration, and the other screens do not enter; so $c$ factors through
$\overline\Mo^{\,\mathrm{DM}}_\ell$. The image of the stratum of $T$ is as stated
because $\sigma(x,v_*)$ has pattern $\pi$ and, as $x$ runs over the stratum, the screen
at $v_*$ runs over all configurations of its branches modulo similarity, which is
$X_\pi/\Mob$ by \cite[Proposition~4.14(2a)]{PartI}; when $\pi$ is not stable the image
is $\pt_{A|A^c}$ by \cite[Lemma~4.6(3)]{PartI} again. Surjectivity: every stable
partition $\pi$ is the branch partition at the centroid of the tree whose family is
$\{B\in\pi: |B|\ge2,\ \ell\notin B\}\cup\{B_0^c\}$, $B_0$ the block containing $\ell$
(omit $B_0^c$ if $|B_0|=1$), and every $\pt_{A|A^c}$ is hit by the tree with the single
edge $e_A$.

Continuity. Let $x$ be a boundary point with family $\mathcal A$, tree $T$ and centroid
$v_*$, let $\mathcal U_{\mathcal A}$ be its corner chart (Definition~\ref{D:AS}), and
let $y_{v_*}(\sigma,t)$ be the node position of $v_*$ (with $y_r=0$). For
$(\sigma,t)\in\mathcal U_{\mathcal A}$ with all $t_A>0$ put
\[
z_i(\sigma,t):=\frac{y_i(\sigma,t)-y_{v_*}(\sigma,t)}{\epsilon_{\le v_*}}\quad(i<\ell),
\qquad z_\ell:=\infty,
\]
the configuration \eqref{eq:recon} recentred at the node of $v_*$ and rescaled to the
size of its screen. It is similar to $y(\sigma,t)$, so
$q(\psi(z(\sigma,t)))=[y(\sigma,t)]$. If $i\in A_{v_*}$, then by \eqref{eq:recon}
\[
z_i=\sum_{s:\,A_{v_s}\subseteq A_{v_*}}\Big(\prod_{A\supseteq A_{v_s},\ A\not\supseteq A_{v_*}}t_A\Big)\,
a^{(v_s)}_{b_{v_s}(i)}
=a^{(v_*)}_{b_{v_*}(i)}+\big(\text{terms divisible by some }t_A\big),
\]
a polynomial in $t$, which extends smoothly to $\mathcal U_{\mathcal A}$. If
$i\notin A_{v_*}$, let $u$ be the vertex at which $i$ and the vertex $v_*$ lie in
different branches. As in the proof of Lemma~\ref{L:exp},
\[
z_i=\frac{a^{(u)}_{b_u(i)}-a^{(u)}_{b_u(v_*)}+r_i(\sigma,t)}{\mu_i(t)},\qquad
\mu_i(t):=\prod_{A\supseteq A_{v_*},\ A\not\supseteq A_u}t_A ,
\]
where $r_i$ is smooth and vanishes at $t=0$, and the numerator does not vanish on
$\mathcal U_{\mathcal A}$. Since $(w,\mu)\mapsto\psi(w/\mu)$ is continuous on
$\{w\ne0,\ \mu\ge0\}$, with value $\psi(\infty)$ at $\mu=0$, the map $\psi\circ z$
extends continuously to $\mathcal U_{\mathcal A}$ as a map to $X$. At $t=0$ its value
is $\sigma(x,v_*)$: $z_i=a^{(v_*)}_b$ for $i$ in a finite branch $b$ at $v_*$, and
$\psi(z_i)=\psi(\infty)$ for $i$ in the branch at infinity. At a point $x'$ of a face
of the chart it is, up to similarity, $\sigma(x',v'_*)$, where $v'_*$ is the image of
$v_*$ in the tree of $x'$, a centroid of that tree by Lemma~\ref{L:centroid}. All
these values lie in the open set $\Xss$, and $c=q\circ\psi\circ z$ on
$\mathcal U_{\mathcal A}$. As $q$ is continuous, so is $c$ on the chart, and
continuity on the open interior is clear.
\end{proof}

\begin{Remark}\label{R:ASDM}
For $n=2$ the automorphism group of $\mathbf P^1$ fixing the node is the complex affine
group, the group of orientation-preserving similarities, so the DM bubble is the
screen modulo similarity and the AS face is a circle bundle over the DM divisor: the
classical relation between the real oriented blow-up and the ordinary one. In general
the AS faces have real codimension one and the DM strata real codimension
$\dim\mathrm O(n)+1$; the two spaces have the same tree-indexed strata and the same
screens up to the group used, but they are not the same. The mass matrix is built from
distances and does not see the relative rotation of a bubble against its parent
(Theorem~\ref{T:bv}(c)): for $(\ell,k,n)=(6,3,3)$ the face $\partial_A$ has dimension
$7$, its image under $\eta^\ren$ has dimension $\dim\mathcal H_4+\dim\mathcal H_4=4$,
and the $3$-dimensional fibres are the orbits of $SO(3)$. So the mass data
depend only on the image in the DM quotient; the rotation is needed to glue metrics back together, not to describe their
limits.
\end{Remark}

\subsection{Why not the polystable quotient}
By Theorem~\ref{T:recentred}, the component of $\eta^\ren$ at a vertex $v$ under
the gluing isomorphism of Theorem~\ref{T:gluing} is
$\rho_v(\eta^\ren)=\Xi_{\ell_v}(\hat\sigma_v)$, the mass matrix of the screen at $v$
with its node adjoined. It vanishes if $\ell_v=3$, since $\mathcal H_3=0$. If
$\ell_v\ge4$ it determines the screen at $v$ up to similarity, because $\Xi_{\ell_v}$
is injective \cite[Theorem~3.6]{PartI} and two configurations
$\{\psi(a_b)\}\cup\{\psi(\infty)\}$ are M\"obius equivalent, with $\psi(\infty)$
fixed, exactly when the screens are similar.

\begin{Proposition}\label{P:notPS}
Let $c$ be the contraction of Proposition~\ref{P:three}(2).
\begin{enumerate}[label=\textup{(\arabic*)},leftmargin=*]
\item If $\ell\le5$, every vertex $v$ of a bubble tree with $\ell_v\ge4$ is a
centroid. If $x,x'$ lie in one open boundary stratum and $c(x)=c(x')$, then
$\rho_v(\eta^\ren(x))=\rho_v(\eta^\ren(x'))$ for every internal vertex $v$.
\item If $\ell\ge6$, there are points $x\ne x'$ of one open boundary stratum with
$c(x)=c(x')$ and $\rho_v(\eta^\ren(x))\ne\rho_v(\eta^\ren(x'))$ for some vertex $v$.
\end{enumerate}
\end{Proposition}

\begin{proof}
(1) For $\ell=4$ every bubble tree has one internal edge and two vertices of
valence $3$, so all $\rho_v(\eta^\ren)$ vanish. For $\ell=5$, a tree with two internal
edges has three vertices of valence $3$; a tree with one internal edge has a vertex of
valence $3$ and a vertex of valence $4$ whose branches carry $2,1,1,1$ leaves, so the
latter is the centroid, which is unique because $\ell$ is odd. Thus only
$\rho_{v_*}(\eta^\ren)$ can be nonzero, for the centroid $v_*$, and only when
$\ell_{v_*}=4$. In that case $\pi$, the partition of $[\ell]$ into the branches at
$v_*$, is stable, and $c(x)=q(\sigma(x,v_*))\in X_\pi/\Mob$. Under the homeomorphism
$X_\pi/\Mob\cong\Mo_{|\pi|}(S^n)$ of \cite[Proposition~4.14(2a)]{PartI}, which records
the distinct values labelled by the blocks, $c(x)$ is the class of
$\hat\sigma_{v_*}$ with its points labelled by the branches. Hence $c(x)$ determines
$\Xi_{\ell_{v_*}}(\hat\sigma_{v_*})=\rho_{v_*}(\eta^\ren(x))$.

(2) Let $\ell\ge6$ be even and $A=\{1,\dots,\ell/2\}$. The tree of $\{A\}$ has two
vertices, both of valence $\ell/2+1\ge4$, and both are centroids, so $c$ maps the
whole open stratum to the point $\pt_{A|A^c}$ (Proposition~\ref{P:three}(2)). The
screen at the vertex $v_A$ ranges over $C_{\ell/2}(\R^n)$ independently of the root
screen, and $\Mo_{\ell/2+1}(S^n)$ has more than one point, so
$\rho_{v_A}(\eta^\ren)$ takes more than one value on the stratum.

Let $\ell\ge7$ be odd and $A=\{1,2,3\}$. The vertex $v_A$ has the three leaves of
$A$ and the parent branch, which carries $\ell-3>\ell/2$ leaves, so $\ell_{v_A}=4$
and $v_A$ is not a centroid. The root has the branch $A$ with $3\le\ell/2$ leaves and
$\ell-3$ single leaves, so the root is the centroid, and $c(x)$ depends only on the
root screen. Fixing the root screen and varying the screen at $v_A$ in
$C_3(\R^n)$ keeps $c(x)$ fixed and changes
$\rho_{v_A}(\eta^\ren)=\Xi_4(\hat\sigma_{v_A})$.
\end{proof}

Proposition~\ref{P:notPS} proves Theorem~\ref{thm:mainF}(iii). In geometric terms, for
$\ell\ge6$ the contraction to $\MopsL$ forgets the screens of the bubbles away from the
centroid, and with them the mass data of the corresponding pieces of the limit
(Theorem~\ref{T:geom}, Remark~\ref{R:two}).

\begin{Remark}\label{R:cautions}
There are two cautions. First, it is the fixed-parameter
Brill--Lindquist data, not the canonical metrics, for which the
rescaled metrics converge to a bubble with the node as an extra end:
with the canonical normalization the weights $c_i$ scale with the
degeneration, the node term is subdominant by the factor
$\epsilon^{\kp(\ell-k-1)/(\ell-2)}$ (Lemma~\ref{L:dom}), and the node
closes up. Second, the geometric limit at a vertex without punctures
depends on the path to the corner (Remark~\ref{R:leafless}); the
object attached to a boundary point is $\eta^\ren$, and the limit
metrics are read off it where they are canonical.
\end{Remark}
\begin{Remark}\label{R:snow}
  Topology of the moduli spaces $\mathcal{M}_\ell(S^n)$ and of their
compactifications is addressed in \cite{BS1,BS2}.
\end{Remark}
\section{Questions}\label{s:questions}

\begin{Question}\label{Q:leafless}
At a vertex of the bubble tree without punctures the geometric limit
depends on the path to the corner (Remark~\ref{R:leafless}), while
$\eta^\ren$ does not. Is there a canonical object attached to such a
vertex --- a family of limits parametrized by the ratios $N_b/N_{b'}$
of its children --- and does it have a natural place in the
compactification?
\end{Question}

\begin{Question}\label{Q:DM}
Is the space $\overline\Mo^{\,\mathrm{DM}}_\ell$ of
Proposition~\ref{P:three}(1) Hausdorff, and does the mass matrix,
which factors through it, extend to a continuous map from it to a
natural target built from the arrangement $\{W_{\mathcal
  A}/V_{\mathcal A}\}$?
\end{Question}

\begin{Question}\label{Q:operad}
The gluing \eqref{eq:gamma} is symmetric under $A\leftrightarrow A^c$
($\E(A)=\E(A^c)$, $W_A=W_{A^c}$, and $\rho_v$ treats the branch at
infinity like any other), and it is compatible with collapsing an edge,
which is what Lemma~\ref{L:rigid} uses: the shape of a cyclic operad,
with the passage to reduced families as the choice of a root. The
would-be composition $\mathcal H_{k+1}\oplus\mathcal
H_{\ell-k+1}\oplus\R\to W_A\subset\Hh$ is, however, not a map to $\Hh$
with the right equivariance. Is there a precise operadic statement, in
which the ``one real parameter per composition'' is built into the
objects, that has Theorem~\ref{T:gluing} as its associativity?
\end{Question}

\end{document}